\pdfoutput=1
\documentclass[11pt]{article}

\usepackage{etex}
\reserveinserts{28}

\usepackage{amsmath,amssymb,amsfonts}
\usepackage[noconfig]{refstyle}
\usepackage[dvipsnames]{xcolor}
\usepackage{array,tabularx,booktabs,geometry,graphicx,longtable}
\usepackage[bf]{caption}
\usepackage{tikz}
\usetikzlibrary{shapes,arrows}
\usepackage{graphicx,color}
\usepackage{cite}

\usepackage[all,cmtip]{xy}
\usetikzlibrary{matrix, arrows}

\makeatletter
\newenvironment{tablehere}
  {\def\@captype{table}}
  {}

\makeatother

\numberwithin{equation}{section}
\newcommand{\eref}[1]{(\ref{#1})}

\newcommand{\eeq}{\end{equation}}
\newcommand{\beq}{\begin{equation}}
\newcommand{\ba}{\begin{array}}
\newcommand{\ea}{\end{array}}

\newcommand{\nn}{\nonumber}

\newcommand{\cF}{{\cal F}}

\newcommand{\rk}{{\rm rk}}

\newcommand{\cO}{{\cal O}}
\newcommand{\cA}{{\cal A}}
\newcommand{\IP}{\mathbb P}

\newcommand{\be}{\begin{equation}}
\newcommand{\ee}{\end{equation}}
\newcommand{\bea}{\begin{equation}\begin{aligned}}	
\newcommand{\eea}{\end{aligned}\end{equation}}		

\newcommand{\iddots}{\mathinner{\mkern2mu\raise1pt\hbox{.}\mkern2mu \raise4pt\hbox{.}\mkern2mu\raise7pt\hbox{.}\mkern1mu}}

\providecommand{\id}{\leavevmode\hbox{\small$\mathrm{1}$\kern-3.8pt\normalsize$\mathrm{1}$}}
\def\fnote#1#2{\begingroup\def\thefootnote{#1}\footnote{#2}
     \addtocounter{footnote}{-1}\endgroup}

\begin{document}

\vspace{1cm}

\title{
       \vskip 40pt
       {\huge \bf New Evidence for ${\bf (0,2)}$ Target Space Duality }}

\vspace{2cm}

\author{Lara B. Anderson${}^{1}$ and He Feng${}^{1}$}
\date{}
\maketitle
\begin{center} {\small ${}^1${\it The Department of Physics, Robeson Hall, Virginia Tech, Blacksburg, VA 24061, USA}}\\
\fnote{}{lara.anderson@vt.edu, fenghe@vt.edu}
\end{center}

\begin{abstract}
\noindent
In the context of $(0,2)$ gauged linear sigma models, we explore chains of perturbatively dual heterotic string compactifications. The notion of target space duality originates in non-geometric phases and can be used to generate distinct GLSMs with shared geometric phases leading to apparently identical target space theories. To date, this duality has largely been studied at the level of counting states in the effective theories. We extend this analysis to the effective potential and loci of enhanced symmetry in dual theories. By engineering vector bundles with non-trivial constraints arising from slope-stability (i.e. D-terms) and holomorphy (i.e. F-terms) the detailed structure of the vacuum space of the dual theories can be explored. Our results give new evidence that GLSM target space duality may provide important hints towards a more complete understanding of $(0,2)$ string dualities.
\end{abstract}

\thispagestyle{empty}
\setcounter{page}{0}
\newpage

\tableofcontents

\section{Introduction} \label{intro}

The goal of this work is to further explore heterotic target space duality, with the hope of moving towards a more fully developed notion of dualities involving the $(0,2)$ heterotic string. At present, many heterotic string dualities -- including any complete notion of $(0,2)$ mirror symmetry \cite{Blumenhagen:1996vu,Adams:2003zy,Melnikov:2010sa,Melnikov:2012hk} -- remain relatively unexplored compared to their counterparts in $(2,2)$ string theories. Heterotic mirror symmetry, for instance, must necessarily exhibit an even more intricate structure than $(2,2)$ (Calabi-Yau) Mirror symmetry which interchanges a pair of manifolds (and has sparked a rich interchange of ideas with mathematics, including the field of homological mirror symmetry\footnote{See \cite{homological_mirror} for a review.}.). In the case of heterotic $(0,2)$ theories it has been established that duality between $(0,2)$ sigma models and their supergravity limits can involve more complicated relationships between the geometry of Calabi-Yau manifolds and vector bundles over them. In many cases, rather than mirror pairs, heterotic dualities could involve whole chains of compactification geometries leading to the same effective theory. One concrete step towards understanding such correspondences has been taken in the direction of ``Target Space Duality" originating in $(0,2)$ Gauged Linear Sigma Models in \cite{Distler:1995bc} and further studied in \cite{Blumenhagen:1997vt,Blumenhagen:1997cn,Blumenhagen:2011sq,Rahn:2011jw}.

Within $(0,2)$ GLSMs, the target space duality of \cite{Distler:1995bc} is manifested by the existence of a shared sub-locus in the moduli space of the paired theories. More precisely each of the ``dual" $(0,2)$ GLSMs shares a non-geometric phase (commonly a Landau-Ginzburg phase). In this web of theories, the geometric phases -- consisting of a Calabi-Yau threefold, $X$, and a holomorphic bundle $\pi: V \to X$ over it -- are connected through a wealth of geometric correspondences, including geometric (i.e. conifold) transitions between the base CY geometries $(X, \widetilde{X})$. However, the relationship between the vector bundles $(V, \widetilde{V})$ is more subtle and no systematic geometric characterization of the full paired geometries, $(X,V), (\widetilde{X},\widetilde{V})$, has yet been provided. However, one essential feature of this target space duality is that the correspondence between two different geometric phases preserves the net number of moduli of the theory:
\beq\label{mod_count}
h^{1,1}(X)+h^{2,1}(X)+h^1(X, End_0(V))=h^{1,1}(\widetilde{X})+h^{2,1}(\widetilde{X})+h^1(X, End_0(\widetilde{V)})~.
\eeq
Unlike the simple interchange of $h^{1,1} \leftrightarrow h^{2,1}$ of CY mirror symmetry, the distribution of the total degrees of freedom given above between the moduli of the base CY manifold and those of the vector bundle can change dramatically across a $(0,2)$ dual pair.

In recent work \cite{Blumenhagen:2011sq,Rahn:2011jw}, systematic landscape studies of $(0,2)$ target space duals were undertaken. In particular, the connected geometric phases were explored in detail and it was found that the gauge symmetry, moduli and charged matter spectrum were generically identical in the geometric phases. This last was explored in the context of vector bundles with $SU(n)$ structure group, for which the charged matter is counted by 
\beq\label{chargmat}
h^i(X, \wedge^k V)=h^i(\widetilde{X}, \wedge^k \widetilde{V})
\eeq
for $k=0, \ldots , n-1$. In the extensive studies carried out in \cite{Blumenhagen:2011sq,Rahn:2011jw}, the complete charged and singlet particle spectra were found to correspond exactly in almost all $(0,2)$ target space dual pairs\footnote{In a handful of cases ($< 2\%$ of roughly $83,000$ models analyzed) in the landscape scan, these spectra did not fully agree. The authors conjecture that this is due to the fact that several technical simplifications/assumptions had been made in the scan. See \cite{Blumenhagen:2011sq} for details.}.

Since the first observation of target space duality, it has been an open question whether or not the relationship described above corresponds to a true isomorphism of the target space theories/GLSMs (or indeed full non-linear sigma models) or rather represents a transition -- i.e. a shared locus in moduli space for two distinct theories (perhaps similar to conifold transitions in type II theories \cite{Strominger:1995cz})? To establish which of these possibilities is correct requires a more a detailed study of the ``dual" theories and a clear map between them -- which could determine whether they can be identified everywhere in their moduli spaces or only at some sublocus. 

In the present work, our goal is to move towards this goal and test the duality further. Thus far, the study of such pairs has focused primarily on the degrees of freedom (i.e. particle spectra) of the theories as described above. In the present work, we hope to extend this further to explore the \emph{vacuum spaces} of target space dual pairs. This is in general a non-trivial task, since for $(0,2)$ theories the full form of the matter field K\"ahler potential and hence, the full scalar potential is  largely unknown. However, recent work has made progress in explicitly connecting the underlying geometry of $(X, V)$ with the form of the D-term and F-term contributions to the $4$-dimensional ${\cal N}=1$ potential \cite{Anderson:2009nt, Anderson:2009sw,Anderson:2010ty,Anderson:2013qca, Anderson:2011ty}. Briefly, explicit obstructions to the slope-stability of the bundle correspond to D-terms (associated to Green-Schwarz anomalous $U(1)$ symmetries \cite{Lukas:1999nh,Blumenhagen:2005ga,Sharpe:1998zu,Anderson:2009sw,Anderson:2009nt}) and holomorphy conditions on $V$ arise as F-terms in the ${\cal N}=1$ effective theory \cite{Anderson:2011ty,Anderson:2013qca}.

With this link between bundle geometry and the effective potential in mind, the goals of the present work are:
\begin{itemize}
\item Generate examples of target space dual pairs in which the vector bundles lead to non-trivial D-terms and/or F-terms in their effective potentials (i.e. bundles that are not everywhere slope stable or holomorphic in the moduli space of the base CY geometry). 
\item To build examples of such bundles, we must extend the class of monad bundles (see \cite{horrocks, beilinson}) that have been the primary focus of $(0,2)$ GLSMs to date. Thus far, the bundles appearing in the majority of the literature have been of the class of so-called ``positive" monad bundles. These consist of bundles defined via a complex built of \emph{ample} line bundles
\beq
0 \to \bigoplus_{k} {\cal O}({\bf a}_k) \stackrel{E}{\longrightarrow} \bigoplus_i {\cal O}({\bf b}_i) \stackrel{F}{\longrightarrow} \bigoplus_j {\cal O}({\bf c}_j) \to 0\label{first_mon}
\eeq
where $f \cdot g=0$ and $a_k,b_i, c_j \geq 0$ for all $i, j, k$. One motivation for focusing on this simple class of bundles is that they are frequently (though not always) stable with respect to the entire K\"ahler cone of $X$ \cite{Anderson:2008ex}. In this work, we will extend this class to involve some negative degree entries in \eref{first_mon} and demonstrate that although the bundles are not everywhere stable in K\"ahler moduli space, none-the-less the associated monad bundles correspond to good $(0,2)$ GLSMs. These more general monad bundles have already been considered in a variety of contexts in heterotic vacua and the extension of the class may be an important step towards applying sigma model tools to bundles of interest in string phenomenology \cite{Anderson:2009mh,Anderson:2008uw,Anderson:2007nc,Anderson:2008ex}.
\item At special points in the vacuum space of the theories, enhanced symmetries can arise. These include enhancements of what would in the full sigma model be worldsheet $(0,2)$ to $(2,2)$ supersymmetry (when $V$ is deformable into the holomorphic tangent bundle $TX$) and the loci in moduli space corresponding to ``stability walls" in which the bundle $V$ becomes reducible ($V \to V_1 \oplus V_2 \oplus \ldots$) and enhanced Green-Schwarz massive $U(1)$ symmetries are present in the effective theory. In the latter case, such loci can lead to complicated branch structure in the vacuum space of the theory in which many different bundles over the same $X$ are connected \cite{Anderson:2010ty}.

 It is our goal in this context to explore the existence of such loci on each side of a target space dual pair and to explore whether a) such enhancement points always correspond across the target space dual pair? and b) whether the vaccum branch structure is identical in both halves of the pair?
\item Finally, to build example geometries $(X, V)$ with non-trivial D-term and F-term potentials that fit well into GLSMs a number of new technical tools will be needed. In particular, it will be necessary to describe such bundles and their possible deformations via monads. A systematic study of such bundles involves generalizations of the target space duality procedure and developing multiple ways of describing the same bundle to make branch structure in the effective theory evident.
\end{itemize} 

In the following sections we will approach the problems/goals outlined above. It should be noted that this list raises a number of interesting questions in the GLSMs themselves -- most prominently how the large volume constraints of bundle holomorphy and stability are realized in the GLSM, and whether the corresponding D- and F-term lifting of moduli in the ${\cal N}=1$ effective theory extends to other non-geometric phases of the theory? For now, we will leave these important questions unanswered and focus primarily on the structure of the geometric phases and target space theories themselves. It is our goal to test target space duality itself via the geometry and potentials described above.

In the first two sections, the two primary starting points of this work are reviewed. In Section \ref{Sec:Review} this is the $(0,2)$ target space duality procedure as developed in \cite{Distler:1995bc,Blumenhagen:1997vt,Blumenhagen:1997cn,Blumenhagen:2011sq,Rahn:2011jw} and in Section \ref{potential_hym}, the correspondence between bundle stability/holomorphy and non-trivial D-terms and F-terms in the effective theory as studied in \cite{Sharpe:1998zu,Anderson:2009nt, Anderson:2009sw,Anderson:2010ty,Anderson:2013qca, Anderson:2011ty}. In Section \ref{dterms} we provide an example of a bundle with non-trivial D-term constraints and track those conditions through the possible target space dual chains of geometries. In particular, this analysis allows for the exploration of the branch structure arising in the vacuum space in the neighborhood of a stability wall on both sides of a target space dual pair. In addition, examples are given of possible perturbative/non-perturbative dual pairs and examples in which disconnected components of the moduli space are reproduced in the dual theory. In Section \ref{Fterm_sec} this procedure is repeated for a bundle with a non-trivial F-term constraint. To probe the target space dual correspondence at yet another point of enhanced symmetry, in Section \ref{tan_defs} dual pairs are constructed in which the starting bundle is constructed as a deformation of the holomorphic tangent bundle (i.e. the $(0,2)$ theory contains a $(2,2)$ locus). Finally, our results are summarized and future directions outlined in Section \ref{conclusions}. Section \ref{AppendixA} contains details of bundle stability, while Appendix \ref{appendixB} provides a comprehensive list of target space dual chains for several examples.

To begin, in the following two sections, we review briefly the necessary ingredients for this work - namely a) The basic ideas of target space duality in heterotic $(0,2)$ GLSMs and b)  the relationship between the Hermitian Yang-Mills equations and non-trivial D-/F-terms in the effective theory.

\section{A Brief Review of Target Space Duality}\label{Sec:Review}

The notion of $(0,2)$ target space duality is well-studied in the literature \cite{Distler:1995bc,Blumenhagen:1997vt,Blumenhagen:1997cn,Blumenhagen:2011sq,Rahn:2011jw} (see \cite{Blumenhagen:2011sq} for a thorough review) and in this section we briefly summarize the basic ingredients as well as clarify the simple class of geometries we will consider in order to explore the vacuum structure of the ${\cal N}=1$ effective theories.

\subsection{${\bf (0,2)}$ Gauged Linear Sigma Models}\label{sec:GLSM}
Gauged linear sigma models (GLSMs) are massive $2$-dimensional theories which can flow in the infrared (under suitable conditions) to true superconformal theories (i.e. string sigma models). The famous ``phase" structure of GLSMs \cite{Witten:1993yc} links the vacuum structure of the $2$-dimensional theory to the geometric data of toric geometry and string compactifications. The freedom to vary a Fayet-Illiopolos parameter leads to a variety of distinct solutions including non-linear sigma models (and their associated target space geometries -- include CY manifolds), Landau-Ginzburg orbifolds, and a rich variety of hybrid models.

In this work we will restrict our consideration to the simplest, Abelian class of GLSMs. The field content will consist of multiple $U(1)$ gauge fields $A^{(\alpha)}$ with $\alpha = 1,...,r$. Following standard notation in the literature, we will label two sets of chiral superfields as $\{X_i|i=1,...,d\}$ with $U(1)$ charges $Q_i^{(\alpha)}$, and $\{P_l|l=1,...,\gamma\}$ with $U(1)$ charges $-M_l^{(\alpha)}$. The essential field content will also include two sets of Fermi superfields: $\{\Lambda^a|a=1,...,\delta\}$ with charges $N_a^{(\alpha)}$, and $\{\Gamma^j|j=1,...,c\}$ with charges $-S_j^{(\alpha)}$.

The $U(1)$ charges appearing in the GLSM are frequently taken to satisfy positivity conditions that are compatible with realizing Calabi-Yau manifolds and stable, holomorphic vector bundles over them in some geometric phase. In this work, we will take the charges $Q_i^{(\alpha)} \geq 0$ and require that for each $i$, there exists at least one $r$ such that $Q_i^{(\alpha)}>0$. This assumption of (semi-) positivity will hold in addition for the charges $S_j^{(\alpha)}$ and $M_l^{(\alpha)}$. However, in some cases we will consider solutions in which charges $N_a^{(\alpha)}$ may be negative (the suitability of such special cases will be further discussed in Sections \ref{dterms} and \ref{Fterm_sec}).

The key field content and charges of the GLSM can be concisely summarized by a ``charge matrix" of the form:

\begin{equation}
\begin{aligned}
&\begin{array}{|c||c|}
\hline
x_i & \Gamma^j \\
\noalign{\hrule height 1pt}
\begin{array}{cccc}
 Q^{(1)}_1 & Q^{(1)}_2 & \ldots & Q^{(1)}_d\\
 Q^{(2)}_1 & Q^{(2)}_2 & \ldots & Q^{(2)}_d\\
 \vdots  &    \vdots  &  \ddots  &  \vdots  \\  
 Q^{(r)}_1 & Q^{(r)}_2 & \ldots & Q^{(r)}_d
\end{array}
&
\begin{array}{cccc}
-S^{(1)}_1 & -S^{(1)}_2 & \ldots & S^{(1)}_c\\
-S^{(2)}_1 & -S^{(2)}_2 & \ldots & S^{(2)}_c\\
 \vdots  &    \vdots  &  \ddots  &  \vdots  \\  
-S^{(r)}_1 & -S^{(r)}_2 & \ldots & S^{(r)}_c\\
\end{array}\\
\hline
\end{array}
\\[0.1cm]
&\begin{array}{|c||c|}
\hline
\Lambda^a & p_l \\
\noalign{\hrule height 1pt}
\begin{array}{cccc}
 N^{(1)}_1 & N^{(1)}_2 & \ldots & N^{(1)}_\delta\\
 N^{(2)}_1 & N^{(2)}_2 & \ldots & N^{(2)}_\delta\\
 \vdots  &    \vdots  &  \ddots  &  \vdots  \\  
N^{(r)}_1 & N^{(r)}_2 & \ldots & N^{(r)}_\delta
\end{array}
&
\begin{array}{cccc}
-M^{(1)}_1 & -M^{(1)}_2 & \ldots & -M^{(1)}_\gamma\\
-M^{(2)}_1 & -M^{(2)}_2 & \ldots & -M^{(2)}_\gamma\\
 \vdots  &    \vdots  &  \ddots  &  \vdots  \\ 
-M^{(r)}_1 & -M^{(r)}_2 & \ldots & -M^{(r)}_\gamma\\
\end{array}\\
\hline
\end{array}
\end{aligned}\label{charge_matrix}
\end{equation}

Familiar GLSM anomaly cancellation conditions (including gauge and gravitational anomalies) impose the following linear and quadratic constraints on the $U(1)$ charges to hold for all $\alpha, \beta = 1, ..., r$:
\begin{eqnarray}\nonumber \label{anomalies}
\sum_{a=1}^\delta N_a^{(\alpha)} = \sum_{l=1}^\gamma M_l^{(\alpha)} \qquad&& \sum_{i=1}^d Q_i^{(\alpha)} = \sum_{j=1}^c S_j^{(\alpha)}\\
\sum_{l=1}^\gamma M_l^{(\alpha)} M_l^{(\beta)} - \sum_{a=1}^\delta N_a^{(\alpha)} N_a^{(\beta)} &=& \sum_{j=1}^c S_j^{(\alpha)} S_j^{(\beta)} - \sum_{i=1}^d Q_i^{(\alpha)} Q_i^{(\beta)}
\end{eqnarray}
The structure of the GLSM is further determined by a non-trivial scalar potential and superpotential. The latter is given by
\begin{equation} \label{superpotential}
S = \int d^2z d\theta \bigg[\sum_j \Gamma^j G_j (x_i) + \sum_{l,a} P_l \Lambda^a F_a^l (x_i) \bigg]
\end{equation}
where the functions $G_j$ and $F_a^l$ are quasi-homogeneous polynomials whose multi-degrees are determined by $U(1)$ gauge invariance to be:
\begin{equation}
\begin{aligned}
&\begin{array}{|c|}
\hline
 G^j \\
\noalign{\hrule height 1pt}
\begin{array}{cccc}
 S_1 & S_2 & \ldots & S_c
\end{array}\\
\hline
\end{array}
\\[0.1cm]
&\begin{array}{|c|}
\hline
F_a{}^l  \\
\noalign{\hrule height 1pt}
\begin{array}{cccc}
 M_1-N_1 & M_1-N_2 & \ldots & M_1-N_\delta\\
 M_2-N_1 & M_2-N_2 & \ldots & M_2-N_\delta\\
 \vdots  &    \vdots  &  \ddots  &  \vdots  \\  
M_\gamma-N_1 & M_\gamma-N_2 & \ldots & M_\gamma-N_\delta
\end{array}\\
\hline
\end{array}
\end{aligned}
\end{equation}
Furthermore, the function $F_a^l$ will be chosen to satisfy a \emph{transversality condition} such that $F_a^l(x)=0$ only when $x_i=0$

The scalar potential includes contributions from F-terms:
\begin{equation}
V_F = \sum_j \big |G_j(x_i) \big|^2 + \sum_a\big |\sum_l p_lF_a^l(x_i) \big|^2
\end{equation}
and D-terms
\begin{equation}
V_D = \sum_{\alpha=1}^r \bigg( \sum_{i=1}^d Q_i^{(\alpha)} |x_i|^2 - \sum_{l=1}^\gamma M_l^{(\alpha)}|p_l|^2 - \xi^{(\alpha)} \bigg)^2
\end{equation}
In this latter expression the $\xi^{(\alpha)} \in \mathbb R$ is the Fayet-Iliopoulos (FI) parameter which determines the structure of the vacuum. In particular the sign of each of the FI terms (for $\alpha=1, \ldots r)$, determine whether the solution of theory is a geometric, Landau-Ginzburg, or hybrid phase.

This can be easily illustrated in the case of a single $U(1)$ symmetry: For $\xi > 0$ the D-term implies that not all $X_i$ are zero simultaneously, thus not all $F_a$ are zero, and the F-term implies $G_j(x_i) = 0$ and $<p> = 0$. This ``geometric" phase contains the data of a $(0,2)$ non-linear sigma-model on a generally singular complete intersection, $X$, in a weighted projective space $\mathbb P_{Q_1,...,Q_d}[S_1,...,S_c]$. Moreover, the superpotential in \eref{superpotential} leads to a mass term of the form $\sum_a \pi \lambda^a F_a$, which makes massive one linear combination of the $\lambda^a$. The details of fermionic gauge symmetries, etc can be found in \cite{Distler:1995mi,Distler:1996tj,Blumenhagen:2011sq}. In brief summary, the remaining massless combinations of the left-moving fermions $\lambda_a$ couple to a vector bundle  over it defined by the monad
\begin{equation}\label{full_monad}
0 \to \mathcal O_{X}^{\oplus r_{\mathcal V}} \xrightarrow{E_i^a} \bigoplus_{a=1}^\delta \mathcal O_{X}(N_a) \xrightarrow{F_a^l} \bigoplus_{l=1}^\gamma \mathcal O_{X}(M_l) \to 0
\end{equation}
on the complete intersection CY manifold $X=\cap_{j=1}^{c} G_j$. The rank $(\delta - \gamma - r_{\cal V})$ vector bundle $V \to X$ is defined as
\begin{equation}
V = \frac{ker(F_a^l)}{im(E_i^a)}
\end{equation}
where $E_i^{a}$ arises from $r_{{\cal V}}$ additional fermionic gauge symmetries that have been introduced leading to a deviation in chirality of ${\overline {\cal D}}\Lambda^a=\sqrt{2} \Sigma^i E_{i}^{a}$. Note that although the new, neutral, chiral superfields, $\Sigma^i$, contribute to the scalar potential, they will play no role in the analysis undertaken here and will be omitted. In general, $E_i^{a}$ obeys a composition rule with $F_a^l$: $E \circ F=0$ on $X$. Furthermore, in this work, we will consider the simple case that $E_i^{a}=0$ and $V$ is defined 
as a kernel: $V = ker(F_a^l)$ defined by a short exact sequence
\begin{equation}
0 \to V \to \bigoplus_{a=1}^\delta \mathcal O_{\mathcal M}(N_a) \xrightarrow{F_a^l} \bigoplus_{l=1}^\gamma \mathcal O_{\mathcal M}(M_l) \to 0
\end{equation}

The simplest possible ``non-geometric" phase can be realized with $\xi < 0$. In this case, the D-term implies that  $<p> \neq 0$ thus all $<x_i> = 0$. This vacuum corresponds to a Landau-Ginzburg orbifold with a superpotential:
\begin{equation}\label{LG_superpotential}
\mathcal W(x_i, \Lambda^a, \Gamma^i) = \sum_j\Gamma^jG_j(x_i) + \sum_a \Lambda^aF_a(x_i)
\end{equation}
Finally, when $r>1$, the possibility of hybrid phases -- with some $x^{(\alpha)}$ less than and some great than zero -- naturally arises.

The simple form of \eref{LG_superpotential} first inspired the observation in \cite{Distler:1995bc} of target space duality. By inspection of \eref{LG_superpotential} it is clear that an exchange/relabeling of the functions $G_j$ and $F_a$ will not affect the Landau-Ginzburg model, as long as anomaly cancellation conditions are satisfied. The natural observation is that two distinct GLSMs could ``share" a non-geometric phase in which the original (large volume) role of  $G_j$ and $F_a$ is obscured. This notion of duality has been extended in a variety of ways including by allowing for the resolution of singularities in the geometry \cite{Chiang:1997kt} and extending the duality with a trivial re-writing which allows for a change in the number of $U(1)$ symmetries in the dual pair \cite{Distler:1996tj,Blumenhagen:2011sq}.

In the following section we review and illustrate the target space duality algorithm as laid out clearly in \cite{Blumenhagen:2011sq,Rahn:2011jw} and extend it still further by applying new redundant descriptions of the bundle geometry that make new target space duals manifest.

\subsection{Target Space Duality}\label{the_algorithm}

Given a starting point of a $(0,2)$ GLSM, the procedure for generating a target space dual model was laid out succinctly in \cite{Blumenhagen:2011sq,Rahn:2011jw}. We follow the conventions of \cite{Blumenhagen:2011sq} by taking $||X||$ to indicate the charge of the field $X$ and $||G||$ to denote the multi-degree of a homogeneous function $G(X)$. For completeness we summarize the algorithm here: \\

{\bf The Algorithm}:
\begin{enumerate}
\item Find all phases of a smooth $(0,2)$ GLSM, including a geometric phase with geometry $(X,V)$.
\item Construct a phase in which at least one field $p_l$ (for illustration, take $p_1$) is required to have a non-zero vacuum, $\langle p_1 \rangle \neq 0$.
\item Rescale $k$ Fermi superfields by the constant vev, $\langle p_1 \rangle$, and exchange the role of some $\lambda^a$ and $\Gamma^j$:
\beq
\tilde{\Lambda}^{a_i}:=\frac{\Gamma^{j_i}}{\langle p_1 \rangle} ~,~~~~\tilde{\Gamma}^{j_i}=\langle p_1 \rangle \Lambda^{a_i}~,~~~ \forall ~i=1, \ldots k,
\eeq
For consistency, this relabeling/rescaling can only be done when $\sum_i ||G_{j_1} || = \sum_i || F_{a_i}^{1} ||$ for anomaly cancellation.
\item Tune the bundle moduli (i.e. morphisms $F_a^l$) so that $\Lambda^{a_i}$ appear only in terms with $P_1$ for all $i$. i.e. choose
\beq
F_{a_i}^{l}=0 ~,~~~~ \forall~ l \neq 1,~~i=1, \ldots k
\eeq
\item Move away from this shared non-geometric phase and define the Fermi superfields of the new GLSM such that each term in the superpotential is gauge invariant:
\beq
|| \tilde{\Lambda}^{a_i} ||= || \Gamma^{j_i} || - ||P_1 ||~~~~,~~~|| \tilde{\Gamma}^{j_i} ||= || \Lambda^{a_i} || + || P_1 ||
\eeq
\item Allow the moduli to move to a generic point and define a new (dual) $(0,2)$ GLSM with a distinct geometric phase corresponding to a new Calabi-Yau and vector bundle, $(\widetilde{X}, \widetilde{V})$.
\end{enumerate}

It was proved in \cite{Blumenhagen:2011sq} that for the algorithm as described above if the initial $(0,2)$ GLSM satisfies the anomaly cancellation conditions in \eref{anomalies} then so does the constructed dual. Moreover, as mentioned in the Introduction, it was observed in the landscape scan of \cite{Blumenhagen:2011sq} that throughout this process, in the target space theories the net multiplicities of charged matter, and the total number of massless gauge singlets are preserved for almost all of the examples, where the individual number of complex, K\"ahler and bundle moduli are interchanged as
\begin{eqnarray}\nonumber
h^*(X,\wedge^{k}V) &=& h^*(\widetilde X, \wedge^{k} \widetilde V) \\
h^{2,1}(X) + h^{1,1}(X) + h^1_X(End_0(V)) &=& h^{2,1}(\widetilde X) + h^{1,1}(\widetilde X) +  h^1_X(End_0(\widetilde V))
\end{eqnarray}
On a technical note, the dimensions of the cohomology groups above can be computed using standard tools such as \cite{Blumenhagen:2010pv,cicy_package}. This procedure is illustrated explicitly below.

\subsubsection{Illustration of the duality procedure}

For concreteness, it is easiest to understand the target space duality algorithm in the context of an example. In principle an arbitrary number of $G$'s and $F$'s can be interchanged in the non-geometric phase. Below we will consider the simplest case in which two such maps play a role (note that in this case the quadratic anomaly cancellation conditions are known to be satisfied in the dual theory \cite{Blumenhagen:2011sq}).

In the non-geometric phase, we will make explicit the two terms to be exchanged in the corresponding superpotential (here  the first two terms with $a=1,2$):

\begin{equation}
\mathcal W = \sum_{j=1}^c\Gamma^jG_j + \sum_{a=1}^2 P_1 \Lambda^aF_a^1 +  \sum_{a=3}^\delta P_1 \Lambda^aF_a^1+ \sum_{l=2}^\gamma \sum_{a=1}^\delta P_l \Lambda^aF_a^l
\end{equation}

In the GLSM, not all the $<p>$'s are allowed to vanish simultaneously. Choosing $p_1$ to obtain a vev, the superpotential can be written as

\begin{equation}\label{eg_switch1}
\mathcal W = \Gamma^1G_1 + \Gamma^2G_2 + \sum_{j=3}^c\Gamma^jG_j 
+ <p_1> \Lambda^1F_1^1 + <p_1> \Lambda^2F_2^1 + \sum_{a=3}^\delta <p_1> \Lambda^aF_a^1+ \sum_{l=2}^\gamma \sum_{a=1}^\delta P_l \Lambda^aF_a^l
\end{equation}
Performing the interchange of the first two $G_j$'s and $F_a$'s as prescribed above, the equivalent superpotential can be expressed by $\tilde G_j$'s and $\tilde F_a$'s
\begin{equation}\label{eg_switch2}
\mathcal W = \tilde \Gamma^1\tilde G_1 + \tilde \Gamma^2\tilde G_2 + \sum_{j=3}^c\Gamma^jG_j \\
+ <p_1> \tilde \Lambda^1\tilde F_1^1 + <p_1> \tilde \Lambda^2\tilde F_2^1 + \sum_{a=3}^\delta <p_1> \Lambda^aF_a^1+ \sum_{l=2}^\gamma \sum_{a=1}^\delta P_l \Lambda^aF_a^l ~.
\end{equation}
In this case, the rescalings are
\begin{eqnarray}\nonumber
&\tilde{\Gamma}^1 := <p_1>\Lambda^1,  \tilde{\Gamma}^2 := <p_1>\Lambda^2, \tilde{\Lambda}^1 := \frac{\Gamma^1}{<p_1>}, \tilde{\Lambda}^2 := \frac{\Gamma^2}{<p_1>}&\\
&\tilde G_1 :=\ F_1^1,\ \tilde G_2 :=\ F_2^1,\ \tilde F_1^1 :=\ G_1,\ \tilde F_2^1 :=\ G_2 &
\end{eqnarray}
and $||G_1|| + ||G_2|| = ||F_1^1|| + ||F_2^1||$. From this point, a return to a geometric phase would yield a new configuration corresponding to a geometric pair $(\tilde X, \tilde V)$.

\subsubsection{Target space duals with an additional $U(1)$}\label{tsd_extrau1}
As outlined above, it is clear that the ``relabeling" of fields at the shared Landau-Ginzburg point can mix the degrees of freedom in $h^{2,1}(X)$ and $h^1(X, End_0(V))$ in the target space dual. More general possibilities are possible however that can \emph{also} change the dimension of $h^{1,1}$. As one particular example of such a change, we will be interested here in geometries $X, \widetilde{X}$ which are related by geometric (i.e. conifold-type) transitions.

To change the dimension of $h^{1,1}$ across the target space dual it is necessary to re-write the initial GLSM in an equivalent/redundant way. Stated briefly, it is always possible to introduce into the GLSM a new coordinate (i.e. a new Fermi superfield) $y_1$ with multi-degree ${\mathcal B}$ and a new hypersurface (i.e. a Chiral superfield with opposite charge to the new Fermi superfield) $G^{\cal B}$ corresponding to a homogeneous polynomial of multi-degree ${\cal B}$. Denoting the initial bundle/manifold in terms of its GLSM charges as
\beq\label{starting_point}
V_{N_1, \ldots N_{\delta}}[M_1, \ldots, M_{\gamma}] \longrightarrow \mathbb{P}_{Q_1, \ldots Q_d}[S_1, \ldots, S_c]
\eeq
the above addition can be written
\beq\label{same_base}
V_{N_1, \ldots N_{\delta}}[M_1, \ldots, M_{\gamma}] \longrightarrow \mathbb{P}_{Q_1, \ldots Q_d,{\cal B}}[S_1, \ldots, S_c,{\cal B}]
\eeq
As argued in \cite{Blumenhagen:2011sq}, this addition leaves the GLSM and the corresponding target space geometry completely unchanged. However, as we will demonstrate, this redundant description can play a non-trivial role in the target space duality re-labeling.

Following \cite{Blumenhagen:2011sq}, suppose that in an example as in the previous Subsection, there are two chosen map elements $F_1^1$ and $F_2^1$ that have been chosen to be interchanged with a defining relation $S_1$. In this case we can choose the redundant new coordinate, $y_1$, to have charge
\begin{equation}
||{\cal B}|| = ||F_1^1|| + ||F_2^1|| - S_1
\end{equation}
and under the re-labelings required in the algorithm of Section \ref{the_algorithm} it is possible to choose, for example,
\beq\label{u1_field_defs}
\tilde{N}_1=M_1-S_1=0~,~\tilde{N}_2=M_2 - {\cal B}~,~\tilde{S}_1 = ||F_1^1||~,~\tilde{{\cal B}}=||F_2^1 ||
\eeq
That is, a Fermi superfield has now been generated with apparently vanishing charge. This is misleading however because as we will see, this field may be thought of as carrying an additional $U(1)$ charge. To see the origin of this new $U(1)$, it should be noted that the redundancy/relabeling described above can, unfortunately, lead to singular geometries. As pointed out in \cite{Distler:1996tj}, to avoid such singularities in the new model and have a valid, smooth large volume limit $(\tilde X, \tilde V)$, it is possible to add an additional $U(1)$ gauge symmetry under which the formally uncharged Fermi-superfield carries a non-vanishing charge. Geometrically this corresponds with introducing one further coordinate $y_2$ and identifying the two new coordinates as the homogeneous coordinates on a $\mathbb{P}^1$ (i.e. a blow-up of the original geometry). So the charges now read

\begin{tablehere}
\begin{center}
\begin{tabular}{|ccccc|cccc||cccc|cccc| }
  \hline                       
   $x_1$ & \ldots & $x_d$ & $y_1$ & $y_2$  &$\Gamma^1$ & \dots & $\Gamma^c$ & $\Gamma^B$ & $\Lambda^1$ & $\Lambda^1$ & \ldots & $\Lambda^\delta$ & $p_1$ & $p_2$ & \ldots & $p_\gamma$\\
 \noalign{\hrule height 1pt}
     0   & \ldots &   0   &   1   &    1       &    0     & \ldots &     0      &    $-1$      & 0   & 0   & \ldots & 0 & $-1$ & 0 & \ldots & 0  \\
   $Q_1$ & \ldots & $Q_d$ &   ${\cal B}$ &    0       & $-S_1$   & \ldots & $-S_c$     &     $-{\cal B}$     & $N_1$ & $N_2$ & \ldots & $N_\delta $ & $-M_1$ & $-M_2$ &  \ldots & $-M_\gamma$  \\
  \hline  
\end{tabular}
\hfill
\end{center}
\end{tablehere}
where once again, this configuration is equivalent to the initial one in \eref{starting_point}. This equivalence can be verified by noting that the constraint $G^{\cal B} = y_1 = 0$ can be used to eliminate the coordinate $y_1$. Further, $y_2$ can be fixed by the additional $U(1)$ gauge symmetry and associated D-term. Thus, this configuration is simply the original space $\times$ a single point. Use the same method discussed before, we can perform rescalings to this new configuration, resulting in a new target space dual with an additional $U(1)$.

Applying the field redefinitions in \eref{u1_field_defs} we arrive at last to the new configuration
{\scriptsize
\begin{tablehere}
\begin{center}
\begin{tabular}{|ccccc|cccc||cccc|cccc| }
  \hline                       
   $x_1$ & \ldots & $x_d$ & $y_1$ & $y_2$  &$\tilde{\Gamma}^1$ & \dots & $\Gamma^c$ & $\tilde{\Gamma}^B$ & $\tilde{\Lambda}^1$ & $\tilde{\Lambda}^1$ & \ldots & $\Lambda^\delta$ & $p_1$ & $p_2$ & \ldots & $p_\gamma$\\
 \noalign{\hrule height 1pt}
     0   & \ldots &   0   &   1   &    1       &    $-1$     & \ldots &     0      &    $-1$      & 1   & 0   & \ldots & 0 & $-1$ & 0 & \ldots & 0  \\
   $Q_1$ & \ldots & $Q_d$ &   ${\cal B}$ &    0       & $-(M_1-N_1)$   & \ldots & $-S_c$     &     $-(M_1-N_2)$     & $0$ & $M_2-{\cal B}$ & \ldots & $N_\delta $ & $-M_1$ & $-M_2$ &  \ldots & $-M_\gamma$  \\
  \hline  
\end{tabular}
\hfill
\end{center}
\end{tablehere}}
For simplicity, in this work all examples will be chosen so that the target space duals satisfy ${\cal B}=0$ which will guarantee that the theories stay within the class of simple complete intersection Calabi-Yau manifolds in products of projective spaces \cite{Candelas:1987kf,Hubsch:1986ny}.

\subsubsection{Redundant bundle geometry and more general target space duals}\label{repeated_line_bundles}

It is useful to observe here that the procedure of inducing a redundant description of the manifold, $X$, illustrated above can also be applied to the vector bundle, $V$. At the level of the GLSM, once again we introduce a new pair consisting of a chiral/fermi superfields of opposite charge, but this time to the GLSM data corresponding to the bundle $V$, rather than to $X$ as in the previous section. Once again denoting the bundle in terms of its GLSM charges as in \eref{starting_point}
this time the addition of fields takes the form
\beq\label{repeated_ent}
V_{N_1, \ldots N_\delta, {\cal B}}[M_1,\ldots M_{\gamma},{\cal B}] \to \mathbb{P}_{Q_1,\ldots Q_d}[S_1, \ldots, S_c]
\eeq
Note that this is exactly the redundancy utilized in \eref{same_base}, only here it is applied to $V$ rather than to $X$.

At the level of the bundle geometry, the procedure involves a simple modification of the monad defining $V$. Beginning with the short exact sequence
\beq\label{monadfirst}
0 \to V \to B \stackrel{F}{\longrightarrow} C \to 0
\eeq
with $B,C$ sums of line bundles, we can realize the addition in \eref{repeated_ent} by adding the same line bundle $L$ to both $B$ and $C$:
\beq\label{add_line_bundle}
0 \to V' \to B \oplus L \stackrel{F'}{\longrightarrow} C \oplus L \to 0
\eeq
It is clear that the bundles $V'$ and $V$ share a locus in moduli space for which the map $F'$ takes the form
\beq\label{simple_map}
F'=\left(\begin{array}{cc}
F & \bf{0} \\
\bf{0} & \mathbb{C}
\end{array}\right)
\eeq
where $F$ is the original fiberwise morphism in \eref{monadfirst} and $\mathbb{C}$ represents the constant map from $L \to L$. For this choice of $F'$ the bundles $V$ and $V'$ are isomorphic. It is also straightforward to verify that the topology of $V$ and $V'$ are identical. Their agreement at a general point in moduli space can be probed via Lemmas \ref{OSS_lemma_thm} and \ref{OSS_corollary_thm} in Appendix \ref{AppendixA}.

For more general maps of the form
\beq\label{complicated_map}
F'=\left(\begin{array}{cc}
F & \bf{\alpha} \\
\bf{\beta} & \mathbb{C}
\end{array}\right)
\eeq
where $\alpha \in H^0(X, L^{\vee} \otimes C)$ and $\beta \in H^0(X, B^{\vee} \otimes L)$ are non-trivial, it is possible to probe a broader region of the shared moduli space. As we will see below, adding $L$ such that $\alpha, \beta \neq 0$ can \emph{lead to new target space dual theories} and is a generalization of the procedure systematically explored in \cite{Blumenhagen:2011sq}. Moreover, demanding that there exist generic non-trivial maps, $\alpha, \beta \neq 0$, bounds the set of possible line bundles $L$ that can be added to produce new target space duals to a finite set. Roughly, these conditions amount to the constraint that the entries $O(a,b,\ldots)$ in $L$ are ``small enough" to map into $C$ but ``big enough" that some entries in $B$ can be mapped non-trivially into $L$. 
For example, starting with the geometry
\begin{equation}\begin{aligned}&\begin{array}{|c|c| |c|c|}
\hline
x_i & \Gamma^j & \Lambda^a & p_l \\ \noalign{\hrule height 1pt}\begin{array}{cccccc}
1& 1& 0& 0& 0& 0  \\
0& 0& 1& 1& 1& 1 
\end{array}&\begin{array}{c}
-2  \\
-4 
\end{array}&\begin{array}{cccccccc}
 0& 0& 1& 1& 1& -1& 2& 2  \\
 1& 1& 0& 0& 3&  1& 1& 2 
\end{array}&\begin{array}{ccc}
 -3& -1& -2 \\
 -2& -4& -3 
\end{array}\\
\hline
\end{array} \label{eq:bundle3a}\end{aligned}\end{equation}
where $V$ is defined via the monad
\begin{equation}
\begin{split}
0 \to V & \to \mathcal O(0,1)^{\oplus2} \oplus \mathcal O(1,0)^{\oplus2} \oplus \mathcal O(1,3) \oplus \mathcal O(-1,1) \oplus \mathcal O(2,1) \oplus \mathcal O(2,2) \\
&\stackrel{F}{\longrightarrow} \mathcal O(3,2) \oplus \mathcal O(1,4) \oplus \mathcal O(2,3) \to 0
\end{split}
\end{equation}
For a component $B_i = \mathcal O(b_1, b_2)$ mapping into $C_j = \mathcal O(c_1, c_2)$, $h^0(B_i^*\otimes C_j) = h^0(\mathcal O(c_1-b_1, c_2-b_2)) \neq 0$. Which means $c_1 \geq b_1, c_2 \geq b_2$, or $c_1 - b_1 = -1, c_2 - b_2 \geq 4$.

In this case, $L$ can only be one of the set
\begin{align}\label{eg_set}
&\{\mathcal O(-1,2),~\mathcal O(-1,3),~\mathcal O(-1,4), ~\mathcal O(0,2),
 ~\mathcal O(0,3), ~\mathcal O(0,4), ~\mathcal O(1,1), ~\mathcal O(1,2), \nonumber \\
& ~\mathcal O(1,3), ~\mathcal O(2,0), ~\mathcal O(2,1), ~\mathcal O(2,2), ~\mathcal O(3,0), ~\mathcal O(3,1) \}
\end{align}

It should be noted finally that another concern is whether each of the line bundles appearing in \eref{eg_set} can appear in \eref{add_line_bundle} an arbitrary number of times? Having identified the range of possible line bundles in \eref{add_line_bundle} such that the maps are non-trivial, we may inquire whether they could each be added an arbitrary (i.e. infinite) number of times to the monad? In principle, such an infinite repetition of a line bundle leads to an equivalent monad, however as we will see in future sections, such additions do not increase the number of non-trivial target space dual theories. For the examples in consideration in this work, we begin with at most two defining relations in the geometric realization of the CY manifold (i.e. the index on the fields, $\Gamma^j$, runs only over $j=1,2$ in the present examples). Thus, at most $3$ line bundles can play a non-trivial role in the algorithm (i.e. $\Lambda^1$, $\Lambda^2$ and $P_1$ as in \eref{eg_switch1} and \eref{eg_switch2}). So there nothing to be gained by adding more than $3$ additional line bundles, $\mathcal L_i$. The utility of adding ``repeated entires" to the monad bundle will be further illustrated in Appendix \ref{appendixB}.

\subsubsection{Summary of choices for the target space dual geometries considered here}
It is worth briefly summarizing the classes of target space duals we will consider in this work. We will require that
\begin{itemize}
\item All target space duals constructed will increase the number of $U(1)$ symmetries in the GLSM by $1$ and take ${\cal B}=0$ in the notation of Section \ref{tsd_extrau1} to remain within the class of CICYs \cite{Candelas:1987kf}. This simple choice furthermore ensures that $X$ and $\widetilde{X}$ are related by geometric transitions (in general, so-called ``effective splits" for CICY configuration matrices \cite{Hubsch:1992nu}).
\item For the initial geometric pairs, $(X,V)$, we take $c_2(V)=c_2(TX)$ and it was established in \cite{Blumenhagen:2011sq} that this will guarantee that $c_2(\widetilde{V})=c_2(T\widetilde{X})$ in the target space dual geometry. The structure group of $V$ is embedded into a single $E_8$ factor of the $E_8 \times E_8$ theory. The second, unbroken $E_8$ factor to the gauge symmetry will not play a role in our analysis.
\item We choose bundles, $V$, defined as kernels: $V=Ker(f)$ and will allow for repeated line bundle entries in the monad, subject to the constraints described in the previous Section.
\end{itemize}

We turn now to a brief review of the structure of the ${\cal N}=1$ effective potential in the $4$-dimensional heterotic theory.

\section{The Target Space Potential: Supersymmetry Conditions and Vector Bundle Geometry}\label{potential_hym}

To preserve $\mathcal N = 1$ supersymmetry in the 4-dimensional effective theory, the geometry of the pair $(X,V)$ is constrained. The simplest class of solutions are found when $X$ is a Calabi-Yau threefold and the vector bundle $\pi: V \to X$ satisfies the Hermitian Yang-Mills equations \cite{Green:1987mn}:
\bea \label{HYM} 
g^{a \bar{b}} F_{a \bar{b}}=0 \;,\; F_{ab}=0 \;,\;
F_{\bar{a}\bar{b}}=0 \ .  
\eea
where $F$ is the gauge field strength
associated with the vector bundle $V$, and $a$ and $\bar{b}$ are holomorphic and anti-holomorphic indices on the
Calabi-Yau manifold. These equations involve all three types of singlet ``moduli" described in the previous section ($h^{1,1}(X)$,$h^{2,1}(X)$ and $h^1(X, End_0(V))$) and in general not all possible values of the CY and bundle moduli will correspond to a solution of \eref{HYM}.

It is straightforward to see that a solution to \eref{HYM} is directly linked to the minimum of the potential in the $4$-dimensional effective theory \cite{Witten:1985bz}. The argument below was given in \cite{Anderson:2011ty} and we briefly summarize it here:

The structure of the $4$-dimensional effective potential can be seen by focusing on several terms in the ten-dimensional effective action,
\begin{eqnarray}
\label{spartial}
S_{\textnormal{partial}} = -\frac{1}{2 \kappa_{10}^2} \frac{\alpha'}{4}\int_{{\cal M}_{10}} \sqrt{-g} \left\{ \textnormal{tr} (F^{(1)})^2+\textnormal{tr} (F^{(2)})^2 - \textnormal{tr} R^2  + \ldots \right\}
\end{eqnarray}
The notation here is standard \cite{Green:1987mn} with the field
strengths $F^{(1)}$ and $F^{(2)}$ being associated with the two $E_8$
factors in the gauge group. In addition to the action, the ten-dimensional
Bianchi Identity holds,
\begin{eqnarray}
  dH = -\frac{3 \alpha'}{\sqrt{2}} \left( \textnormal{tr} F^{(1)} \wedge F^{(1)} +\textnormal{tr} F^{(2)} \wedge F^{(2)} - \textnormal{tr} R \wedge R \right) .
\end{eqnarray}
The associated integrability condition can be phrased as
\begin{eqnarray}
\label{intcond}
\int_{M_{6}} \omega  \wedge \left( \textnormal{tr} \;F^{(1)} \wedge F^{(1)} + \textnormal{tr} \; F^{(2)} \wedge F^{(2)} - \textnormal{tr}\; R \wedge R \right) = 0 \;,
\end{eqnarray}
where $\omega$ is the K\"ahler form.  Working
to lowest order, with a Ricci flat metric on a manifold of $SU(3)$
holonomy, equation \eref{intcond} then leads to
\begin{eqnarray}
  \int_{M_{10}} \sqrt{-g} \left( \textnormal{tr} (F^{(1)})^2+\textnormal{tr} (F^{(2)})^2 - \textnormal{tr} R^2  +2 \,\textnormal{tr} (F^{(1)}_{a \bar{b}} g^{a \bar{b}})^2   +2\, \textnormal{tr} (F^{(2)}_{a \bar{b}} g^{a \bar{b}})^2  \right. \\ \nonumber \left.-4 \, \textnormal{tr} (g^{a \bar{a}} g^{b \bar{b}} F^{(1)}_{ab} F^{(1)}_{\bar{a}\bar{b}})  -4 \, \textnormal{tr} (g^{a \bar{a}} g^{b \bar{b}} F^{(2)}_{ab} F^{(2)}_{\bar{a}\bar{b}})\right)= 0~.
\end{eqnarray}
Using this relation in \eref{spartial}, the result is
\begin{eqnarray}
\label{finalintro}
S_{\textnormal{partial}}= - \frac{1}{2 \kappa_{10}^2} \alpha' \int_{{\cal M}_{10}} \sqrt{-g} \left\{ -\frac{1}{2} \textnormal{tr} (F^{(1)}_{a \bar{b}} g^{a \bar{b}})^2  - \frac{1}{2} \textnormal{tr} (F^{(2)}_{a \bar{b}} g^{a \bar{b}})^2 \right. \\ \nonumber \left. + \textnormal{tr} (g^{a \bar{a}} g^{b \bar{b}} F^{(1)}_{ab} F^{(1)}_{\bar{a}\bar{b}}) +  \textnormal{tr} (g^{a \bar{a}} g^{b \bar{b}} F^{(2)}_{ab} F^{(2)}_{\bar{a}\bar{b}}) \right\}~.
\end{eqnarray}
By inspection, the terms included in \eref{finalintro} are a part of the ten-dimensional
theory which does not contain any four-dimensional derivatives. Thus, under dimensional reduction, they will contribute to the potential energy of the four-dimensional theory. In the case of a supersymmetry preserving field configuration, the terms in the integrand of \eref{finalintro}
vanish, corresponding exactly to a geometric moduli choice which solves \eref{HYM}.  In this case, no potential is generated. 

The detailed form of the potential generated in the ${\cal N}=1$, $4$-dimensional effective theory has been developed in recent work \cite{Anderson:2009sw,Anderson:2009nt,Anderson:2013qca,Anderson:2011ty}. By a theorem of Donaldson-Uhlenbeck and Yau \cite{duy1,duy2}, the first condition in \eref{HYM} is equivalent to the condition of Mumford (slope) poly-stability of $V$. The characterization of slope stability as well as tools to test whether a given bundle is stable are provided in Appendix \ref{AppendixA}. The condition of stability (equivalently \eref{HYM}) manifestly depends on the K\"ahler moduli through the presence of the metric. The K\"ahler class can in principle be smoothly varied so that the bundle transitions from being fully stable in one region of moduli space to poly-stable along a higher co-dimensional boundary (a so-called ``stability wall" \cite{Anderson:2009sw}) to full unstable (and hence supersymmetry breaking) for some region of moduli space. In the effective theory, this is associated with \emph{$D$-term breaking }of supersymmetry at stability walls. Along the stability walls the condition of poly-stability forces the bundle $V$ to become reducible (i.e. $V \to {\cal F} \oplus V/{\cal F}$), leading to the presence of enhanced, Green-Schwarz massive $U(1)$ symmetries in the effective theory. See \cite{Sharpe:1998zu,Anderson:2009nt,Anderson:2009sw,Anderson:2010ty}. In the neighborhood of a stability wall, the D-term can be written in the effective theory as
\beq\label{dterm_first}
D^{U(1)} \sim FI(t) - \frac{1}{2} \sum_{i} Q_iG_{P\bar{Q}} {C_i}^P{\bar{C}_i}^{\bar Q}
\eeq
where the first term is a moduli-dependent Fayet-Iliopolous parameter (depending on the K\"ahler moduli, $t^r$, $r=1,\ldots h^{1,1}(X)$ of $X$) and the second arises from matter fields, $C_i$, charged under the anomalous $U(1)$. Briefly, the form of the FI term to leading order is\footnote{See \cite{Anderson:2009nt} for a discussion of the one-loop correction to this FI term and its geometric interpretation.}
\beq\label{FI_slope}
FI(t) \sim \frac{\mu({\cal F})}{Vol(X)}
\eeq
where $Vol(X)$ is the volume of the CY 3-fold, $Vol(X)= \int_X J \wedge J \wedge J$, with $J$ is the K\"ahler form on $X$. Here $\mu({\cal F})=\frac{1}{rk({\cal F})}\int_X c_1({\cal F}) \wedge J \wedge J$ is a quasi-topological number called the  ``slope" \cite{huybrechts_book},  associated to a crucial (geometrically de-stabilizing \cite{Anderson:2009nt}) sub-sheaf, ${\cal F}$. This last object, ${\cal F}$, is defined by the fact that near the stability wall, $V$ must be decomposed as $V \to {\cal F} \oplus \frac{V}{{\cal F}}$. See \cite{Anderson:2009nt} for a review and Appendix \ref{AppendixA} for further details.

\begin{figure}
\centering \includegraphics[width=0.6\textwidth]{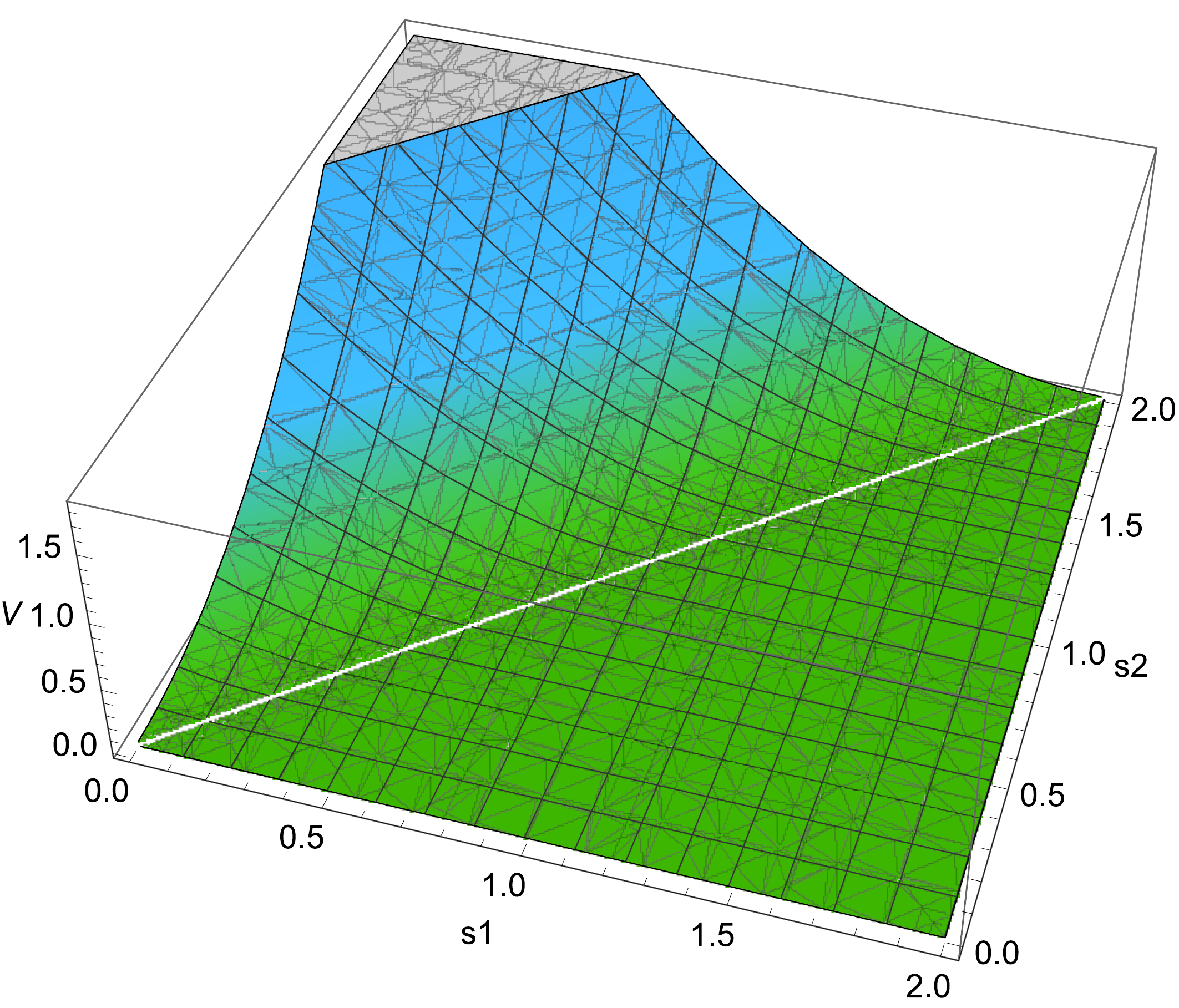} \caption{\emph{An illustration of the D-term effective potential corresponding to \eref{dterm_first} (and a part of \eref{finalintro}) and geometrically realized by the bundle failing to be slope-stable for a sub-cone of K\"ahler moduli space. The white boundary line is referred to as a ``stability wall" in the K\"ahler cone. }} \label{fig:wall}
\end{figure}

Likewise, the second conditions in \eref{HYM} correspond geometrically to the condition that the bundle is holomorphic. It is clear that the conditions $F_{ab}=F_{\bar{a}\bar{b}}=0$ depend on the definition of holomorphic and anti-holomorphic coordinates on $X$. This choice clearly depends on the complex structure. For a fixed topology of the gauge fields, specific changes in the complex structure of the Calabi-Yau threefold may be such that these holomorphy equations no longer have a solution \cite{Anderson:2010mh,Anderson:2011cza}. In the effective theory, this obstruction is associated with the \emph{$F$-term breaking} of supersymmetry by the complex structure moduli. Geometrically, the flat directions in moduli space corresponding to holomorphic deformations of the complete geometry (manifold $+$ bundle) are described by the ``Atiyah class" of $V$ \cite{atiyah}. More precisely, the infinitesimal complex moduli of the heterotic theory are not the independent bundle moduli ($h^1(X, End_0(V))$) and complex structure moduli ($h^{2,1}(X)$) but rather the elements of the cohomology group
\beq\label{h1q}
H^1(X, {\cal Q})
\eeq
which describes the tangent space to \emph{the infinitesimal deformation space of the pair $(X,V)$}. Here ${\cal Q}$ is defined by \cite{atiyah}:
\beq
0 \to End_0(V) \to {\cal Q} \to TX \to 0
\eeq
and 
\beq\label{mod_def}
h^1(X, {\cal Q})=h^1(X, End_0(V)) \oplus ker\{H^1(TX) \stackrel{[F^{1,1}]}{\longrightarrow} H^2(End_0(V))\}
\eeq
In general, the moduli in \eref{mod_def} contain the bundle moduli, but only a subset of the complex structure moduli of $X$. The scale at which the moduli are ``lifted" is generally the compactification scale (and thus, they should be excluded from the start in the zero-mode count given in \eref{mod_count}), however in some special cases this scale is lowered so that the moduli fixing due to holomorphy of the pair $(X,V)$ can be realized in the effective theory by the Gukov-Vafa-Witten superpotential \cite{Gukov:1999ya}:
\beq
W = \int_X H \wedge \Omega
\eeq
and its associated F-terms
\beq
F_{C_i}= \frac{\partial W}{\partial C_i} \sim \int_X \frac{\partial \omega^{3YM}}{\partial C_i}
\eeq
where $\omega^{3YM}= tr(F \wedge A - \frac{1}{3} A \wedge A \wedge A)$.
See \cite{Anderson:2010mh,Anderson:2011ty,Anderson:2011cza,Anderson:2013qca} for further details and computational tools. These moduli considerations have recently been extended to include heterotic non-K\"ahler vacua \cite{Anderson:2014xha,delaOssa:2014cia,delaOssa:2015maa,delaOssa:2015rhu} and a more detailed understanding of moduli space metrics \cite{Candelas:2016usb,McOrist:2016cfl}. The geometry and field theory of these F-terms and holomorphy are explored further in Section \ref{Fterm_sec}.

\section{D-terms, Bundle Stability, and Duality}\label{dterms}

In this section we will begin to explore the role of a non-trivial D-term condition in target space dual chains. One complication in this process is that we do not have a concrete map (i.e. an explicit isomorphism) between the two moduli spaces of the $(0,2)$ theories. In general, the presence of a non-trivial D-term potential does not lift entire flat directions (i.e. change the dimension of the vacuum space) but merely change its shape (see Figure \ref{fig:wall}). At the level of the effective theory, this is a realization of the fact that the vacuum space can be written as a solution to the F-terms, modulo D-term constraints \cite{Luty:1995sd}. In the bundle geometry, this same notion plays a crucial role in the definition of bundle moduli spaces, geometric invariant theory (GIT), quotients, etc \cite{huybrechts_book}.

To simplify our analysis then, we will take as a first example an extreme case in which the bundle is only stable for a sub-cone of the K\"ahler moduli space of \emph{strictly smaller dimension}. This will make it possible to simply count and observe whether or not there is a drop in the total degrees of freedom on both sides of the duality. For the details of how to analyze the slope-stability of a vector bundle and how to engineer interesting monad examples, see Appendix \ref{AppendixA}. Using these tools, we can select a pair, $(X,V)$, of CY manifold and vector bundle $\pi: X \to V$ with a D-term which ``lifts" whole directions of K\"ahler moduli space. Below, we will start with a simple example with $h^{1,1}=2$ and a bundle over it which contains \emph{two} destabilizing sub-sheaves, ${\cal F}_1, {\cal F}_2$. These sheaves, as described in the previous Section, appear in the associated D-terms as in \eref{FI_slope}  and in the slope-stability analysis of $V$ as described in Appendix \ref{AppendixA}. Here, the sub-sheaves are chosen such that
\beq\label{twosheaves}
c_1({\cal F}_1)= - c_{1}({\cal F}_2)
\eeq
Since a bundle is only stable if $\mu(V)> \mu({\cal F})$ for all ${\cal F} \subset V$ (see Appendix \ref{AppendixA}), it is clear that with two de-stabilizing sub-sheaves as in \eref{twosheaves}, and $c_1(V)=0$, $V$ cannot be properly stable at all. Indeed it can only solve the Hermitian-Yang-Mills equations, \eref{HYM}, on the locus in K\"ahler moduli space for which
\beq\label{slopey_cond}
\mu({\cal F}_1)= - \mu({\cal F}_2)=\mu(V)=0
\eeq
where $V$ itself must become reducible as a poly-stable bundle $V \to V_1 \oplus V_2\oplus \ldots$.

For the CY manifold, $X$, consider the simple degree $\{2,4\}$ hypersurface in $\mathbb P^1 \times \mathbb P^3$ with Hodge numbers $h^{1,1}(X)=2$ and $h^{2,1}(X)=86$. Expanding the second Chern class of the manifold in a basis of $\{1,1\}$ forms as $c_2(TX)_r=d_{rst}{c_2(TX)}^{st}$ (where $d_{rst}$ are the triple intersection numbers and $r=1, \ldots h^{1,1}$) we have $c_2(TX)^r=\{24,44\}$ (see Appendix \ref{bundle_man_top}). Finally, to complete the geometry, a rank $5$ monad bundle, $V$, can be chosen over $X$, written as a kernel:
\beq\label{moneg}
0 \to V \to B \stackrel{F}{\longrightarrow} C \to 0
\eeq
where $V=ker(F)$ and $B$ and $C$ are sums of line bundles over $X$ as in \eref{full_monad}. In GLSM language, the manifold and bundle are given by the following data:

\begin{equation}\begin{aligned}&\begin{array}{|c|c| |c|c|}
\hline
x_i & \Gamma^j & \Lambda^a & p_l \\ \noalign{\hrule height 1pt}\begin{array}{cccccc}
1& 1& 0& 0& 0& 0  \\
0& 0& 1& 1& 1& 1 
\end{array}&\begin{array}{c}
 -2  \\
 -4 
\end{array}&\begin{array}{cccccccc}
 1& -1& 0& 0& 2& 1& 1& 2  \\
 -1&  1& 1& 1& 1& 2& 2& 2 
\end{array}&\begin{array}{ccc}
 -3& -1& -2 \\
 -2& -4& -3 
\end{array}\\
\hline
\end{array} \label{bundle2a}\end{aligned}\end{equation}
By inspection, $c_1(V)=0$ and using the tools of \cite{Anderson:2008ex} it can be verified that 
\beq
c_2(V)=c_2(TX)=\{24,44\}
\eeq
and the third Chern class is $c_3(V)=-80$. Finally, a straightforward, but lengthy calculation yields the following naive total for the uncharged (geometric) moduli of the theory:
\beq\label{naive_count}
dim({\cal M}_{0})=h^{1,1}(X)+h^{2,1}(X)+h^1(X,End_0(V))=2+86+340=428
\eeq
Thus far we are in familiar territory. As described in Section \ref{potential_hym}, it is now possible to go further and explore the theory in more depth by considering the structure of its effective potential. As will be shown in the next subsections, the bundle in \eref{bundle2a} is exactly of the form required to generate a non-trivial D-term constraint and a corresponding reduction in the number of flat directions. We will begin by analyzing the effective theory arising from this geometry along the stable ray in the K\"ahler cone (the ``stability wall" \cite{Sharpe:1998zu,Anderson:2009sw,Anderson:2009nt}).

\begin{figure}
\centering \includegraphics[width=0.4\textwidth]{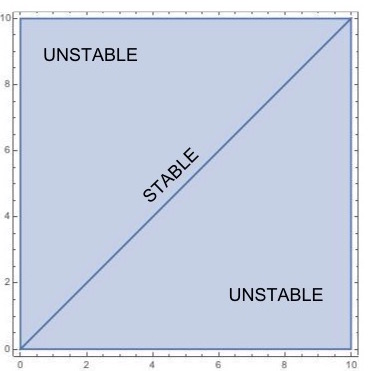} \caption{\emph{The stable ray in K\"ahler moduli space shown for the monad bundle in \eref{bundle2a}. Along this co-dimension 1 subcone the bundle takes the poly-stable form shown in \eref{decomp_on_wall}.}} \label{fig:firstwall}
\end{figure}

The non-vanishing triple intersection numbers of the CY 3-fold are $d_{122}=4$ and $d_{222}=2$. Expanding the K\"ahler form in a basis of $\{1,1\}$ forms as $J=t^r J_r$ (with $t^r$ the K\"ahler moduli), it is clear that the condition in \eref{slopey_cond} restricts the theory to the locus in K\"ahler moduli space for which $-8t^1 t^2+2 (t^2)^2=0$. Choosing perturbative solutions in the interior of the K\"ahler cone requires $t^r >0$ for all $r$, and thus the ``stability wall" in this case is the one-dimensional ray in K\"ahler moduli space where 
\beq\label{constrained_t}
t^2=4 t^1~~~\text{equivalently}~~~ s_2=s_1
\eeq
in the dual K\"ahler coordinates with $s_r=d_{rst}t^st^t$. The stable subcone for $V$ is shown in Figure \ref{fig:firstwall}.

This bundle is an example of one where the naive degree-of-freedom counting given in \eref{mod_count} fails to capture important structure. In fact, the dimension of the singlet moduli space is not the sum given in \eref{naive_count}.  Instead, with the stability constraint in \eref{constrained_t} it is reduced to 
\beq\label{real_count}
dim({\cal M}_1)=dim({\cal M}_{0})-1=427
\eeq

This type of example provides a direct new probe of target space duality. Will the dual theories also reflect this reduction in degrees of freedom? What is the structure of the potential in the dual theory and will the moduli lifting happen through D-terms or F-terms? Before answering these questions however, we will see below that in fact the detailed structure of the theory on the stability wall provides many new features that can be compared.

\subsection{Effective theory near the stability wall}
As described in Appendix \ref{AppendixA}, the presence of each of the line bundles with negative entries in $B$ (the middle term of the monad) in \eref{moneg} implies the existence of a particular sub-sheaf of $V$. In this case, following \eref{mon_subsheaf} we see that $V$ can be written in terms of its de-stabilizing sub-sheaves as 
\begin{align}
&0 \to Q_4 \to V \to \cO(1,-1) \to 0 \\
&0 \to \tilde{Q}_4 \to V \to \cO(-1,1) \to 0
\end{align}
where the rank $4$ sheaves, $Q_4$ and $\tilde{Q}_4$ are described via the monads 
\begin{align}\label{qdefs}
& 0 \to Q_4 \to \cO(-1,1)\oplus \cO(0,1)^{\oplus 2}\oplus \cO(2,1) \oplus \cO(1,2)^{\oplus 2} \oplus \cO(2,2) \to \cO(3,2) \oplus \cO(1,4) \oplus \cO(2,3) \to 0 \nonumber \\
& 0 \to \tilde{Q}_4 \to \cO(1,-1)\oplus \cO(0,1)^{\oplus 2}\oplus \cO(2,1) \oplus \cO(1,2)^{\oplus 2} \oplus \cO(2,2) \to \cO(3,2) \oplus \cO(1,4) \oplus \cO(2,3) \to 0 
\end{align}
Since $Q_4$ and $\tilde{Q}_4$ are sub-sheaves of $V$, with non-trivial first Chern classes, it is clear that supersymmetry will only be preserved if the bundle is poly-stable. This implies a decomposition of $V$ into each de-stabilizing sub-sheaf as $V \to {\cal F} \oplus V/{\cal F}$. In this case, for the two subsheaves of interest:
\begin{align}
& V \to Q_4 \oplus \cO(1,-1) \\
&V \to \tilde{Q}_4 \oplus \cO(-1,1)
\end{align}
Moreover, these descriptions are only compatible if $Q_4$ and $\tilde{Q}_4$ decompose still further as
\begin{align}
Q_4 \to U_3 \oplus \cO(-1,1) \\
\tilde{Q}_4 \to U_3 \oplus \cO(1,-1)
\end{align}
for \emph{the same} rank $3$ sheaf $U_3$ with $c_1(U_3)=0$. By inspection of \eref{qdefs} it is straightforward to verify that such a reducible locus in the moduli space of $Q_4$ and $\tilde{Q}_4$ does indeed exist and here $U_3$ is in fact locally free (i.e. an $SU(3)$ bundle) itself described as a monad:
\beq
 0 \to U_3 \to  \cO(0,1)^{\oplus 2}\oplus \cO(2,1) \oplus \cO(1,2)^{\oplus 2} \oplus \cO(2,2) \to \cO(3,2) \oplus \cO(1,4) \oplus \cO(2,3) \to 0 
\eeq
One less obvious observation can be made using the tools of Appendix \ref{AppendixA}: the bundle $U_3$ is properly stable for the stability wall, $t^2=4t^1$. Thus, in summary, the only locus in the moduli space of the monad bundle $V$, given in \eref{bundle2a} for which the bundle is poly-stable and supersymmetry is preserved is when the map $f$ is chosen so that the bundle is reducible:
\beq\label{decomp_on_wall}
V \to U_3 \oplus L \oplus L^{\vee}
\eeq
with $L=\cO(1,-1)$. 

The consequences of this decomposition are immediate for the effective target space theory. Since we began with a seemingly indecomposable $SU(5)$ bundle the natural expectation would have been to find a $4$-dimensional $SU(5)$ effective theory (since $SU(5) \times SU(5)$ is a maximal subgroup of $E_8$). However, once the D-term is taken into account we see that the structure group of the bundle on the stability wall has decreased to $S[U(1) \times U(1)] \times SU(3) \simeq U(1) \times SU(3)$ which has a commutant within $E_8$ of $SU(6) \times U(1)$. The $U(1)$ symmetry is in general Green-Schwarz massive \cite{Lukas:1999nh} and as described in Section \ref{potential_hym}, leads to a non-trivial D-term constraint. Thus \emph{the presence of a stability wall in the theory has led not only to a reduction in K\"ahler moduli, but also to a non-Abelian enhancement of the $4$-dimensional gauge symmetry}.

The charged matter spectrum can be determined by beginning with the decomposition \cite{Slansky:1981yr}
\begin{equation}
 {\bf 248}_{E_8}\rightarrow \left[({\bf 3},{\bf 1},{\bf 1})\oplus ({\bf 1},{\bf 8},{\bf 1})\oplus ({\bf 1},{\bf 1},{\bf 35})\oplus ({\bf 1},{\bf 3},\overline{\bf 15})
 \oplus ({\bf 1},\overline{\bf 3},{\bf 15})
 \oplus ({\bf 2},{\bf 3},{\bf 6})\oplus ({\bf 2},\overline{\bf 3},\overline{\bf 6})\oplus ({\bf 2},{\bf 1},{\bf 20})\right]\label{248dec}
\end{equation} 
of the adjoint of $E_8$ into $SU(2) \times SU(3) \times SU(6)$. Then the further reduction of $SU(2) \to S[U(1) \times U(1)]$ allows us to read off the final charged matter content of the $SU(6) \times U(1)$ theory\footnote{Note that locally $S[U(n) \times U(1)] \simeq SU(n) \times U(1)$. Globally, however, there is a different normalization on the $U(1)$ which is comprises part of the commutant of this symmetry within $E_8$. Because it will not affect the vacuum analysis in this section, for simplicity we will follow the local charge normalization conventions of \cite{Slansky:1981yr} for all $U(1)$ charges in this work.}. The result is given in Table \ref{table1}.

\begin{table}
\begin{centering}
\begin{tabular}{|c|c|c|c|c|c|}\hline
Field & Cohom. & Multiplicity & Field & Cohom. & Multiplicity \\ \hline
${\bf 1}_{+2}$ & $H^1(L \otimes L)$ & 0 &${\bf 1}_{-2}$ & $H^1(L^{\vee} \otimes L^{\vee})$ & $10$  \\ \hline
${\bf 15}_0$ & $H^1({U_3}^{\vee})$ & 0 & $\overline{ {\bf 15}}_0$ & $H^1(U_3)$ & 80 \\ \hline
${\bf 20}_{+1}$ & $H^1(L)$ & 0 & ${\bf 20}_{-1}$ & $H^1(L^{\vee})$ & 0 \\ \hline
${\bf 6}_{+1}$ & $H^1(L \otimes U_3)$ & 72 & ${\bf 6}_{-1}$ & $H^1(L^{\vee} \otimes U_3)$ & 90 \\ \hline
${\bf \overline{6}}_{+1}$ & $H^1(L \otimes {U_{3}}^{\vee})$ & 0 & $\overline{\bf{6}}_{-1}$ & $H^1(L^{\vee} \otimes {U_3}^{\vee})$ & 2 \\ \hline
${\bf 1}_0$ & $H^1(U_3 \otimes {U_3}^{\vee})$ & 166 & & & \\
\hline
\end{tabular}
\caption{\it Particle content of the $SU(6) \times U(1)$ theory associated to the bundle \eref{bundle2a} along its reducible and poly-stable  locus $V= \cO(-1,1)\oplus \cO(1,-1) \oplus U_3$ (i.e. on the stability wall given by \eref{constrained_t} and shown in Figure \ref{fig:firstwall}).}\label{table1}
\end{centering}
\end{table}
In addition to the D-terms associated to the non-Abelian $SU(6)$ gauge symmetry there is a D-term associated to the Green-Schwarz massive $U(1)$ which is self commuting in the $S[U(1) \times U(1)]$ factor in $E_8$. Written in the expansion parameters of heterotic M-theory, this takes the schematic form (see \cite{Anderson:2009nt} for details):
\beq
D^{U(1)} \sim \frac{3}{16}\frac{\epsilon_S {\epsilon_R}^2 \mu(\cal{F})}{{\kappa_4}^2 Vol(X)}- \frac{1}{2} \sum_{i} Q_iG_{P\bar{Q}} {C_i}^P{\bar{C}_i}^{\bar Q}
\eeq
where $Vol(X)$ is the volume of the CY threefold $X$, $G_{P\bar{Q}}$ is the field space metric, and $C_i$ denotes any of the matter fields charged under the Green-Schwarz massive $U(1)$ symmetry (with charge $Q_i$) appearing in Table \ref{table1}. Once again, we note the key feature of this example: because of the non-trivial D-term above, when $\langle C \rangle =0$ (corresponding to the monad in \eref{bundle2a}), the D-term non-trivially constrains the K\"ahler moduli through the condition that $\mu({\cal F})=0$ and the moduli of the theory are not those counted in \eref{naive_count} but rather the $dim({\cal M})=427$ moduli in \eref{real_count}.

We will explore the vacuum branch structure of this theory in more detail in Section \ref{branch_sec}, but for now we turn to the target space dual theories.

\subsection{Target Space Duals of Theories with D-terms}
Following the target space duality algorithm as described in Section \ref{the_algorithm} (including repeated line bundles as in Section \ref{repeated_line_bundles}) it is possible to generate 16 target space duals for the pair $(X,V)$ given in \eref{bundle2a} in which the new manifolds, $\widetilde{X}_i$, are still described as CICYs. The complete chain of target space dual theories is given in Appendix \ref{appendixB}. To investigate these results, a simple starting point is given below -- a dual for which the Calabi-Yau manifold, $X$, is related to the original threefold by a geometric transition.

\begin{equation}\begin{aligned}&\begin{array}{|c|c| |c|c|}
\hline
x_i & \Gamma^j & \Lambda^a & p_l \\ \noalign{\hrule height 1pt} \begin{array}{cccccccc}
0& 0& 0& 0& 0& 0& 1& 1 \\
1& 1& 0& 0& 0& 0& 0& 0 \\
0& 0& 1& 1& 1& 1& 0& 0 
\end{array}&\begin{array}{cc}
 -1& -1  \\
 -2&  0 \\
 -2& -2 
\end{array}&\begin{array}{cccccccc}
  0&  0&  1& 0& 0& 0& 0& 0  \\
 1& -1&  0& 0& 2& 1& 1& 2  \\
 -1&  1& -1& 1& 3& 2& 2& 2 
\end{array}&\begin{array}{ccc}
  0&  0& -1 \\
 -3& -1& -2 \\
 -2& -4& -3 
\end{array}\\
\hline
\end{array}\end{aligned}\label{bundle2a.1}\end{equation}

In view of the results of the previous subsection, obvious questions include:
\begin{enumerate}
\item Does $(\widetilde{X},{\widetilde V})$ give rise to a stability wall?
\item Is $dim(\widetilde{\cal{M}}_{1})$ also reduced in dimension compared to $dim(\widetilde{\cal{M}}_{0})$? Are the number of lifted moduli the same?
\item Does the structure group of ${\widetilde V}$ also reduce, leading to a $4$-dimensional non-Abelian enhancement of symmetry (i.e. $SU(6)$)?
\item Do the charged matter spectra of the two theories match?
\end{enumerate}
We will address each of these questions in turn.

First, the complete intersection CY (CICY) manifold and rank $5$ vector bundle given in \eref{bundle2a.1} satisfies $h^{1,1}(\tilde{X})=3$, $h^{2,1}(\tilde{X})=55$ and
\beq
c_2(\tilde{V})=c_2(T\tilde{X})=\{24,24,44\}
\eeq
and the third Chern class is $c_3(\tilde{V})=-80$ as expected. The comparison to \eref{bundle2a} can begin with the uncharged (geometric) moduli of the theory:
\beq\label{naive_count_mirror}
dim({\cal \widetilde{M}}_{0})=h^{1,1}(\widetilde{X})+h^{2,1}(\widetilde{X})+h^1(X,End_0(\widetilde{V}))=3+55+370=428
\eeq
Once again, it is important to observe that this naive moduli count may not be the correct zero mode count for the theory. As illustrated in the previous Section, this count is only preliminary since we have not yet taken into account any non-trivial D-term potential associated to the geometry above. 

By inspection, the negative charges in \eref{bundle2a.1} indicate that the bundle will possess stability walls (see Appendix \ref{AppendixA}). For the bundle in \eref{bundle2a.1} it can be verified that there are now \emph{three} de-stabilizing subsheaves rather than the two of \eref{bundle2a}. Explicitly, these are given by
\begin{align}
& 0 \to {\cal \tilde{F}}_1 \to \tilde{V} \to \cO(0,1,-1) \to 0 & c_1({\cal \tilde{F}}_1)=(0,-1,1) \\
& 0 \to {\cal \tilde{F}}_2 \to \tilde{V} \to \cO(0,-1,1) \to 0 & c_1({\cal \tilde{F}}_2)=(0,1,-1) \\
& 0 \to {\cal \tilde{F}}_3 \to \tilde{V} \to \cO(1,0,-1) \to 0 & c_1({\cal \tilde{F}}_3)=(-1,0,1)
\end{align}
To proceed, it can be noted that the non-vanishing triple intersection numbers for $\widetilde{X}$ are 
\beq
d_{123}=d_{133}=d_{233}=4~~~,~~~d_{333}=2
\eeq

\begin{figure}
\centering \includegraphics[width=0.4\textwidth]{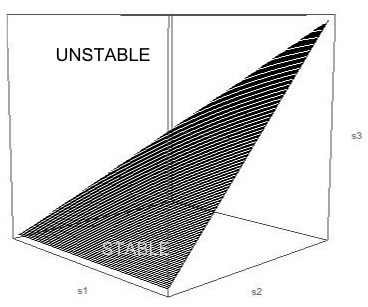} \caption{\emph{The $2$-dimensional stable sub-cone in K\"ahler moduli space shown for the monad bundle in \eref{bundle2a.1}. Along this co-dimension 1 sub-cone the bundle takes the poly-stable form shown in \eref{mirror_walltheory}.}} \label{fig:secondwall}
\end{figure}

Combining this with the stability conditions that $\mu({\cal \tilde{F}}_i) \leq 0$ for $i=1,2,3$ leads to the constraint that (for $t^i >0$ in the interior of the K\"ahler cone) that $t^2>t^1$ and $t^3= 2t^2+2\sqrt{t^1 t^2 + (t^2)^2}$. In the dual K\"ahler variables this takes the form of a simple plane the 3-dimensional K\"ahler cone of $\widetilde{X}$:
\beq\label{mirror_wall}
s_3 < s_1~~~\text{and}~~~s_2=s_3
\eeq
The 2-dimensional stable plane is shown in Figure \ref{fig:secondwall}. 

Although the number of de-stabilizing subsheaves differs in the target space dual theory, the total number of moduli has been precisely matched! That is, in this case we have
\beq
dim({\cal \widetilde{M}}_1)=dim({\cal \widetilde{M}}_0)-1=427
\eeq
exactly as required for a correspondence of the target space theories. Thus, the first two questions posed in the previous section are answered: \emph{The presence of the stability wall (i.e. the locus of enhanced symmetry) and the reduction in moduli due to the D-term constraint are both exactly mirrored in the target space dual theory.} 

To see if the enhancement of non-Abelian symmetry is the same as that for the bundle given in \eref{bundle2a} it should be noted that the bundle in \eref{bundle2a.1} has a poly-stable decomposition along the $2$-dimensional stability wall given by \eref{mirror_wall} as
\beq\label{mirror_walltheory}
\tilde{V} \to (\tilde{L} \oplus {\tilde{L}}^{\vee})\oplus \widetilde{U}_3
\eeq
with $\tilde{L}=\cO(0,1,-1)$ and the rank $3$ bundle $\tilde{U}_3$ is given by the monad
{\small \beq
0 \to \tilde{U}_3 \to \cO(1,0,-1)\oplus \cO(0,0,1) \oplus \cO(0,2,3)\oplus \cO(0,1,2)^{\oplus 2} \oplus \cO(0,2,2) \to \cO(0,3,2) \oplus \cO(0,1,4) \oplus \cO(1,2,3) \to 0
\eeq}
The presence of the negative line bundle, $\cO(1,0,-1)$ in $\widetilde{U}_3$ indicates that there is a sub-locus in moduli space where $\widetilde{V}$ can be decomposed still further (namely, where $\widetilde{U}_3 \to \cO(1,0,-1) \oplus \widetilde{U}_2$). However, unlike the decomposition in \eref{mirror_walltheory} which is \emph{required} by stability of $\widetilde{V}$, this further decomposition is not generic and would correspond to some further tuning in bundle moduli of \eref{bundle2a} in the dual theory.

We are now in a position to answer yet another of our questions. From the form of \eref{mirror_walltheory} it is clear that the \emph{non-Abelian symmetry enhancement is identical}. Once again, on the wall, the $4$-dimensional gauge symmetry enhances from $SU(5) \to SU(6)$. Furthermore, in this case the number of Green-Schwarz massive $U(1)$ symmetries appears to to be identical\footnote{Although the Green-Schwarz massive $U(1)$ symmetry is identical in this example, in general one would not expect that a discrepancy in the number of such symmetries should signal a breakdown of duality. The mass scale of the massive $U(1)$ symmetries is only just below the compactification scale \cite{Anderson:2009sw,Anderson:2009nt} and care should be taken with how they are included in the effective theories. In general, while the presence of the Green-Schwarz massive $U(1)$ can affect the structure of the theory (for example in Yukawa textures and the structure of the matter field K\"ahler potential \cite{Anderson:2010tc,Anderson:2011ns,Anderson:2013xka,Anderson:2012yf,Anderson:2014hia}), it should not be expected that the massive $U(1)$ gauge bosons themselves should be visible in the low-energy effective theory and the difference noted above does not indicate a failure of duality.}.

\begin{table}
\begin{centering}
\begin{tabular}{|c|c|c|c|c|c|}\hline
Field & Cohom. & Multiplicity & Field & Cohom. & Multiplicity \\ \hline
${\bf 1}_{+2}$ & $H^1(\tilde{L} \otimes \tilde{L})$ & 0 &${\bf 1}_{-2}$ & $H^1(\tilde{L}^{\vee} \otimes \tilde{L}^{\vee})$ & $10$  \\ \hline
${\bf 15}_0$ & $H^1({\tilde{U}_3}^{\vee})$ & 0 & $\overline{ {\bf 15}}_0$ & $H^1(\tilde{U}_3)$ & 80 \\ \hline
${\bf 20}_{+1}$ & $H^1(\tilde{L})$ & 0 & ${\bf 20}_{-1}$ & $H^1(\tilde{L}^{\vee})$ & 0 \\ \hline
${\bf 6}_{+1}$ & $H^1(\tilde{L} \otimes \tilde{U}_3)$ & 72 & ${\bf 6}_{-1}$ & $H^1(\tilde{L}^{\vee} \otimes \tilde{U}_3)$ & 90 \\ \hline
${\bf \overline{6}}_{+1}$ & $H^1(\tilde{L} \otimes {\tilde{U}_{3}}^{\vee})$ & 0 & $\overline{\bf{6}}_{-1}$ & $H^1(\tilde{L}^{\vee} \otimes {\tilde{U}_3}^{\vee})$ & 2 \\ \hline
${\bf 1}_0$ & $H^1(\tilde{U}_3 \otimes {\tilde{U}_3}^{\vee})$ & 196 & & & \\
\hline
\end{tabular}
\caption{\it Particle content of the $SU(6) \times U(1)$ theory associated to the bundle \eref{bundle2a.1} along its reducible and poly-stable locus $V=\tilde{L} \oplus {\tilde{L}}^{\vee} \oplus  \tilde{U}_3$ (i.e. on the \emph{stability wall} given by \eref{mirror_wall} and shown in Figure \ref{fig:secondwall}).}\label{table2}
\end{centering}
\end{table}

To address the final question, that of whether the $SU(6)$-charged matter spectra of the theories agree, the cohomology of the reducible bundle in \eref{bundle2a.1} (equivalently \eref{mirror_walltheory}) can be computed. The results are given in Table \ref{table2}. By inspection, it can be verified that the number of fields in the ${\bf 15}, {\bf 20}, {\bf 6}$, etc representations is identical to the particle content given in Table \ref{table1}. Moreover, the $U(1)$-charges under the GS massive $U(1)$ symmetries are identical and would appear to lead to identical constraints from gauge invariance on the ${\cal N}=1$ $4$-dimensional theory!\footnote{It would be illuminating to probe the effect of higher-codimensional loci such as that determined by $\widetilde{U}_3 \to \cO(1,0,-1) \oplus \widetilde{U}_2$ and its effect on Yukawa couplings and compare this with special loci in the moduli space of the target space dual. At the moment however, this is difficult to accomplish without an explicit moduli map between $(X,V)$ and $(\widetilde{X},\widetilde{V})$.}.

Beyond the geometry listed in \eref{bundle2a.1}, there are another $15$ target space dual theories in the chain related to \eref{bundle2a}. These are listed explicitly in Appendix \ref{appendixB} and share many of the features of the dual theory described above. In each case we can verify the existence of a stability wall in the new theory and in each case
\beq
dim({\cal \widetilde{M}}_1)=427
\eeq
However, in 6 out of 16 we find a surprising feature which is that the stability condition forces the theory to the boundary of the K\"ahler cone. Such a point is a metric singularity and corresponds to a perturbative/non-perturbative duality and will be explored in Section \ref{non_pert_duals}.

\subsection{Branch Structure in the vacuum space of the dual theories}\label{branch_sec}

An intriguing aspect of theories with stability walls is the rich branch structure in the vacuum space of the theories. As described in \cite{Anderson:2009nt,Anderson:2009sw,Anderson:2010ty} the distinct branches of the vacuum moduli space near the wall correspond to distinct infinitesimal deformations at the level of the bundle geometry. These walls also provide non-trivial information about discrete symmetries and Yukawa textures in the effective target space theories away from the wall \cite{Kuriyama:2008pv,Anderson:2010tc}.

As a result of this, heterotic theories with stability walls provide a further detailed check of the target space duality correspondence. We have seen in the previous sections that a stability wall in one theory, $(X,V)$ is reproduced in its target space dual $(\widetilde{X},\widetilde{V})$ and that the charged and singlet matter of the two wall theories matches exactly. In the next sections we briefly explore the branch structures in these two moduli spaces away from the stability walls (i.e. away from the locus of enhanced symmetry. To begin, consider the bundle given in \eref{bundle2a} and its associated $SU(6) \times U(1)$ theory with the particle spectrum given in Table \ref{table1}. 

We will consider deformations away from the stability wall that lead to indecomposable, properly stable, rank $5$ vector bundles. Such a deformation of bundle geometry would correspond in the $4$-dimensional theory to a Higgsing $SU(6) \to SU(5)$ of the effective theory along a flat direction. Several such directions can be readily identified. First, there are two non-trivial D-term constraints on the theory corresponding to the Green-Schwarz massive $U(1)$, and to the $U(1)$ associated with $SU(5) \times U(1) \subset SU(6)$ (${\bf 6} \to {\bf 1}_{-5} + {\bf 5}_{+1}$). Note we will take the Vevs of all fields charged under $SU(5) \subset SU(6)$ to be vanishing. 

\subsubsection{Branches for $(X,V)$}\label{branch_in_v}
We begin with the geometry in \eref{bundle2a}. Taking into account the spectrum of Table \ref{table1} and adding a second $U(1)$ charge subscript for the $U(1) \subset SU(6)$ described above, the D-terms take the schematic form:
\begin{align}
&D^{U(1)}_{GS} \sim \frac{3}{16}\frac{\epsilon_S {\epsilon_R}^2 \mu(L^{\vee})}{{\kappa_4}^2 {\cal V}}- \frac{1}{2}\left((-2)|C_{-2,0}|^2+(+1)|C_{+1,-5}|^2+(-1)|C_{-1,-5}|^2+(-1)|C_{-1,+5}|^2\right)\label{du1} \\
&D^{U(1)}_{SU(6)}\sim \frac{1}{2}\left((-5)|C_{+1,-5}|^2+(-5)|C_{-1,-5}|^2+(+5)|C_{-1,+5}|^2\right) \label{dsu6}
\end{align}
In the above, sums over multiplicities of each type of term are suppressed and for each category of charged matter field in Table \ref{table1} we abbreviate $\sum_{i} Q_iG_{P\bar{Q}} {C_i}^P{\bar{C}_i}^{\bar Q}\simeq Q|C|^2$. In considering the vacuum structure, note that \eref{dsu6} leads to a condition of the form
\beq\label{vevsbranch1}
\langle C_{-1,+5} \rangle^2 \sim \langle C_{-1,-5} \rangle^2+ \langle C_{+1,-5} \rangle^2
\eeq
However, it is also clear that substituting this constraint into to \eref{du1} we find that D-flatness requires that
\beq\label{vevsbranch2}
\frac{3}{16}\frac{\epsilon_S {\epsilon_R}^2 \mu(L^{\vee})}{{\kappa_4}^2 Vol(X)}+\frac{1}{2} \left(\langle C_{-2,0}\rangle^2  +\langle C_{-1,-5} \rangle^2\right) \sim 0
\eeq
Since this involves only matter fields with negative charge under the first (GS massive) $U(1)$ symmetry, it is clear that \emph{all vacuum solutions breaking $SU(6) \to SU(5)$ with $\langle C \rangle \neq 0$ must involve a non-trivial FI term with $\mu(L^{\vee})<0$}. 

\begin{figure}
\centering \includegraphics[width=0.4\textwidth]{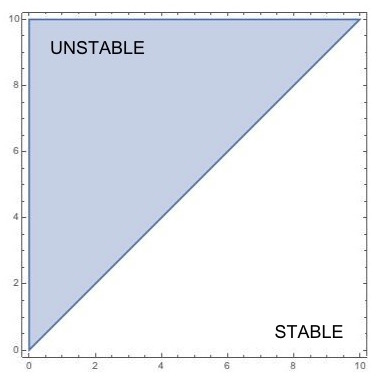} \caption{\emph{The stable sub-cone in K\"ahler moduli space shown for the $SU(5)$ monad bundle in \eref{branch2a}. This corresponds to the branch of vacuum space described by \eref{vevsbranch1} and \eref{vevsbranch2}.}} \label{fig:branchchamber1}
\end{figure}

The indecomposable, $SU(5)$ bundle corresponding to this deformation will be properly slope-stable only in the region of K\"ahler moduli space for which $\mu(L^{\vee}) <0$. This chamber structure is depicted in Figure \ref{fig:branchchamber1}. 

 To describe this bundle as a monad (and hence, build the associated GLSM) we need two facts: 1) the new monad bundle corresponding to this branch of the vacuum space must admit a point in its moduli space for which it can be decomposed as in \eref{decomp_on_wall} and 2) By the results of Appendix \ref{AppendixA}, if both $\cO(1,-1)$ and $\cO(-1,1)$ appear in the monad, it cannot be stable in the region shown in Figure \ref{fig:branchchamber1}.
A useful technical observation is that to accommodate both these facts, it is clear that one of the line bundles $L$ or $L^{\vee}$ must be ``replaced" in the monad with an alternate description. To that end, we observe that $\cO(-1,1)$ itself can be written as a monad:
\beq\label{lnew}
0 \to L_{new} \to \mathcal O(0,1)^{\oplus 2} \stackrel{g}{\longrightarrow} \mathcal O(1,1) \to 0
\eeq
This monad describes $L_{new}$ as a rank 1 sheaf given by the kernel of the map $g$, with $c_1(L_{new})=-J_1 + J_2$. Since line bundles on CY 3-folds are classified by their first Chern classes, it remains to show only that \eref{lnew} describes a locally free sheaf. To this end, recall that the kernel of a monad sequence defines a locally free sheaf only if the map satisfies the transversality condition described in Section \ref{sec:GLSM} (i.e. the condition that the rank of $g$ does not drop anywhere on $X$). In this case, the map, $g$, can be written as a vector of two linear functions of the ambient $\mathbb{P}^1$ coordinates which manifestly do not vanish simultaneously on $X$, thus $L_{new}\simeq \cO(-1,1)$.

As a result, the new branch, corresponding to deforming off of the stability wall is given by the following monad:
\begin{equation}\begin{aligned}&\begin{array}{|c|c| |c|c|}
\hline
x_i & \Gamma^j & \Lambda^a & p_l \\ \noalign{\hrule height 1pt}\begin{array}{cccccc}
1& 1& 0& 0& 0& 0  \\
0& 0& 1& 1& 1& 1 
\end{array}&\begin{array}{c}
 -2  \\
 -4 
\end{array}&\begin{array}{ccccccccc}
 1& 0& 0& 0& 0& 2& 1& 1& 2  \\
 -1&  1& 1& 1& 1& 1& 2& 2& 2 
\end{array}&\begin{array}{cccc}
-1& -3& -1& -2 \\
-1& -2& -4& -3 
\end{array}\\
\hline
\end{array} \label{branch2a}\end{aligned}\end{equation}
The $SU(5)$ charged matter spectrum associated to this bundle can be computed directly and is shown in Table \ref{table3}. Note that here
\beq\label{branchy_total}
h^{1,1}(X)+h^{2,1}(X)+h^1(End_0(V))=2+86+338=426
\eeq
In this case \emph{there are no obstructions in moduli} and $dim({\cal M}_1)=dim ({\cal M}_0)=426$ since despite the presence of the stability wall forming a sub-chamber of K\"ahler moduli space, there are still two free degrees of freedom associated to the K\"ahler moduli. More importantly, the count in Table \ref{table3} matches exactly with what is expected from the Higgsing in the vacuum analysis above with precisely two degrees of freedom are removed by the D-terms above. This agrees with comparison to \eref{real_count} on the branch in moduli space corresponding to $\langle C \rangle \neq 0$ (i.e. ``gluing" the reducible bundle in \eref{decomp_on_wall} back into the irreducible $SU(5)$ bundle given in \eref{branch2a}). In addition, in the Higgsing $SU(6) \to SU(5)$ a number of ${\bf 5},{\bf \bar{5}}$ pairs can become massive\footnote{We thank J. Gray for useful suggestions on computer algebra methods \cite{Gray:2008zs} used in the computation of the particle spectrum in Table \ref{table3}.}. The exact number depends on the detailed form of the effective potential, for example, the number of non-vanishing couplings of the form
\beq
\langle 1_{+1,-5} \rangle {\bf \bar{5}}_{0,+4} {\bf 5}_{-1,+1}~~\text{and}~~ \langle 1_{-1,-5} \rangle {\bf \bar{5}}_{0,+4} {\bf 5}_{+1,+1}
\eeq
induced from the ${\bf \overline{15}}_0{\bf 6}_{-1}{\bf 6}_{+1}$ couplings of the original $SU(6) \times U(1)$ theory. These moduli-dependent mass terms are realized geometrically in Table \ref{table3} by the dependence of the ${\bf 5},{\bf \bar{5}}$ spectrum on the monad map.

 The fact that we can ``see" the complete branch structure of the theory on the stability wall (and off of it) both in terms of the stability and D-term analysis and as monad bundles leads to a complete and clear picture. It only remains to compare these results to those in the target space dual theories.
\begin{table}
\begin{centering}
\begin{tabular}{|c|c|c|c|c|c|}\hline
Field & Cohom. & Multiplicity & Field & Cohom. & Multiplicity \\ \hline
${\bf 1}$ & $H^1(V \otimes V^{\vee})$ & 338 & &  &   \\ \hline
${\bf 10}$ & $H^1(V^{\vee})$ & 0 & $\overline{ {\bf 10}}$ & $H^1(V)$ & $80$ \\ \hline
${\bf 5}$ & $H^1(\wedge^2 V)$ & $81+x$ & $\overline{{\bf 5}}$ & $H^1(\wedge^2 V^{\vee})$ & $1+x$ \\ \hline
\end{tabular}
\caption{\it Particle content of the $SU(5)$ theory associated to the indecomposable $SU(5)$ bundle \eref{branch2a} -- the branch of the theory in the stable chamber of K\"ahler moduli space. See Figure \ref{fig:branchchamber1}. The ambiguity (denoted by $x$) in the ${\bf 5},{\bf \bar{5}}$ spectrum arises from its dependence on the monad map in \eref{branch2a}.}\label{table3}
\end{centering}
\end{table}

\subsubsection{Branches for $(\widetilde{X},\widetilde{V})$}\label{branch_in_dual}
In the case of the dual geometry in \eref{bundle2a.1}, the symmetry $SU(6) \times U(1)$ and the charged matter content is precisely identical (see Table \ref{table2}). The only difference arises in the count of the uncharged singlets $h^1(\widetilde{X},\tilde{U}_3 \otimes \tilde{U}_3^{\vee})=196$. Thus, the vacuum structure of the theory under the Higgsing $SU(6) \to SU(5)$ is also \emph{manifestly the same}.

As in the previous Section, the vacuum equations, \eref{vevsbranch1} and \eref{vevsbranch2} indicate that there will be a solution when $\mu(\tilde{L}^{\vee}) <0$. It only remains to understand how to geometrically realize this deformation as a monad.

From here, the analysis exactly mirrors that of the previous theory. There is only one branch to the theory leaving the stability wall and breaking $SU(6) \to SU(5)$. As before we can replace one negative entry line bundle in the monad (in this case, $\cO(0,-1,1)$) as
\beq
0 \to \tilde{L}_{new} \to \mathcal O(0,0,1)^{\oplus 2} \stackrel{\tilde{g}}{\longrightarrow} \mathcal O(0,1,1) \to 0
\eeq
This leads at last to the bundle
\begin{equation}\begin{aligned}&\begin{array}{|c|c| |c|c|}
\hline
x_i & \Gamma^j & \Lambda^a & p_l \\ \noalign{\hrule height 1pt} \begin{array}{cccccccc}
0& 0& 0& 0& 0& 0& 1& 1 \\
1& 1& 0& 0& 0& 0& 0& 0 \\
0& 0& 1& 1& 1& 1& 0& 0 
\end{array}&\begin{array}{cc}
 -1& -1  \\
 -2&  0 \\
 -2& -2 
\end{array}&\begin{array}{ccccccccc}
  0&  0& 0&  1& 0& 0& 0& 0& 0  \\
 1& 0& 0& 0& 0& 2& 1& 1& 2  \\
 -1&  1&1& -1& 1& 3& 2& 2& 2 
\end{array}&\begin{array}{cccc}
 0& 0&  0& -1 \\
-1& -3& -1& -2 \\
 -1&-2& -4& -3 
\end{array}\\
\hline
\end{array}\end{aligned}\label{branch2a.1}\end{equation}

The charged matter of this $SU(5)$ theory matches exactly to that given in Table \ref{table3}. Once again, the only difference arises in the number of uncharged moduli. In this case we find
\beq
h^{1,1}(\widetilde{X})+h^{2,1}(\widetilde{X})+h^1(\widetilde{X},End_0(\widetilde{V}))=3+55+368=426
\eeq
In agreement with \eref{branchy_total}.

\begin{figure}
\centering \includegraphics[width=0.4\textwidth]{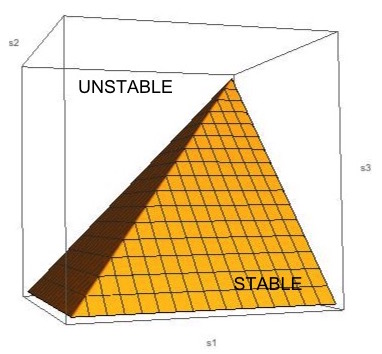} \caption{\emph{The stable sub-cone in K\"ahler moduli space shown for the $SU(5)$ monad bundle in \eref{branch2a.1}. This corresponds to the Higgsed branch of the target space dual theory in \eref{bundle2a.1} and Figure \ref{fig:secondwall}.}} \label{fig:branchchamber2}
\end{figure}

This bundle is stable in a 3-dimensional chamber of the K\"ahler cone of $\widetilde{X}$ determined by
\beq
\mu(\cO(0,-1,1))<0~~\text{and}~~\mu(\cO(-1,0,1)<0~~\Leftrightarrow~~ s_3<s_2~~\text{and}~~s_3<s_1
\eeq 
This region is shown in Figure \ref{fig:branchchamber2}.

With this monad bundle in hand we have formed a complete picture of the local deformation space near the stability wall in both the original theory and its target space dual. It is clear from the above analysis that not only do the loci of enhanced symmetry correspond in the two theories, but their vacuum branch structure is also identical. In the end we have a commutative diagram of geometric deformations and target space duals (i.e. ``transgressed" \cite{Candelas:2007ac} geometries):
\begin{equation}
\begin{array}{lllll}
&V_1&\stackrel{dual}{\longrightarrow}&\widetilde{V}_1&\\
 \langle C \rangle &\downarrow&&\downarrow&\langle \tilde{C} \rangle \\
&V_2 &\stackrel{dual}{\longrightarrow}&\widetilde{V}_2&
\end{array}
\label{first_def}
\end{equation}
It can be verified that the deformed manifolds/bundles (labeled as $V_2$ and $\widetilde{V}_2$ above) given in \eref{branch2a} and \eref{branch2a.1} also form a target space dual pair, as expected. Thus, the correspondence given in \eref{first_def} is complete.

\subsection{Perturbative/Non-perturbative duality}\label{non_pert_duals}
Of the $17$ bundles in the target space dual chain associated to \eref{bundle2a} (see Appendix \ref{appendixB}), the majority of them exhibit the same essential features seen in the previous example, including: a) new stability walls in the bundle geometry, b) a matching reduction of dimension arising from the associated D-terms, c) Agreement in the degrees of freedom (both charged and uncharged) in the target space effective theory and d) perfect correspondence between the vacuum branch loci near the stability wall as described in \eref{first_def}.

There are however, several interesting exceptions that arise and we turn to one such now. The following target space bundle
\begin{equation}\begin{aligned}&\begin{array}{|c|c| |c|c|}
\hline
x_i & \Gamma^j & \Lambda^a & p_l \\ \noalign{\hrule height 1pt}\begin{array}{cccccccc}
0& 0& 0& 0& 0& 0& 1& 1  \\
1& 1& 0& 0& 0& 0& 0& 0 \\
0& 0& 1& 1& 1& 1& 0& 0  
\end{array}&\begin{array}{cc}
 -1& -1  \\
 -2&  0  \\
 -3& -1 
\end{array}&\begin{array}{ccccccccc}
  1&  0& 0& 0& 0& 0& 0& 0& 0  \\
  1& -1& 0& 0& 2& 1& 1& 2& 3  \\
 -2&  1& 1& 1& 1& 2& 2& 2& 2  
\end{array}&\begin{array}{cccc}
 -1&  0&  0&  0 \\
 -3& -1& -2& -3 \\
 -2& -4& -3& -1 
\end{array}\\
\hline
\end{array}\end{aligned}\label{non_pert_dual}\end{equation}
Once again exhibits a stability wall and shows the exact reduction in Kahler moduli expected by the D-term condition. However, the new feature in this case is that the stability wall is \emph{forced to the boundary of the K\"ahler cone.} That is, the target space dual leads to a theory that is necessarily at a singular point in moduli space. For the CY 3-fold given in \eref{non_pert_dual}, the K\"ahler cone is given by the positive octant $t^a >0$, $a=1,2,3$ and the D-term condition on the stability wall constrains the Ka\"ahler moduli of $X$ to
\beq\label{weird_wall_cond}
t^1=0~,~~t^2 \geq 0~,~~t^3=2t^2-2\sqrt{(t^2)^2}~~~\text{or}~~~t^1>0~,~~t^2 = 0~,~~t^3=\frac{3}{2}\left(t^1 \pm \sqrt{(t^1)^2} \right)
\eeq
Although the overall volume is still finite when \eref{weird_wall_cond} holds, the presence of small cycles leads us to expect the appearance of non-perturbative physics. Indeed, a closer inspection of \eref{non_pert_dual} suggests that the metric singularities in $X$ must play an important role in the effective physics. On the stability wall, the bundle in \eref{non_pert_dual} splits as
\beq
V_5 \to L_1 \oplus L_2 \oplus U_3
\eeq 
with $L_1=\cO(1,1,-2)$, $L_2=\cO(0,-1,1)$ and $c_1(U_3)=(-1,0,1)$. The structure group of the bundle on the stability wall is $S[U(1)\times U(1) \times U(3)]$ which has commutant $SU(5) \times U(1) \times U(1)$ within $E_8$. Thus, this bundle naively does not lead to the same non-Abelian enhancement of symmetry (i.e. $SU(6)$) seen in the initial bundle \eref{bundle2a}. This is not wholly unexpected however, since it is well-known that singularities in the base manifold, $X$, can force bundles (and more generally small instantons) to exhibit non-Abelian enhancements of symmetry (see for example \cite{Berglund:1998ej}). We leave a full exploration of this interesting class of examples to further work.

\subsection{Isomorphic geometry in target space duality}\label{isomorphic_CYs}
An interesting possibility within the target space duality algorithm includes cases in which the base manifold $X$ is mapped to an identical or related geometry. As an example, consider the following target space dual to the geometry given in \eref{bundle2a}:
\begin{equation}\begin{aligned}&\begin{array}{|c|c| |c|c|}
\hline
x_i & \Gamma^j & \Lambda^a & p_l \\ \noalign{\hrule height 1pt}\begin{array}{cccccccc}
0& 0& 0& 0& 0& 0& 1& 1  \\
1& 1& 0& 0& 0& 0& 0& 0 \\
0& 0& 1& 1& 1& 1& 0& 0 
\end{array}&\begin{array}{cc}
 -1& -1  \\
 -1& -1  \\
 -4& \ 0 
\end{array}&\begin{array}{ccccccccc}
  1&  0& 0& 0& 0& 0& 0& 0& 0  \\
  0& -1& 0& 0& 2& 1& 1& 2& 2  \\
 -1&  1& 1& 1& 1& 2& 2& 2& 3 
\end{array}&\begin{array}{cccc}
  0&  0& -1&  0 \\
 -3& -1& -2& -1 \\
 -2& -4& -3& -3 
\end{array}\\
\hline
\end{array}\end{aligned}\label{bundle2a.11m_study}\end{equation}
 It is well known that many CICY threefolds can have redundant descriptions \cite{Candelas:1987kf} and it is straightforward to verify that the CY threefold given in \eref{bundle2a.11m_study} is in fact \emph{the same manifold} as the $\{2,4\}$ hypersurface in $\mathbb{P}^1 \times \mathbb{P}^3$ given in \eref{bundle2a} (this is an example of a so-called ``ineffective split" \cite{Hubsch:1992nu}). Solving the linear constraint in the defining relation of \eref{bundle2a.11m_study} leads to $X \times pt$ with $X$ defined in \eref{bundle2a}. Interestingly, even when $X = \widetilde{X}$ the action on the associated monad bundle is not manifestly trivial. To test whether or not the bundles are also isomorphic, it is possible to employ a useful Lemma (Lemma \ref{OSS_corollary_thm}, Appendix \ref{AppendixA}) for isomorphisms of stable bundles.
 
 Finally, another quantitatively different possibility can arise which $X$ and $\widetilde{X}$ are presented as distinct complete intersection manifolds in different ambient products of projective spaces but are ``redundant" manifolds from the point of view of their topological data. Wall's Theorem \cite{wallthm} says that two (simply connected) CY 3-folds which share the same total Chern classes and triple intersection numbers are of the same homotopy type and from the point of view the effective $4$-dimensional physics, equivalent. For example, the following pair of CY 3-folds have equivalent data in the sense of Wall's theorem:
 \beq
X_1=\left[\begin{array}{c|cc}
\mathbb{P}^1 & 1& 1  \\
 \mathbb{P}^1& 0 & 2  \\
 \mathbb{P}^3 & 2& 2 
\end{array}\right]~~~~~~~~~~,~~~~~~~~X_2=\left[\begin{array}{c|cc}
\mathbb{P}^1 & 1& 1  \\
 \mathbb{P}^1& 1 & 1  \\
 \mathbb{P}^3 & 2& 2 
\end{array}\right]
 \eeq
 Both of these manifolds appear in the list of target space duals to \eref{bundle2a} given in Appendix \ref{appendixB}. Interestingly, the number of bundles appearing over these manifolds in the target space duality procedure can apparently differ (starting with the bundle in \eref{bundle2a} the target space duality algorithm generates 3 bundles over $X_1$ and 5 over $X_2$). This indicates that the target space duality procedure probes some properties of the manifolds beyond their topology.

\subsection{On S-Equivalence Classes}\label{sequivalence_sec}
Given the strong role played by stability walls in the chain of examples in the previous Sections, it is natural to wonder whether all bundles that ``share" a wall must somehow be connected in moduli space? The mathematics associated with this question includes the notion of a so-called ``S-Equivalence Class" \cite{huybrechts_book} this is the unique decomposition of a semi-stable bundle (on the stability wall) into a poly-stable sum
\beq
V \to V_1 \oplus V_2 \oplus \ldots
\eeq
This decomposition on the wall characterizes the physical theory and as we will see below bundles with different S-equivalence classes are generally unconnected in the vacuum space of the ${\cal N}=1$, $4$-dimensional theory. The mathematics of S-equivalence classes and their relationship to stability walls and the decomposition into poly-stable bundles (such as \eref{decomp_on_wall} and \eref{mirror_walltheory}) can be found in Appendix \ref{AppendixA}.

In this section we provide an example to demonstrate that even for bundles with the same topology and a shared stability wall, there can be disconnected components to the moduli space, ${\cal M}_{bundle}(rk, c_1,c_2,c_3)$ of (semi-) stable bundles with fixed rank and total Chern class.

To illustrate this possibility, consider the following pair of bundles on the bi-cubic hypersurface, $\{3,3\}$ in $\mathbb P^2 \times \mathbb P^2$:
\begin{equation}\begin{aligned}&\begin{array}{|c|c| |c|c|}
\hline
x_i & \Gamma^j & \Lambda^a & p_l \\ \noalign{\hrule height 1pt}\begin{array}{cccccc}
1& 1& 1& 0& 0& 0  \\
0& 0& 0& 1& 1& 1 
\end{array}&\begin{array}{c}
-3  \\
 -3 
\end{array}&\begin{array}{cccccccc}
-2& 1& 1& 2& 2& 2& 2& 0 \\
1& 0& 1& 2& 0& 0& 0& 2 
\end{array}&\begin{array}{ccc}
 -4& -3& -1 \\
 -1& -2& -3 
\end{array}\\
\hline
\end{array}\end{aligned}\label{bundle5a}\end{equation}

and

\begin{equation}\begin{aligned}&\begin{array}{|c|c| |c|c|}
\hline
x_i & \Gamma^j & \Lambda^a & p_l \\ \noalign{\hrule height 1pt}\begin{array}{cccccc}
1& 1& 1& 0& 0& 0  \\
0& 0& 0& 1& 1& 1 
\end{array}&\begin{array}{c}
 -3  \\
 -3 
\end{array}&\begin{array}{cccccccc}
 2& 4& 1& 0& 0& 0& 0& 1 \\
-1& 1& 1& 1& 1& 1& 2& 0 
\end{array}&\begin{array}{ccc}
 -4& -3& -1 \\
 -1& -2& -3 
\end{array}\\
\hline
\end{array}\end{aligned}\label{bundle5b}\end{equation}
The topology for each bundle is identical, both satisfying $c_2(V)=c_2(TX)$ and $Ch_3(V)=-63$. In both cases, the charged matter spectrum is $h^*(V) = (0, 63, 0, 0)$. Labeling the bundle in \eref{bundle5a} as $V_a$ and that in \eref{bundle5b} as $V_b$, we see that they are each stable in complementary chambers of K\"ahler moduli space (determined by whether $\mu(\cO(-2,1))$ is $>0$ or $<0$). Along the shared stability wall they decompose as 
\begin{align}
& V_a \to \cO(-2,1) \oplus U_a \\
& V_b \to \cO(2,-1) \oplus U_b
\end{align}
with 
\begin{align}
& 0 \to U_a \to \cO(1,0)\oplus \cO(1,1) \oplus \cO(2,2) \oplus \cO(2,0)^{\oplus 3}\oplus \cO(0,2) \to \cO(4,1) \oplus \cO(3,2) \oplus \cO(1,3) \to 0 \\
&0 \to U_b \to \cO(4,1)\oplus \cO(1,1)\oplus \cO(0,1)^{\oplus 3}\oplus \cO(0,2) \oplus \cO(1,0) \to \cO(4,1) \oplus \cO(3,2) \oplus \cO(1,3) \to 0
\end{align}
Direct calculation yields that for $V_a$, $H^*(X, \cO(-2,1) \otimes V_a)=(0,99,0,0)$. Thus, only one possible ``direction" of deformation exists to move the bundle away from its reducible locus and off of the stability wall (corresponding to only one sign of charged bundle moduli in \eref{dterm_first}, \cite{Anderson:2009sw,Anderson:2009nt}). Furthermore, on the reducible locus, $Hom(U_a, U_b)=0$. Thus, there is no deformation of $V_a$ can lead to a stable bundle in the region determined by $V_b$ and the bundles live in different S-equivalence classes.

Just as we saw in Section \ref{branch_in_dual}, where connected branches of moduli space were carried faithfully through the target space duality (see \eref{first_def}), it can also be verified that disconnected components of bundle moduli space on $X$ lead to disconnected components on $\widetilde{X}$. 

For each bundle, there are many possible target space dual theories, but to illustrate that the ``disconnected" bundles persist in the dual theories, the following pair is sufficient to illustrate the phenomenon described above.

A target space dual to \eref{bundle5a} is
\begin{equation}\begin{aligned}&\begin{array}{|c|c| |c|c|}
\hline
x_i & \Gamma^j & \Lambda^a & p_l \\ \noalign{\hrule height 1pt}\begin{array}{cccccccc}
0& 0& 0& 0& 0& 0& 1& 1 \\
1& 1& 1& 0& 0& 0& 0& 0 \\
0& 0& 0& 1& 1& 1& 0& 0 
\end{array}&\begin{array}{cc}
 -1& -1  \\
 -2& -1  \\
 -1& -2 
\end{array}&\begin{array}{cccccccc}
 0& 0&  1& 0& 0& 0& 0& 0 \\
 -2& 1&  0& 2& 2& 2& 3& 0 \\
 1& 0& -1& 2& 0& 0& 2& 2 
\end{array}&\begin{array}{ccc}
 0& -1&  0 \\
 -4& -3& -1 \\
 -1& -2& -3 
\end{array}\\
\hline
\end{array}\end{aligned}\label{bundle5a.1m}\end{equation}

and the following target space dual to \eref{bundle5b}
\begin{equation}\begin{aligned}&\begin{array}{|c|c| |c|c|}
\hline
x_i & \Gamma^j & \Lambda^a & p_l \\ \noalign{\hrule height 1pt}\begin{array}{cccccccc}
0& 0& 0& 0& 0& 0& 1& 1  \\
1& 1& 1& 0& 0& 0& 0& 0  \\
0& 0& 0& 1& 1& 1& 0& 0 
\end{array}&\begin{array}{cc}
 -1& -1  \\
 -2& -1  \\
 -1& -2 
\end{array}&\begin{array}{ccccccccc}
 0& 0&  1& 0& 0& 0& 0& 0& 0  \\
 2& 4&  0& 0& 0& 0& 0& 1& 3 \\
 -1& 1&  -1& 1& 1& 1& 2& 0& 2 
\end{array}&\begin{array}{cccc}
 0& -1&  0&  0 \\
 -4& -3& -1& -2 \\
 -1& -2& -3& \ 0 
\end{array}\\
\hline
\end{array}\end{aligned}\label{bundle5b.1m}\end{equation}
Once again, we see that the stability wall structure persists into the target space dual theory and here a similar analysis to that shown above demonstrates that $\widetilde{V}_a$ and $\widetilde{V}_b$ again form a disjoint pair of bundles that share a stability wall.

Finally, a note of caution should be raised that it is possible that seemingly distinct monads can in fact correspond to the same elements of the S-equivalence class:

To see this, consider the following rank $5$ bundle defined on the $\{2, 4 \}$ hypersurface in $\mathbb P^1 \times \mathbb P^3$:
\begin{equation}\begin{aligned}&\begin{array}{|c|c| |c|c|}
\hline
x_i & \Gamma^j & \Lambda^a & p_l \\ \noalign{\hrule height 1pt}\begin{array}{cccccc}
1& 1& 0& 0& 0& 0  \\
0& 0& 1& 1& 1& 1  \\
\end{array}&\begin{array}{c}
 -2  \\
 -4 
\end{array}&\begin{array}{ccccccc}
 0& 0& 0& 0&  1& 1& 1  \\
 1& 1& 1& 2& -1& 1& 2
\end{array}&\begin{array}{cc}
  -1& -2 \\
 -4& -3 
\end{array}\\
\hline
\end{array}\end{aligned}\label{bundle2b_again}\end{equation}
This bundle shares \emph{identical topology} with the bundle given in \eref{branch2a} in Section \ref{branch_in_v}. Moreover it shares a stability wall (given by the condition that $\mu(\cO(1,-1)=0$) and is stable in the same sub-cone in K\"ahler moduli space. However, on the stability wall it decomposes as
\beq
V_5 \to \cO(1,-1) \oplus {\cal Q}_4
\eeq
where 
\beq
0 \to {\cal Q}_4 \to \cO(0,1)^{\oplus 3} \oplus \cO(0,2) \oplus \cO(1,1) \oplus \cO(1,2) \to \cO(1,4) \oplus \cO(2,3) \to 0
\eeq
We can now ask: is this decomposition the same as the one arising on the wall for the $SU(5)$ bundle given in \eref{branch2a}?
There, the bundle data of \eref{branch2a} makes it clear that the bundle decomposes as $V_5 \to \cO(1,-1) \oplus U_4$, with 
\beq
0 \to U_4 \to \cO(0,1)^{\oplus 4} \oplus \cO(2,1) \oplus \cO(1,2)^{\oplus 2} \oplus \cO(2,2) \to \cO(1,1) \oplus \cO(3,2) \oplus \cO(1,4) \oplus \cO(2,3) \to 0
\eeq
At first pass, ${\cal Q}_4$ and $U_4$ appear to be different rank $4$ monad bundles. However, a direct calculation yields that
\beq
dim (Hom({\cal Q}_4, U_4))=1
\eeq
and once again it is possible to apply Lemma \ref{OSS_corollary_thm} in Appendix \ref{AppendixA} to determine that these are, in fact, isomorphic stable bundles. Interestingly, the different monad realizations of isomorphic bundles seem in general to lead to new target space dual theories. This effect and the corresponding list of target space dual theories is given in Appendix \ref{appendixB}.

\section{F-Terms, Bundle Holomorphy, and Duality}\label{Fterm_sec}

\subsection{F-terms in the effective theory}
In this section we will explore examples of $(0,2)$ GLSMs with non-trivial F-term contributions to their $4$-dimensional, ${\cal N}=1$ potential. At the level of geometry, this corresponds to obstructions to geometric deformations arising from the constraint of bundle holomorphy.

In general, it is exceedinly challenging computationally to calculate directly the Atiyah obstruction (largely due to the difficulty in determining the the class of the Atiyah map ($[F^{1,1}]$) in \eref{mod_def}. To find tractable/computable examples a series of approaches were developed in \cite{Anderson:2009sw,Anderson:2009nt,Anderson:2011ty,Anderson:2013qca}. In particular, one simple way to generate examples employs methods of constructing bundles that \emph{manifestly depend on the complex structure moduli of $X$}. One such tool is the well-known ``jumping" phenomena of bundle-valued cohomology on a CY manifold, $X$. This was applied to build examples of monad and extension bundles that presented non-trivial holomorphy obstructions. For a thorough treatment of the deformation theory associated to this idea, see \cite{Anderson:2011ty}. Here we will briefly review the relevant techniques for monads.

In the case of monad bundles, a simple class of holomorphy obstructions can be realized in the dependence on $F_a^l(x^i)$ on the complex structure of $X$. In general, elements of $F_a^l(x^i)$ may be generically zero and only non-vanishing for certain higher co-dimensional sub-loci of complex structure moduli space. Suppose we consider a rank $n$ monad of the form
\beq
0 \to V \to \cO({\bf b}_1)\oplus \ldots \cO({\bf b}_{n+1}) \stackrel{F}{\longrightarrow} \cO({\bf c}) \to 0
\eeq
with the property that a given map element, say $Hom(\cO({\bf b}_1),\cO({\bf c}))$ is generically vanishing. That is,
\beq
h^0(X, \cO({\bf c- b_1}))=0
\eeq
Such a monad does not define a good, slope stable bundle since in general it will be reducible (since $\cO({\bf b}_1) \in ker(F)$ generically):
\beq
\cO({\bf b_1}) \oplus V'
\eeq
Note that in general if $\cO({\bf b_1})$ is of purely positive or negative degree, then the sum above \emph{cannot} be poly-stable within the K\"ahler cone. However, on special loci in complex structure moduli space, line bundle cohomology can ``jump" to a non-zero value. We will denote this locus as ${\cal CS}_{jump}$. On that locus, $h^0(X, \cO({\bf c- b_1}))\neq 0$ and we find new global holomorphic sections to $\cO({\bf c- b_1})$ that can be used to define a stable, holomorphic bundle, $V=ker(F)$. As an example, consider the following line bundle on the $\{2,4\}$ hypersurface in $\mathbb{P}^1 \times \mathbb{P}^3$:
\beq\label{nocohom_gen}
h^0(X,\cO(-2,4))=0~~\text{for generic values of complex structure}
\eeq
However, as computed in \cite{Anderson:2011ty} on a $53$-dimensional sub-locus of the $86$-dimensional complex structure moduli space, this cohomology can ``jump" to
\beq\label{cohom_jump}
\text{On}~{\cal CS}_{jump},~~~h^0(X, \cO(-2,4))=1
\eeq
In the following section, we will make use of this calculation to build a monad bundle with a non-trivial holomorphy obstruction and construct its target space dual theory. Before we begin however, two notes of caution are in order.

First, it should be noted that the inclusion of map elements such as the one above $F \in H^0(X, \cO(-2,4)$ introduce more dramatic negative charges into the definition of the GLSM then those we previously considered (only negative charges for some $\Lambda^a$ and all maps, $F_a^l$ positive in the D-term examples of the previous Sections). This can change the vacuum structure of the GLSM and care must be taken interpreting these both in the geometric regime and for other phases. For this section, we will consider only examples that we can confirm correspond to good geometric phases (i.e. large volume, perturbative, etc). We leave the important and intriguing question of how many negative degrees can be included\footnote{Note this is closely related to the new ``Generalized Complete Intersection" (gCICY) method of constructing Calabi-Yau manifolds \cite{Anderson:2015yzz,Anderson:2015yzz,Berglund:2016yqo}.} in the GLSM to a future publication \cite{to_appear} and focus in this Section only on the well-defined target space theories.

Second, the calculations of holomorphy obstructions in the following sections will be done at the level of the compactification geometry. Caution must be used in describing this reduction in moduli from the naive count in \eref{naive_count} in the effective theory. In the examples that follow below, the mass scale of the holomorphy obstructions is too high to be included in the effective theory. Instead, the geometric obstructions are directly calculated at the compactification scale (as is done to compute \eref{chargmat} in the charged matter) and the resulting massless modes (corresponding to the solution of \eref{mod_def}) are presented. For some steps towards realizing holomorphy obstructions at the level of sigma models, see \cite{Melnikov:2011ez}.

As a final comment, it should be noted that compared to the D-term computations of Section \ref{dterms}, the holomorphy computations required here are significantly more computationally intensive and introduce a number of difficult problems in geometry/deformation theory. As a result, we focus on one example geometry and leave a broader scan for examples to future work.

We turn now to a simple example of a monad bundle inducing an obstruction in the complex structure of $X$ and its target space dual.

\subsection{A simple example}
In this section we will consider a simple vector bundle which was inspired by an example in \cite{Anderson:2011ty}.  This $SU(3)$ monad bundle:
\begin{equation}\begin{aligned}&\begin{array}{|c|c| |c|c|}
\hline
x_i & \Gamma^j & \Lambda^a & p_l \\ \noalign{\hrule height 1pt}\begin{array}{cccccc}
1& 1& 0& 0& 0& 0  \\
0& 0& 1& 1& 1& 1 
\end{array}&\begin{array}{c}
-2  \\
-4 
\end{array}&\begin{array}{ccccc}
 2& -1& -1 &1  &0 \\
 0& 2& 2 &  0 & 2
\end{array}&\begin{array}{cc}
 0 & -1 \\
 -4 & -2
\end{array}\\
\hline
\end{array}\end{aligned}\label{eq:Fbundle2}\end{equation}
cannot be defined for generic choices of the complex structure of $X$ given in \eref{eq:Fbundle2}. It is clear that by \eref{nocohom_gen}, for generic values of the complex structure the map, $F$ takes the form

\beq\label{two_row_map}
F_a^l=\left(
\begin{array}{ccccc}
f_{(-2,4)} & f_{(1,2)} & f'_{(1,2)} & f_{(-1,4)} &f_{(0,2)}) \\
0 & f_{(1,0)} &  f'_{(1,0)} & f_{(0,2)} & f_{(1,0)}
\end{array}
\right)
\eeq
where $f_{(a,b)}$ is generic polynomials on $X$ of multi-degree $\{a,b\}$. Of note are the maps with ``negative" degree. It is straightforward to verify that on $X$, $h^0(X, \cO(-1,4))=2$ and thus, despite the negative degree this map is an ordinary holomorphic function on $X$. In \cite{Anderson:2015iia} tools were developed to describe such ``non-polynomial" functions explicitly. Here we will not require the explicit description of $f'$ and merely need to know that it is non-trivial. 

As noted above however, the map $f_{(-2,4)}$ is on a different footing since \emph{for generic values of the complex structure} $h^0(X, \cO(-2,4))=0$. When $f_{(-2,4)}$ is not available as a non-trivial map, it is clear that this $F_a^l(x)$ cannot define a good bundle. Here $V$ is reducible as the sum of a line bundle and rank $2$ sheaf:
\beq
V \sim \cO(2,0) \oplus {\cal V}
\eeq
and \emph{cannot be made poly-stable} in the K\"ahler cone of $X$. Thus, in general, this monad cannot provide a good definition of the background gauge fields in the heterotic theory. However, as shown in \cite{Anderson:2011ty} and using \eref{cohom_jump}, on the 53-dimensional sub-locus of complex structure moduli space, there is a new map element available given by $f_{(-2,4)} \in H^0(X, \cO(-2,4))$. Once again, despite the negative degree in the line bundle, for the chosen complex structure this is an ordinary, global, holomorphic section to $\cO(-2,4)$ on $X$. Thus on the locus ${\cal CS}_{jump}$, it is possible to define a stable, \emph{indecomposable} monad, $V$. We will begin there and use this example to construct a target space dual.

We have chosen this example because of its simplicity and the fact that the difficult computational exercise of computing ${\cal CS}_{jump}$ and the associated Atiyah calculation was already completed in \cite{Anderson:2011ty}. However, it should be noted that the monad given above does not satisfy the anomaly cancellation condition alone since $c_2(V)=\{16,32\}$. This can be remedied by choosing to ``complete" this example with another monad bundle. We will embed this second bundle into the second $E_8$ factor. One such good $SU(3)$ bundle is
\begin{equation}\begin{aligned}&\begin{array}{|c|c| |c|c|}
\hline
x_i & \Gamma^j & \Lambda^a & p_l \\ \noalign{\hrule height 1pt}\begin{array}{cccccc}
1& 1& 0& 0& 0& 0  \\
0& 0& 1& 1& 1& 1 
\end{array}&\begin{array}{c}
-2  \\
-4 
\end{array}&\begin{array}{ccccc}
 1& 0& 0 &0 & 0   \\
 0& 1& 1 &1 &1
\end{array}&\begin{array}{cc}
 -1 & 0  \\
 -2 &-2
\end{array}\\
\hline
\end{array}\end{aligned}\label{eq:bundle2_hidden}\end{equation}
with $c_2(V)=\{8,12 \}$ as required so that
\beq
c_{2}(V_{total})=c_2(V_1) + c_2(V_2)=c_2(TX)
\eeq
With this addition, the full $4$-dimensional theory has gauge symmetry $E_6 \times E_6$.

In what follows the bundle in \eref{eq:bundle2_hidden} plays a trivial, spectating role and thus, we will recall that it must be included, but it will not be central to our further analysis. Focusing first on the rank $3$ bundle in \eref{eq:Fbundle2}, naive moduli count associated to the first $SU(3)$ bundle and the CY manifold would have produced
\beq
dim({\cal M}_0)=h^{1,1}+h^{2,1}+h^1(X,End_0(V))=2+86+92=180
\eeq
However, taking into account the constraint that the complex structure lies within the ``jumping locus" in order for the bundle to be well-defined we have
\beq\label{lifted_goodstuff}
dim({\cal M}_1)=dim({\cal M}_0)-33=147
\eeq
See \cite{Anderson:2011ty} for a discussion of the relationship between this jumping cohomology calculation and the Atiyah obstruction. To complete the degree of freedom count, we note that for $V$ in \eref{eq:Fbundle2} the charged matter is 
\begin{align}\label{charged_e6}
& h^1(X,V)=41~~~(\text{no. of }{\bf 27})  \\
& h^1(X, V^\vee)=1~~~(\text{no. of }{\bf \overline{27}})
\end{align}
For completeness, the spectrum of the second $E_6$ factor consists of $12$ ${\bf 27}$s (and no ${\bf \overline{27}}$s) and $38$ singlets. 

Finally, one last interesting note about this geometry: despite the negative line bundle entry ($\cO(-1,2)$) in \eref{eq:Fbundle2}, the $SU(3)$ bundle presented there is stable throughout the \emph{entire} K\"ahler cone since $\mu(\cO(-1,2))<0$ for all K\"ahler moduli, $t^r$, on $X$. Thus, there is no stability wall and no non-trivial D-term anywhere in the moduli space of this theory.

With these calculations in hand, we turn finally to the target space dual theory. Once again, we can ask the relevant questions of the duality:
\begin{itemize}
\item Is a holomorphy obstruction in $(X,V)$ reflected in some (deformation theoretic) obstruction in $(\widetilde{X},\widetilde{V})$?
\item Since in general the dimensions of $h^{2,1}(X)$ and $h^{2,1}(\widetilde{X})$ (and any Atiyah obstructions within these spaces) are different, is the final set of zero modes the same in each theory?
 \end{itemize}
 We will address these questions in turn in the following Section.
 
\subsection{The target space dual}
The target space dual is easy to construct for the rank $3$ bundle given in \eref{eq:Fbundle2}. We obtain
\begin{equation}\begin{aligned}&\begin{array}{|c|c| |c|c|}
\hline
x_i & \Gamma^j & \Lambda^a & p_l \\ \noalign{\hrule height 1pt}\begin{array}{cccccccc}
0& 0& 0& 0& 0& 0& 1& 1  \\
1& 1& 0& 0& 0& 0& 0& 0  \\
0& 0& 1& 1& 1& 1& 0& 0 
\end{array}&\begin{array}{cc}
 -1& -1  \\
 -1&  -1  \\
 -2& -2 
\end{array}&\begin{array}{ccccc}
 0& 1& 0 & 0 & 0  \\
 2& -2& 0 & 1 & 0 \\
 0& 0& 4 & 0 & 2
\end{array}&\begin{array}{cc}
 -1 & 0  \\
 0 & -1 \\
 -4 & -2
\end{array}\\
\hline
\end{array}\end{aligned}\label{fterm_mirror2}\end{equation}
The new CY manifold, $\widetilde{X}$ has Hodge numbers $h^{1,1}=3$ and $h^{2,1}=55$. 

Now, to answer the first question in our list, we can ask whether the bundle in \eref{fterm_mirror2} provides any obvious obstruction to holomorphy? By inspection, the answer is immediately yes! We see that in fact, the map $\widetilde{F}_a^l(x)$ associated to $\widetilde{V}$ takes the form
\beq\label{two_row_map_mirror}
\widetilde{F}_a^l=\left(
\begin{array}{ccccc}
f_{(1,-2,4)} & f_{(0,2,4)} & f_{(1,0,0)} & f_{(-1,1,4)} & 0 \\
0 & f_{(-1,3,2)} &  0 & f_{(0,0,2)} & f_{(0,1,0)}
\end{array}
\right)
\eeq
All the map entries above are generically non-vanishing polynomials of the given degree with the exception of $\widetilde{F}_1^1$ of degree $(1,-2,4)$. As expected, the line bundle cohomology associated to this
\beq
h^0(\widetilde{X},\cO(1,-2,4))=0~~~\text{for generic values of the complex structure}
\eeq
but once again, \emph{on a higher co-dimensional locus in complex structure moduli space, which can be labeled} ,$\widetilde{CS}_{jump}$, $h^0(\widetilde{X} ,\cO(1,-2,4))=1$. Once again, it is possible to define a good, indecomposable rank $3$ monad bundle, but only by tuning the complex structure moduli of the CY manifold. In this case, the locus has dimension
\beq
dim(\widetilde{CS}_{jump})=37
\eeq
That is, the jumping of the relevant monad map fixes $18$ of the $55$ complex structure moduli of $\widetilde{X}$. The initial moduli count for this bundle is then
\beq\label{silly_count}
dim({\cal \widetilde{M}}_0)=3+55+122=180
\eeq
With the obvious constraint above this reduces the total to $dim({\cal \widetilde{M}}_0)-18=162$. Moreover, the charged matter of the $E_6$ theory is identical to that in \eref{charged_e6}:
\begin{align}\label{charged_e6_mirror}
& h^1(\widetilde{X},\widetilde{V})=41~~~(\text{no. of }{\bf 27})  \\
& h^1(\widetilde{X}, \widetilde{V}^\vee)=1~~~(\text{no. of }{\bf \overline{27}})
\end{align}

This is almost in agreement with the final singlet count of $147$ in the original geometry in \eref{lifted_goodstuff}. We are left however, with an apparent discrepancy of $15$ moduli. Could these too be lifted from the count for $(\widetilde{X}, \widetilde{V})$? There are two hints that this is indeed the case. 

The first is that the lengthy calculation of $h^1(\widetilde{X}, End_0(\widetilde{V}))=122$ in \eref{silly_count} was performed by taking all induced monad maps in $\widetilde{V} \otimes \widetilde{V}^{\vee}$ to be generic (a necessary computational simplification). However, it is clear that for the special bundle at hand a more detailed calculation is necessary. Two hints at this structure are significant.

First, the line bundle $\cO(0,-2,4)$ appears in the calculation of $h^1(\widetilde{X}, End_0(\widetilde{V}))$ (corresponding to morphisms of the form $Hom(\widetilde{B},\widetilde{B})$. As for the similar line bundle on $X$, generically this has no global sections. However, if we further tune the complex structure of $\widetilde{X}$ to make $h^0(\widetilde{X},\cO(0,-2,4))$ jump to a non-zero value, this fixes exactly $15$ additional moduli\footnote{This can be verified using the tools of \cite{Anderson:2011ty,Anderson:2013qca}.}. Such a choice might be expected in comparison with the original pair $(X, V)$ since it is exactly these additional moduli that ``pass through" the conifold transition connecting $X$ and $\widetilde{X}$ and are in some sense ``shared" between the CY manifolds. See \cite{Anderson:2013qca} for results on how jumping loci (${\cal CS}_{jump}$) are fixed across conifold transitions.

Finally, as another hint for where the moduli fixing may arise, it should be noted that unlike the starting geometry, the bundle in \eref{fterm_mirror2} \emph{does display a stability wall in K\"ahler moduli space}. In general, the bundle is stable for the $3$ dimensional sub-cone of K\"ahler moduli space determined by
\beq
0<t^1~~,~~2 t^1<t^2 ~~,~~0< t^3 < 2t^2-4t^1
\eeq
On the 2-dimensional stability wall given by $t^3=2t^2-4t^1$ and $2 t^1<t^2$ the bundle decomposes as
\beq
\cO(1,-2,0)\oplus \tilde{U}_2
\eeq
with
\beq
0 \to \tilde{U}_2  \to \cO(0,2,0)\oplus \cO(0,0,4)\oplus \cO(0,1,0) \oplus \cO(0,0,2) \to \cO(1,0,4) \oplus \cO(0,1,2) \to 0
\eeq
At this locus the bundle decomposes as $SU(3) \to S[U(1)\times U(2)]$ and the full matter spectrum is charged under the Green-Schwarz anomalous $U(1)$ symmetry as in Section \ref{dterms}. Of interest to us are the $123$ singlets corresponding to bundle moduli:
\begin{align}
&C_{+3}~~~~h^1(\widetilde{X}, \tilde{L}^{\vee} \otimes \tilde{U}_2)=68 \\
&C_{-3}~~~~h^1(\widetilde{X}, \tilde{L} \otimes {\tilde{U}^{\vee}}_2)=10 \\
&C_{0}~~~~~h^1(\widetilde{X}, \tilde{U}_2 \otimes {\tilde{U}^{\vee}}_2)=45
\end{align}
with $\tilde{L}=\cO(1,-2,0)$. In this description it is clear that the form of $U_2$, itself and the number of fields $C_0$ depends on the choice complex structure of $\widetilde{X}$. Moreover, this description makes it clear that additional bundle moduli can acquire a mass. Near the stability wall, one origin for such mass terms is evident from the fact that there are allowed, gauge-invariant trilinear couplings in the superpotential of the form
\beq\label{singlet_trilinear}
W \sim C_0 C_{+3} C_{-3}
\eeq
By leaving the stability wall with $\langle C_{\pm 3} \rangle \neq 0$, it is clear that mass terms can be generated for singlets which could lift degrees of freedom, including the significant $15$ moduli above. We leave a detailed analysis of the bundle moduli of \eref{fterm_mirror2} and the computation of trilinear couplings such as \eref{singlet_trilinear} for future work.

 To complete this example, it should be noted that as before, another bundle in the second $E_8$ factor is necessary to satisfy anomaly cancellation and the bundle in \eref{eq:bundle2_hidden} trivially carries through the target space duality procedure.
\begin{equation}\begin{aligned}&\begin{array}{|c|c| |c|c|}
\hline
x_i & \Gamma^j & \Lambda^a & p_l \\ \noalign{\hrule height 1pt}\begin{array}{cccccccc}
0& 0& 0& 0& 0& 0& 1& 1  \\
1& 1& 0& 0& 0& 0& 0& 0  \\
0& 0& 1& 1& 1& 1& 0& 0 
\end{array}&\begin{array}{cc}
 -1& -1  \\
 -1&  -1  \\
 -2& -2 
\end{array}&\begin{array}{ccccc}
0 & 0 & 0 & 0 &0 \\
 1& 0& 0 &0 & 0   \\
 0& 1& 1 &1 &1
\end{array}&\begin{array}{cc}
 0 & 0 \\
 -1 & 0  \\
 -2 &-2
\end{array}\\
\hline
\end{array}\end{aligned}\label{eq:bundle2_hidden_mirror}\end{equation}
The charged and singlet matter spectrum for this bundle is identical to that associated to \eref{eq:bundle2_hidden}.

In summary then, we have found that F-term obstructions are non-trivially preserved across this target space dual pair and correspond to a sensitive inter-mixing of complex structure and bundle moduli which are non-trivially redefined across the duality. 

\section{Target space duality, ${\bf (2,2)}$ loci, and tangent bundle deformations}\label{tan_defs}

Having seen that loci of enhanced symmetry (i.e. stability walls, etc) were reflected in the target space duals in previous sections, it is natural to investigate whether or not another special locus is preserved -- namely, the $(2,2)$ locus of a $(0,2)$ theory. Such a locus --realized in the geometric regime by GLSMs in which $V$ is a deformation of the holomorphic tangent bundle of the CY base manifold, $X$ -- play a crucial role in the study of $(0,2)$ GLSMs. In this section, we will provide examples of starting GLSMs with a $(2,2)$ locus and explore the structure of their target space duals.

To begin, we consider perhaps the simplest case available: the rank-changing deformation of the holomorphic tangent bundle to the quintic hypersurface in $\mathbb{P}^4$. The following $SU(4)$ bundle is a non-trivial deformation of $\cO_X \oplus TX$ on the quintic:

\begin{equation}\begin{aligned}&\begin{array}{|c|c| |c|c|}
\hline
x_i & \Gamma^j & \Lambda^a & p_l \\ \noalign{\hrule height 1pt}\begin{array}{ccccc}
1& 1& 1& 1& 1 \\
\end{array}&\begin{array}{c}
 -5
\end{array}&\begin{array}{ccccc}
 1& 1& 1& 1& 1 \\
\end{array}&\begin{array}{c}
 -5
\end{array}\\
\hline
\end{array}\end{aligned}\label{tandef}\end{equation}
Indeed, this $V_4$ is in fact the most general slope-stable, rank $4$ deformation of this type and the entire local moduli space of deformations can be captured in the moduli of the monad (see \cite{Li:2004hx}). Writing the bundle, $V_4$, as 
\beq
0 \to V \to \cO(1)^{\oplus 5} \stackrel{f}{\longrightarrow} \cO(5) \to 0
\eeq
the locus in bundle moduli space for which it decomposes as $\cO_X \oplus TX$ is given by
\beq
f= dP_5
\eeq
where $P_5$ is the homogeneous, degree $5$ defining relation of the quintic in $\mathbb{P}^4$.

An application of the tangent space duality procedure can produce the following geometry, including a new manifold $\widetilde{X}$ which is related to the quintic via a conifold transition:

\begin{equation}\begin{aligned}&\begin{array}{|c|c| |c|c|}
\hline
x_i & \Gamma^j & \Lambda^a & p_l \\ \noalign{\hrule height 1pt}\begin{array}{ccccccc}
1& 1& 0& 0& 0& 0 & 0 \\
0& 0& 1& 1& 1& 1  & 1
\end{array}&\begin{array}{cc}
 -1 & -1 \\
 -4 & -1
\end{array}&\begin{array}{cccccc}
 1& 0& 0& 0& 0& 0 \\
 0& 5& 1& 1& 1& 1
\end{array}&\begin{array}{cc}
 -1 & 0\\
 -5 & -4
\end{array}\\
\hline
\end{array}\end{aligned}\label{quintic_dual}\end{equation}
As required by the target space duality algorithm, this bundle leads to the same effective theory as the original geometry in \eref{tandef}. In particular, it is clear that the Chiral Index (the difference between ${\bf 16}$ and ${\bf \overline{16}}$ representations in the target space $SO(10)$ theory) is given by 
\beq
Ch_3(\widetilde{V})=-100
\eeq
as required by our starting point of $V=TX$ on the quintic.

However, this fact leads immediately to the observation that this bundle \emph{cannot be a standard deformation of $T\widetilde{X}$}, since for the new manifold\footnote{This difference in index is precisely arising from the classic conifold transition between $X$ and $\widetilde{X}$ -- the difference corresponding to resolving the $16$ singular points on the nodal quintic \cite{Candelas:1989js}.} in \eref{quintic_dual}
\beq
Ch_3(T\widetilde{X})=-84~.
\eeq
This result is clearly a general feature of the target space duality procedure. Starting from a bundle near the $(2,2)$ locus on $X$, the algorithm produces a bundle with $ch_2(\widetilde{V})=ch_2(T\widetilde{X})$ but with $ch_3(\widetilde{V}) \neq ch_3(T\widetilde{X})$, as required by the ${\cal N}=1$, $4$-dimensional effective theories being identical.

If however, the target space dual is really isomorphic in some deeper sense (as a $(0,2)$ sigma model), it is natural to expect that the $(2,2)$ locus should still somehow be present. One possibility is that the only deformations leading to the $(2,2)$ locus in the dual geometry of \eref{quintic_dual} force the theory to a non-perturbative regime. A candidate for such a possibility might be the singular shared locus in the moduli space of $X$ and $\widetilde{X}$. Since the two CY 3-folds are related by a conifold transition \cite{Candelas:1989js}, their common locus in moduli space is given by a nodal quintic of the form
\beq
P_5= l_1(x)q_2(x)-l_2(x)q_1(x)=0
\eeq
where $l_i(x)$ and $q_i(x)$ are linear and quartic polynomials, respectively, in the homogeneous coordinates of $\mathbb{P}^4$ (and the CY 3-fold in \eref{quintic_dual} is the resolution of this singular variety).

A second possibility is that even at general points in complex structure moduli space, the bundle given in \eref{quintic_dual} is deformable to the holomorphic tangent bundle through some more general, chirality changing transition. Such possibilities have been previously conjectured in the heterotic literature \cite{Douglas:2004yv}, though explicit geometric realizations have yet to be fully understood (see \cite{Anderson:2015cqy} for some recent, related results). It would be intriguing to explore the geometry of such transitions further in this context.

To conclude, we provide one interesting hint towards the behavior of $(2,2)$ loci under the duality. A useful observation is that it is possible for some target space duals to lead to a new vector bundle on \emph{an isomorphic CY threefold}. To illustrate this, consider the following rank $5$ deformation of the tangent bundle $TX$ of the $\{2,4\}$ hypersurface in $ \mathbb P^1 \times \mathbb P^3$:

\begin{equation}\begin{aligned}&\begin{array}{|c|c| |c|c|}
\hline
x_i & \Gamma^j & \Lambda^a & p_l \\ \noalign{\hrule height 1pt}\begin{array}{cccccc}
1& 1& 0& 0& 0& 0 \\
0& 0& 1& 1& 1& 1 
\end{array}&\begin{array}{c}
 -2 \\
 -4 
\end{array}&\begin{array}{cccccc}
 1& 1& 0& 0& 0& 0 \\
 0& 0& 1& 1& 1& 1
\end{array}&\begin{array}{c}
 -2 \\
 -4
\end{array}\\
\hline
\end{array}\end{aligned}\label{tan_def24}\end{equation}
The bundle above is a stable deformation of $TX \oplus \cO_X \oplus \cO_X$. This starting point admits a target space dual in which the original manifold appears in a redundant CICY description (achieved by adding the repeated entry $\cO(1,4)$):

\begin{equation}\begin{aligned}&\begin{array}{|c|c| |c|c|}
\hline
x_i & \Gamma^j & \Lambda^a & p_l \\ \noalign{\hrule height 1pt}\begin{array}{cccccccc}
0& 0& 0& 0& 0& 0& 1& 1 \\
1& 1& 0& 0& 0& 0& 0& 0 \\
0& 0& 1& 1& 1& 1& 0& 0 
\end{array}&\begin{array}{cc}
  -1& -1 \\
 -1& -1 \\
 -4&  0 
\end{array}&\begin{array}{ccccccc}
 1& 0& 0& 0& 0& 0& 0 \\
 0& 1& 0& 0& 0& 0& 2  \\
 0& 0& 1& 1& 1& 1& 4 
\end{array}&\begin{array}{cc}
 -1&  0 \\
 -2& -1 \\
 -4& -4 
\end{array}\\
\hline
\end{array}\end{aligned}\label{cy_rewrite}\end{equation}\\ \\ 
As described in Section \ref{isomorphic_CYs}, this target space dual exhibits a particularly interesting structure because it has left the CY $3$-fold unchanged. As argued in Section \ref{isomorphic_CYs}, this threefold is equivalent to $X$ in \eref{tan_def24} times a point. Despite the fact that the manifold has not changed, this example is not trivial however, since the vector bundles given in \eref{tan_def24} and \eref{cy_rewrite} are not obviously the same, nor is their apparent dependence on the complex structure of $X =\widetilde{X}$.

To clarify things, it can be noted that although there are three divisors restricted from the ambient hyperplanes in \eref{cy_rewrite}, these divisors are linearly dependent and $h^{1,1}(\widetilde{X})=2$. Furthermore, a simple exploration of linear equivalence of divisors shows that $\cO(a,b,c)$ on $\tilde{X}$ in \eref{cy_rewrite} is the same line bundle as $\cO(a+b,c)$ on the threefold in \eref{tan_def24} for any integers, $a,b,c \in \mathbb{Z}$. With this observation it is possible to re-write the dual theory in \eref{cy_rewrite} as

\begin{equation}\begin{aligned}&\begin{array}{|c|c| |c|c|}
\hline
x_i & \Gamma^j & \Lambda^a & p_l \\ \noalign{\hrule height 1pt}\begin{array}{cccccc}
1& 1& 0& 0& 0& 0  \\
0& 0& 1& 1& 1& 1 
\end{array}&\begin{array}{c}
-2  \\
-4 
\end{array}&\begin{array}{ccccccc}
 1& 1& 0 & 0 & 0& 0 & 2  \\
 0& 0& 1 &1 & 1& 1 &4
\end{array}&\begin{array}{cc}
 -3& -1  \\
 -4& -4
\end{array}\\
\hline
\end{array}\end{aligned}\label{mirrorbun_again}\end{equation}
which is the same manifold $X$ with an apparently distinct monad bundle $\widetilde{V}$. Now we can ask the question: are the bundles on $X$ given in \eref{tan_def24} and \eref{mirrorbun_again} equivalent? To answer this puzzle it is possible to appeal to a useful lemma which establishes when two stable bundles on $X$ can be identified (See Lemmas \ref{OSS_lemma_thm} and \ref{OSS_corollary_thm} in Appendix \ref{AppendixA}). Briefly, the lemma states if there exists a non-trivial map (in $Hom(V, \widetilde{V})$) between two stable bundles on $X$ with the same slope, $\mu(V)$, then they are isomorphic. In this case, a direct calculation leads to 
\beq
dim(Hom(V, \widetilde{V}))=h^0(X, V \otimes \widetilde{V}^\vee)=1
\eeq
Thus, the naively distinct monads given in \eref{tan_def24} and \eref{mirrorbun_again} are in fact equivalent stable, holomorphic vector bundles, complete with the same $(2,2)$ locus (though the locus is clearly most easily realized in the first description).
Thus in this simple case, the $(2,2)$ locus is preserved in the target space dual, though in a subtle way.

\section{Conclusions and Future Directions}\label{conclusions}
In this work we have explored the notion of $(0,2)$ target space duality by studying in detail the $4$-dimensional effective potential in dual theories and the associated vacuum spaces. It has been an open question whether or not $(0,2)$ target space duality may correspond to a true string duality or merely a shared sub-locus in the moduli space of two distinct theories. While a full correspondence of $(0,2)$ heterotic string theories has yet to be established, our results appear to provide evidence for the former possibility. As in \cite{Blumenhagen:2011sq}, we find that the full massless spectrum of the dual theories (in the geometric phase) is preserved
\begin{eqnarray}\nonumber
h^*(X,\wedge^{k}V) &=& h^*(\widetilde X, \wedge^{k} \widetilde V) \\
h^{2,1}(X) + h^{1,1}(X) + h^1(X,End_0(V)) &=& h^{2,1}(\widetilde X) + h^{1,1}(\widetilde X) +  h^1(X,End_0(\widetilde V))
\end{eqnarray}
with $k=1 \ldots rk(V)$ and that the anomaly cancellation conditions and equations of motion are automatically satisfied in the dual theories. Moreover, we make the novel observation that non-trivial geometric obstructions --corresponding to D- and F-term potentials --  lead to the same vacuum space in the low energy theories. The central results of this work include the following

\begin{enumerate}
\item D-term and F-term constraints associated to stability and holomorphy constraints in the pair $(X,V)$ appear to be faithfully reflected in the dual geometry $(\widetilde{X},\widetilde{V})$ for all the examples studied in this work.
\item The structure of the ${\cal N}=1$ vacuum space of the dual theories is identical. That is, beginning at given points in moduli space infinitesimal fluctuations are preserved, leading to commutative diagrams of bundle deformations as in \eref{first_def}.
\item We find that loci of enhanced symmetry -- including stability walls (with enhanced Green-Schwarz massive $U(1)$ symmetries) and $(2,2)$ loci are preserved across the duality (though the identification of such enhancement points can be subtle and not always in the perturbative regime of the theory).
\end{enumerate}
To make the observations above we have developed a number of new technical tools including a generalization of the target space duality procedure itself (in Section \ref{repeated_line_bundles}) and the ability to describe complete infinitesimal bundle deformation spaces as monads (see Sections \ref{branch_in_v} and \ref{branch_in_dual}).

In the present work, our primary goal has been to compare the structure of the target space effective theories. A natural next step is to explore the examples considered here more fully at the level of GLSMS (or indeed full sigma models when possible). In this work, we have not considered the possible contributions from world sheet instantons to the potential (though in many cases for simple CICY GLSMs such as considered here, they may be expected to vanish \cite{Beasley:2003fx,Beasley:2005iu,Bertolini:2014dna}) or indeed the full structure of even the perturbative Yukawa couplings (see \cite{Anderson:2009ge,Anderson:2010tc,Blesneag:2015pvz,Buchbinder:2016jqr} for example). Both should be studied to more fully understand the validity of the effective theories considered here and their possible duality. Other interesting future comparisons might include a calculation of spectra of the target space dual GLSMS in non-geometric phases (for example the Landau-Ginsburg spectra \cite{Kachru:1993pg}), a study of the quantum cohomology rings in some simple cases \cite{Donagi:2012sca} and perhaps even applying tools from localization \cite{Closset:2015ohf} to investigate the detailed structure of the GLSMS themselves in more detail. We hope to address some of these questions in future work.

Finally, another interesting area of investigation, first mentioned in \cite{Blumenhagen:2011sq} would be to understand the consequences of $(0,2)$ target space duality in Heterotic/F-theory duality \cite{Friedman:1997yq}. In particular, if both $X$ and $\widetilde{X}$ admit an elliptic fibration: $\pi: X \to B_2$ and $\tilde{\pi}: \widetilde{X} \to \widetilde{B}_2$ (with distinct base manifolds, $B_2$ and $\widetilde{B}_2$), then it is expected that there will exist dual $4$-dimensional theories arising from F-theory compactificed on $K3$-fibered CY 4-folds. It was speculated in \cite{Blumenhagen:2011sq} that perhaps the dual F-theory geometries may be realizable as multiple elliptic/$K3$ fibrations of the same CY $4$-fold, $Y_4$:
\beq
 \xymatrix{
& Y_4 \ar[ld]^{\pi_{1}}_{\mathbb{E}} \ar[rd]^{\mathbb{E}}_{\pi_{2}} &\\
{\cal B}_{3}& & \widetilde{{\cal B}}_{3}}
\eeq
where the $3$-fold base manifolds ${\cal B}_3$ and $\widetilde{{\cal B}}_3$ are both $\mathbb{P}^1$ fibered as required by compatibility of the elliptic and $K3$ fibrations. That is, there exist maps, $\rho_1: {\cal B}_{3}\stackrel{\mathbb{P}^1}{\longrightarrow}B_2$ and  $\rho_2:  \widetilde{{\cal B}}_{3} \stackrel{\mathbb{P}^1}{\longrightarrow} \widetilde{B}_2$. This too, would be in intriguing topic for further research and could shed further light on recent progress in heterotic/F-theory duality \cite{Anderson:2014gla,Anderson:2016kuf,Cvetic:2016ner,Cvetic:2015uwu}.

\section*{Acknowledgements}
The authors would like to thank Per Berglund, Ralph Blumenhagen, James Gray, Sheldon Katz, Ilarion Melnikov, and Eric Sharpe for useful discussions. LA also thanks the UI Chicago and the UI Urbana-Champaign for hospitality during the conclusion of this work. The work of LA and HF is supported by NSF grant PHY-1417337. This project is part of the working group activities of the 4-VA initiative ``A Synthesis of Two Approaches to String Phenomenology".

\appendix

\section{Appendix A - Tools for Probing Slope-Stability of Vector Bundles}\label{AppendixA}
\subsection{Slope Stability}

The conditions for supersymmetry in the ${\cal N}=1$ $4$-dimensional heterotic theory have been a notorious source of difficulty. With no explicit, non-trivial solutions known\footnote{See \cite{Douglas:2006rr,Braun:2007sn,Headrick:2009jz,Anderson:2010ke} for numerical approaches to the problem of determining Ricci-flat, CY metrics.} for the Ricci-flat, CY metric, $g_{a{\bar b}}$, direct solutions of the Hermitian-Yang-Mills equations
\beq\label{hym_again}
g^{a {\bar b}}F_{a{\bar b}}=0
\eeq
likewise remain elusive. Progress is possible, however, thanks to the Donaldson-Uhlenbeck-Yau Theorem \cite{duy1,duy2} which states that on each \emph{poly-stable} holomorphic vector bundle $V$, there exists a unique connection, $A$, satisfying \eref{hym_again}. 

A holomorphic vector bundle $V \to X$ is called \emph{stable}, if for all subsheaves $\mathcal F \subset V$ with $rk(\mathcal F) < rk(V)$ the following inequality holds:

\begin{equation}
\mu (\mathcal F) < \mu (V) 
\end{equation}
where 
\begin{equation}
\mu(\mathcal F) = \frac{1}{rk(\mathcal F)} \int_X c_1(\mathcal F) \wedge J \wedge J
\end{equation}
where $J$ is the K\"ahler form on $X$. $V$ is semi-stable if $\mu(\mathcal F) < \mu (V)$ for all subsheaves and is poly-stable if it can be decomposed as a direct sum of stable bundles, all with the same slope: 
\beq
V = \oplus_i V_i~~~ \text{with}~~\mu (V_i) = \mu (V)~~\forall~i
\eeq
It is poly-stability which is one-to-one with a solution of \eref{hym_again}. Expanding $J$ in a basis of harmonic $\{1,1\}$ forms as $J = t^iJ_i$, the slope of any sheaf $\mathcal F$ can be written as 
\begin{equation}
\mu(\mathcal F) = \frac{1}{rk(\mathcal F)}d_{ijk}c_1^i(\mathcal F) t^j t^k = \frac{1}{rk(\mathcal F)} s_ic_1^i(\mathcal F)
\end{equation}
where $d_{ijk} = \int _X J_i \wedge J_j \wedge J_k$ are the triple intersection numbers of $X$, $c_1(\mathcal F) = c_1^i(\mathcal F) J_i$ is the first Chern class of $\mathcal F$, and $s_i$ are the so-called ``dual" K\"ahler variables defined by $s_i = d_{ijk} t^j t^k$.

For the bundles in consideration in this work, $c_1(V) = 0$ so $\mu(V) = 0$, thus $V$ is stable if for all $\mathcal F \subset V$:
\begin{equation} \label{eq:stable}
\mu(\mathcal F) = \frac{1}{rk(\mathcal F)} s_ic_1^i(\mathcal F) < 0
\end{equation}
While we have traded a difficult problem in differential geometry (i.e. \eref{hym_again}) for one in algebraic geometry with the notion of stability, there is still a substantial obstacle to be overcome in determining whether or not a bundle is stable: all possible sub-sheaves $\mathcal F \subset V$ must be characterized. To this end, we employ the tools developed in \cite{Anderson:2008ex} and we refer the interested reader there for further details. The tools from \cite{Anderson:2008ex} used in this work are briefly summarized below.

\subsection{Algorithmically testing stability}\label{sec:stability}
If $\mathcal F$ is a sub-sheaf of $V$ then it is always possible to describe $V$ itself in terms of this substructure via the following short exact sequence:
\beq\label{substruc_v}
0 \to \mathcal F \to V \to \frac{V}{\mathcal F} \to 0
\eeq
Since $V$ is a vector bundle it is torsion-free and thus, only subsheaves with rank $0 < rk(\mathcal F) < rk(V)$ need be considered. A central result of \cite{Anderson:2008ex} is that information about all possible sub-sheaves $\mathcal F$ above, can be encoded by considering only \emph{line bundle sub-sheaves of $\wedge^k V$}.

To understand this approach, a good starting point is to note is that for any rank, $k=rk(\mathcal F)$ of $\mathcal F$, the top wedge power, $\wedge^k \mathcal F$ is a rank $1$ sheaf. Moreover, the short exact sequence in \eref{substruc_v} guarantees that the inclusion $ 0 \to \mathcal F \to V$ induces an injection
\beq\label{stuff_in_v}
0 \to \wedge^k (\mathcal F) \to \wedge^k (V)
\eeq
(once again with $k<rk(V)$). Since $\wedge^k \mathcal F$ is rank $1$ and torsion-free, $(\wedge^k \mathcal F)^{\vee\vee}$ is a line bundle \cite{hartshorne} and 
\beq\label{induced_wedges}
\wedge^k \mathcal F \subset (\wedge^k \mathcal F)^{\vee\vee} \subset (\wedge^k V)^{\vee\vee} \simeq \wedge^k V
\eeq
Let us clarify the utility of this rather opaque chain of results: If there exists a de-stabilizing subsheaf, $\mathcal F \subset V$ as in \eref{substruc_v}, this \emph{induces a corresponding line bundle subsheaf}, $\mathcal L = \wedge^k \mathcal F$, s.t.
\beq
\mathcal L \subset \wedge^k V
\eeq
As a result, line bundle subsheaves of $\wedge^k V$ carry information about possibly de-stabilizing subsheaves of $V$ itself. Finally it should be noted that for any rank $n$ bundle, $c_1(\wedge^k \mathcal F) = \binom {n-1}{k-1} c_1(V) $ gives $\mu (\wedge^k V) = 0$. Thus, to prove that a rank $n$, $SU(n)$ bundle $V$ is stable, one needs only to demonstrate that for all  $\mathcal L \subset \wedge^k V$
\beq
\mu(\mathcal L) < \mu(V)=0
\eeq
for all $k<n$. Note that if a bundle passes the test above it is definitely stable. However, if a line bundle sub-sheaf $\mathcal L \subset \wedge^k$ is found with $\mu(\mathcal L) >0$ this does not guarantee that it arose from a de-stabilizing subsheaf $\mathcal F$ as in \eref{substruc_v} (i.e. it is not necessarily unstable and further analysis is required). 

This algorithmic approach to stability was employed to analyze all the bundles found in this work. We illustrate two important applications of this approach in the following Subsections.

\subsection{Negative line bundle entries in monad bundles and stability}
It was a goal of this work to build examples of bundles $V \to X$ which are in fact \emph{not stable for the entire K\"ahler cone of $X$} and instead, give rise to stability walls and cone-substructure. Moreover, since we are constrained to work within the framework of $(0,2)$ GLSMs, we must represent $V$ as a monad bundle, \eref{first_mon}. The majority of monads studies in this context have been within the class of so-called ``positive" monads in which the degrees of the line bundles in \eref{first_mon} have been chosen to be $\geq 0$. Since the condition of bundle stability is not readily visible within the GLSM itself, this class of positive monads has the added benefit that they are frequently stable everywhere in K\"ahler moduli space \cite{Anderson:2007nc,Anderson:2008uw,Anderson:2008ex,Anderson:2009mh}.

As a result of these observations, in this work we have employed more general classes of monad bundles in which $V$ is described as the kernel of a map, $F$:
\beq
0 \to V \to B \stackrel{F}{\longrightarrow} C \to 0
\eeq
and we allow for some non-ample line bundles within $B$. We will demonstrate below that generically \emph{any mixed positive/negative degree line bundle in the middle term of the monad, $B$, will induce a stability wall in $V$}. We will illustrate the possibilities here in the simplest case of $h^{1,1}(X)=2$.

To see this, note that one immediate consequence of the analysis in the previous Section is that $V$ is stable for any choice of K\"ahler moduli where $V$ is. As a result, if $V$ takes the form
\beq
0 \to V \to \cO(-a,b) \oplus B' \stackrel{F}{\longrightarrow} C \to 0
\eeq
the dual sequence is
\beq
0 \to C^\vee \to \cO(a,-b) \oplus B'^\vee \to V^\vee \to 0
\eeq
By inspection, it is clear that $\cO(a,-b)$ has the potential to be a de-stabilizing line-bundle sub-sheaf of $V^*$. It must be checked whether $Hom(\cO(a,-b),V^\vee)\simeq H^0(X, \cO(-a,b) \otimes V^\vee)$ is non-trivial. To this end, note that
\beq
0 \to \cO(-a,b) \otimes C^\vee  \to \cO \oplus \cO(-a,b)\otimes B'^\vee \to \cO(-a,b) \otimes V^\vee \to 0
\eeq
Thus, whenever 
\beq\label{build_unstable}
h^0(X,\cO(-a,b)\otimes B'^\vee)+1 >h^0(X, \cO(-a,b) \otimes C^\vee)
\eeq
$\cO(a,-b)$ injects into $V^\vee$ and de-stabilizes it whenever $\mu(\cO(a,-b) >0$. By construction the relationship in \eref{build_unstable} holds for all the examples in this work with non-trivial D-terms. By \eref{stuff_in_v} and \eref{induced_wedges} (as well as the observation that $\wedge^{N-1} V \simeq V^\vee$ for an $SU(N)$ bundle, $V$) this leads to information on a non-trivial de-stabilizing subsheaf of $V$ itself:
\beq\label{mon_subsheaf}
0 \to {\cal F} \to V \to \cO(-a,b) \to 0
\eeq
Note that this observation only gives information about one de-stabilizing subsheaf. There can be many others and a full stability analysis is necessary to determine the exact structure of the stable subcone of K\"ahler moduli space. We provide an example of such an analysis in the next Section.

\subsection{Example stability analysis}
In this section we employ the techniques described in Section \ref{sec:stability} to test whether or not a monad bundle is stable. Since stability is difficult to probe directly from the GLSM, these tools play a crucial role in verifying the validity of the theory.

Consider an $SU(3)$ vector bundle $V$ defined on the familiar co-dimension one CICY $\begin{array}{c} \mathbb P^1 \\ \mathbb P^3\end{array} \left[ \begin{array}{c} 2\\4\end{array} \right]$:

\begin{equation}\begin{aligned}&\begin{array}{|c|c| |c|c|}
\hline
x_i & \Gamma^j & \Lambda^a & p_l \\ \noalign{\hrule height 1pt}\begin{array}{cccccc}
1& 1& 0& 0& 0& 0  \\
0& 0& 1& 1& 1& 1 
\end{array}&\begin{array}{c}
 -2  \\
 -4 
\end{array}&\begin{array}{ccccc}
 -1&  1& 1& 1& 2  \\
  1& 1& 1& 1& 2
\end{array}&\begin{array}{cc}
 -3& -1 \\
 -4& -2 
\end{array}\\
\hline
\end{array}\end{aligned}\end{equation}\\
The bundle $V$ is given by a short exact sequence (SES):

\begin{equation}\label{monadeg}
0 \to V  \to \mathcal O(-1,1) \oplus \mathcal O(1,1)^{\oplus 3} \oplus \mathcal O(2,2) 
\to \mathcal O(3,4) \oplus \mathcal O(1,2) \to 0
\end{equation}
and satisfies $c_1(V)=0$ and $c_2(TX)=c_2(V)$. For simplicity we will denote this as $0 \to V \to B \stackrel{F}{\longrightarrow} C \to 0$. 

Following the arguments of Section \ref{sec:stability}, since $V$ is a rank $3$ bundle, to test its stability it is necessary to identify all line bundles $\mathcal L$ that destabilize $\wedge^k V$, where $k = 1,2$. We consider each value of $k$ in turn.

First with $k = 1$, take $\mathcal L = \mathcal O(a,b)$. Note that if $a,b<0$ then $\mathcal L$ de-stabilizes $V$ noweher in K\"ahler moduli space (since the intersection numbers $d_{rst}$ for the manifold above are positive) and thus is not of interest. Thus, we are concerned with cases in which $a,b >0$ and where $ab < 0$, (i.e. one of a and b is less than zero). If $\mathcal L$ injects into $V$, then the space of maps $H^0(\mathcal L^\vee \otimes V)$ must be non-vanishing and can be calculated from the SES:
\begin{equation}
0 \to \mathcal L^\vee \otimes V \to \mathcal L^\vee \otimes B \to \mathcal L^\vee \otimes C \to 0
\end{equation}
This sequence in turn induces the long exact sequence in cohomology
\begin{equation}
\begin{split}
0 & \to H^0(\mathcal L^\vee \otimes V) \to H^0(\mathcal L^\vee \otimes B) \to H^0(\mathcal L^\vee \otimes C) \\
& \to H^1(\mathcal L^\vee \otimes V) \to H^1(\mathcal L^\vee \otimes B) \to H^1(\mathcal L^\vee \otimes C) \\
& \to H^2(\mathcal L^\vee \otimes V) \to H^2(\mathcal L^\vee \otimes B) \to H^2(\mathcal L^\vee \otimes C) \\
& \to H^3(\mathcal L^\vee \otimes V) \to H^3(\mathcal L^\vee \otimes B) \to H^3(\mathcal L^\vee \otimes C) \to 0
\end{split}
\end{equation}
Thus, in each case the question is for which values of $a, b$ is 
\beq
ker \{ H^0(\mathcal L^\vee \otimes B) \to H^0(\mathcal L^\vee \otimes C) \}
\eeq
non-trivial? To answer this, we use the structure of line bundle cohomology on $X$ \cite{Hubsch:1992nu,Anderson:2008ex,Blumenhagen:2010pv}. For any line bundle $\cO(\alpha, \beta)$ which descends from the ambient space ${\cal A}=\mathbb{P}^1 \times \mathbb{P}^4$ we have the Koszul sequence:
\beq
0 \to \cO_{\cal A}(-2,4) \otimes \cO_{\cal A}(\alpha, \beta) \to  \cO_{\cal A}(\alpha, \beta) \to  \cO_{X}(\alpha, \beta) \to 0
\eeq
It follows that in this case
\beq
h^0(X, \cO(\alpha, \beta)) \neq 0~~\text{if and only if}~~\begin{cases}
\alpha,\beta \geq 0\\
\text{or} \\
\alpha=-1, \beta \geq 4
\end{cases}
\eeq
With these tools, the bundle in \eref{monadeg} can be analyzed. The only de-stabilizing subsheaves arise as $\mathcal L\subset V^\vee$ (i.e. the $k=2$ case) and take the form
\beq
\cO(1,-1)~~~ \text{or}~~~\cO(2,-p)~~p\geq 5
\eeq
It is clear that the ``maximally de-stabilizing" line bundle subsheaf of $V$ is exactly $\cO(1,-1)$, the one expected from the arguments of the previous sub-section. Describing the K\"ahler cone in terms of dual variables $s_t = d_{trs}t^s t^r$ the sub-cone structure of K\"ahler moduli space is given in Figure \ref{fig:sw} below.

\begin{figure}
\centering \includegraphics[width=0.6\textwidth]{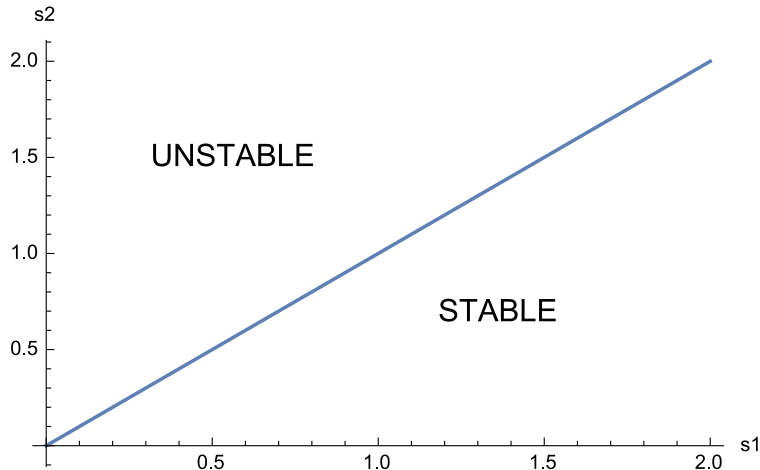} \caption{The stable and unstable regions of the K\"ahler cone associated to the $SU(3)$ bundle in \eref{monadeg}.} \label{fig:sw}
\end{figure}

\subsection{Some useful results on stable bundles}
The following notion of an ``S-Equivalence class" plays a useful role in characterization of semi-stable sheaves and the notion of moduli space (both mathematically and physically)~\cite{Huybrechts}.

\subsubsection{Theorem 1: Harder-Narasimhan}\label{HS_fil}
  \emph{Given a holomorphic bundle $V$ over a closed K\"ahler manifold
  $X$ (with K\"ahler form $\omega$), there is a filtration (called the
  Harder-Narasimhan filtration) by sub-sheaves
  \begin{equation}\label{hs_unstable}
    0=\cF_0 \subset \cF_1 \subset \ldots \cF_{m}=V
  \end{equation}  
  such that $\cF_i/\cF_{i-1}$ are semi-stable sheaves for $i=1,\ldots m$ and the slope of the quotients are ordered
  \begin{equation}
    \mu(\cF_1) > \mu(\cF_2/\cF_1)> \ldots \mu(\cF_{m}/\cF_{m-1}) \ .
  \end{equation}
  If $V$ is semi-stable, then there is a filtration by sub-sheaves (called the Jordan-H\"{o}lder filtration)
  \begin{equation}\label{filt}
    0=\cF_0 \subset \cF_1 \subset \ldots \cF_{m}=V
  \end{equation}  
  such that the quotients $\cF_i/\cF_{i-1}$ are all stable sheaves and have slope $\mu(\cF_{i}/\cF_{i-1})=\mu(V)$. In addition
  \begin{equation}\label{graded_sum}
    Gr(V)=\cF_1 \oplus \cF_{2}/\cF_{1}\oplus \ldots \cF_{m}/\cF_{m-1}
  \end{equation}
  is uniquely determined up to isomorphism (and is called the `graded sum').}

Theorem \ref{HS_fil} plays a crucial role in the notion of a moduli
space of semi-stable sheaves. Two semi-stable bundles, $V_1$ and
$V_2$, are called \emph{S-equivalent} if
$Gr(V_1)=Gr(V_2)$. A moduli space of semi-stable sheaves can only be made Hausdorff if each
point corresponds to an S-equivalence class \cite{Huybrechts}. To compare this with the physical notion of a moduli space near the stability wall, it should be noted that each S-equivalence class contains a unique poly-stable
representative, namely, the graded sum \eref{graded_sum}. Within the effective theory near the stability wall, we have seen in Section \ref{branch_in_v} that the branch structure of the effective theory is visible near the stability wall. That is, all stable bundles that can be constructed at generic points in moduli space {\it by non-trivially ``gluing" together the reducible bundle at the wall} can be characterized by their poly-stable decomposition on the wall -- that is, their S-equivalence class.

As discussed in Section \ref{branch_in_v}, the possible branches of supersymmetric vacua are determined by the charged bundle moduli appearing in $U(1)$ D-terms (see the matter content in Table \ref{table1} and the D-term constraint given in \eref{du1} and \eref{dsu6}) which in the language of Theorem \ref{HS_fil} are described by
\beq\label{graded_ext}
\langle C \rangle \sim Ext^1(\cF_i/\cF_{i-1},\cF_j/\cF_{j-1})
\eeq
for $i<j \leq m$ associated with terms $\cF_i/\cF_{i-1}$ in the graded sum $Gr(V)$ of \eref{graded_sum}.  Thus, for semi- and poly-stable bundles near a stability wall, it is important to observe that the mathematical and physical notions of a moduli space coincide. 

Finally, in exploring collections of target space geometries, the following well-known Lemma and its corollary \cite{okonek}  play a valuable role in identifying isomorphic slope-stable vector bundles described by seemingly distinct monad sequences:
\subsubsection{Morphism Lemma}\label{OSS_lemma_thm}
  \emph{Let $\phi: V_1 \to V_2$ be a non-trivial sheaf homomorphism between semi-stable bundles $V_1,V_2$. If at least one of the bundles is properly stable and $\mu(V_1)=\mu(V_2)$, then $\phi$ is a monomorphism or generically, an epimorphism.}

\subsubsection{Corollary}
  \label{OSS_corollary_thm}
\emph{Let $\phi: V_1 \to V_2$ be a nontrivial sheaf homomorphism between two semistable vector bundles $V_1,V_2$ with $rk(V_1)=rk(V_2)$ and $c_1(V_1)=c_1(V_2)$. Let at least one of the bundles be properly stable. Then $\phi$ is an isomorphism.}

\subsection{A few useful formulas for bundle and manifold topology}\label{bundle_man_top}
For ease of reference, a few standard formulas for the topology of a monad bundle are included here.

As in \eref{first_mon}, define a two-term monad bundle $V=Ker(F)$ via a short exact sequence
\bea
\nn &&0 \to V \to B \stackrel{f}{\longrightarrow} C \to 0\ ,
\mbox{ where} \\
B &=& \bigoplus_{i=1}^{r_B} \cO_X({\bf b}_i) \ , \quad
C = \bigoplus_{j=1}^{r_C} \cO_X({\bf c}_j) \ .
\label{defV}
\eea
are sums of line bundles with ranks $r_B$ and $r_C$, respectively. The rank $N$ of $V$ is easily seen, by exactness of \eref{defV}, to be
\beq
n = \rk(V) = r_B - r_C \ .
\eeq
The topology of the monad is 
\begin{align}
\nn & c_1^r(V) = \sum_{i=1}^{r_B} b^r_i - \sum_{j=1}^{r_C} c^r_j \ ,
\\
& c_{2r}(V) = \frac12  d_{rst} 
   \left(\sum_{j=1}^{r_C} c^s_j c^t_j- 
   \sum_{i=1}^{r_B} b^s_i b^t_i \right) \ , 
\label{chernV} \\
\nn& c_3(V) = \frac13 d_{rst} 
   \left(\sum_{i=1}^{r_B} b^r_i b^s_i b^t_i - \sum_{j=1}^{r_C} c^r_j
   c^s_j c^t_j \right) \ ,
\end{align}
These are equivalent to the GLSM formulae given in \eref{anomalies}. Above $d_{rst}$ is the triple intersection number on $X$, a CY 3-fold:
\beq
d_{rst}=\int_X J_r \wedge J_s \wedge J_t
\eeq
with $J_r$, $r=1, \ldots h^{1,1}$ a basis of K\"ahler forms on $X$.

Finally, we turn to the topology of a CY manifold, $X$, defined as a complete intersection manifold in a product of ordinary projective spaces. Such a manifold is determined by a \emph{configuration matrix}
\beq\label{cy-config}
\left[\ba{c|cccc}
\IP^{n_1} & q_{1}^{1} & q_{1}^{2} & \ldots & q_{1}^{K} \\
\IP^{n_2} & q_{2}^{1} & q_{2}^{2} & \ldots & q_{2}^{K} \\
\vdots & \vdots & \vdots & \ddots & \vdots \\
\IP^{n_m} & q_{m}^{1} & q_{m}^{2} & \ldots & q_{m}^{K} \\
\ea\right]_{m \times K} \ ;
\eeq
The entry $q_{i}^{r}$ denotes the degree of the $j$-th defining
polynomial in the factor $\IP^{n_r}$ and each column corresponds
to one polynomial constraint. In order that the resulting manifold be
Calabi-Yau, the condition
\beq\label{cy-deg}
\sum_{j=1}^K q_{r}^{j} = n_r + 1 \qquad \forall r=1, \ldots, m
\eeq
is imposed. In terms of this data, the topology of $X$ is given by

\beq\label{chernX}
c_1^r = 0, \
c_2^{rs} = \frac12 \left[ -\delta^{rs}(n_r + 1) + 
  \sum_{j=1}^K q^r_j q^s_j \right], \
c_3^{rst} = \frac13 \left[\delta^{rst}(n_r + 1) - 
  \sum_{j=1}^K q^r_j q^s_j q^t_j \right] \ .
\eeq

Integration over $X$ can be done with respect to a measure $\mu_{q^j_r}$ and pulled back to a simpler integration over the ambient space ${\cal A}$:
\beq\label{integration}
\int_X \cdot = \int_{\cA} \mu_{q^j_r} \wedge \cdot \ , \qquad
\mu_{q^j_r} := \wedge_{j=1}^K \left( \sum_{r=1}^m q^j_r J_r \right) \ .
\eeq

\section{Appendix B - Examples of target space dual chains}\label{appendixB}

\subsection{A complete list of target space dual theories}
In this Appendix, we provide the complete mirror chain of the bundle investigated in Section \ref{dterms}. Specifically, we apply the version of the algorithm which enhances the number of $U(1)$ symmetries (see Section \ref{tsd_extrau1}) and choose the field redefinitions such that the target space duals will remain within the class of CICYs \cite{Candelas:1987kf}. To generate the complete list of target space dual theories within this class, the generalized procedure outlined in Section \ref{repeated_line_bundles} (which allows for repeated line bundles in the monad) is applied. 

The bundle given in \eref{bundle2a} was studied in detail in Section \ref{dterms}. Once again, the GLSM data of this starting point is given by
{\footnotesize
\begin{equation}\begin{aligned}&\begin{array}{|c|c| |c|c|}
\hline
x_i & \Gamma^j & \Lambda^a & p_l \\ \noalign{\hrule height 1pt}\begin{array}{cccccc}
1& 1& 0& 0& 0& 0 \\
0& 0& 1& 1& 1& 1 
\end{array}&\begin{array}{c}
 -2  \\
 -4  
 \end{array}&\begin{array}{cccccccc}
  1& -1& 0& 0& 2& 1& 1& 2  \\
 -1&  1& 1& 1& 1& 2& 2& 2  
\end{array}&\begin{array}{ccc}
 -3& -1& -2 \\
 -2& -4& -3 
\end{array}\\
\hline
\end{array}\end{aligned}\label{bundle2a_again}\end{equation}}

As in Section \ref{repeated_line_bundles}, the complete list of repeated line bundles, leading to new target space duals, must be determined. Following the criteria of non-trivial morphisms in \eref{complicated_map} yields the following set:
\beq
\{ \cO(1,0), \cO(2,0), \cO(3,0), \cO(0,1), \cO(1,1), \cO(2,1), \cO(3,1), \cO(0,2), \cO(1,2), \cO(2,2), \cO(0,3), \cO(1,3), \cO(0,4) \}
\eeq

The complete list of dual geometries is given here, categorized by repeated entry:
\begin{enumerate}
\item No repeated entries, analyzed in Section \ref{dterms}.
\begin{equation}\begin{aligned}&\begin{array}{|c|c| |c|c|}
\hline
x_i & \Gamma^j & \Lambda^a & p_l \\ \noalign{\hrule height 1pt} \begin{array}{cccccccc}
0& 0& 0& 0& 0& 0& 1& 1 \\
1& 1& 0& 0& 0& 0& 0& 0 \\
0& 0& 1& 1& 1& 1& 0& 0 
\end{array}&\begin{array}{cc}
 -1& -1  \\
 -2&  0 \\
 -2& -2 
\end{array}&\begin{array}{cccccccc}
  0&  0&  1& 0& 0& 0& 0& 0  \\
 1& -1&  0& 0& 2& 1& 1& 2  \\
 -1&  1& -1& 1& 3& 2& 2& 2 
\end{array}&\begin{array}{ccc}
  0&  0& -1 \\
 -3& -1& -2 \\
 -2& -4& -3 
\end{array}\\
\hline
\end{array}\end{aligned}\label{bundle2a.1m}\end{equation}

\item Repeated entry $\cO(1,0)\oplus \cO(3,0)$

{\footnotesize
\begin{equation}\begin{aligned}&\begin{array}{|c|c| |c|c|}
\hline
x_i & \Gamma^j & \Lambda^a & p_l \\ \noalign{\hrule height 1pt}\begin{array}{cccccccc}
0& 0& 0& 0& 0& 0& 1& 1 \\
1& 1& 0& 0& 0& 0& 0& 0 \\
0& 0& 1& 1& 1& 1& 0& 0 
\end{array}&\begin{array}{cc}
 -1& -1 \\
 -2&  0 \\
 -2& -2  
\end{array}&\begin{array}{cccccccccc}
  0&  0& 0& 0& 0& 0& 0& 0&  1& 0 \\
  1& -1& 0& 0& 2& 1& 1& 2&  1& 3  \\
 -1&  1&  1& 1& 1& 2& 2& 2& -2& 2  
\end{array}&\begin{array}{ccccc}
 -1&  0&  0&  0&  0 \\
 -3& -1& -2& -1& -3 \\
 -2& -4& -3&  0&  0 
\end{array}\\
\hline
\end{array}\end{aligned}\label{bundle2a.2m}\end{equation}}

\item Repeated entry $\cO(1,0)$

\begin{equation}\begin{aligned}&\begin{array}{|c|c| |c|c|}
\hline
x_i & \Gamma^j & \Lambda^a & p_l \\ \noalign{\hrule height 1pt}\begin{array}{cccccccc}
0& 0& 0& 0& 0& 0& 1& 1  \\
1& 1& 0& 0& 0& 0& 0& 0 \\
0& 0& 1& 1& 1& 1& 0& 0 
\end{array}&\begin{array}{cc}
 -1& -1  \\
 -1& -1  \\
 -1& -3  
\end{array}&\begin{array}{ccccccccc}
  0&  0& 0& 0& 0&  1& 0& 0& 0  \\
  1& -1& 0& 0& 2&  0& 1& 2& 2  \\
 -1&  1& 1& 1& 1& -1& 2& 2& 3 
\end{array}&\begin{array}{cccc}
  0&  0& -1&  0 \\
 -3& -1& -2& -1 \\
 -2& -4& -3&  0 \\
\end{array}\\
\hline
\end{array}\end{aligned}\label{bundle2a.3m}\end{equation}

\item Repeated entry $\cO(0,3)$.

\begin{equation}\begin{aligned}&\begin{array}{|c|c| |c|c|}
\hline
x_i & \Gamma^j & \Lambda^a & p_l \\ \noalign{\hrule height 1pt}\begin{array}{cccccccc}
0& 0& 0& 0& 0& 0& 1& 1  \\
1& 1& 0& 0& 0& 0& 0& 0  \\
0& 0& 1& 1& 1& 1& 0& 0  
\end{array}&\begin{array}{cc}
 -1& -1  \\
 -1& -1\\
 -3& -1 
\end{array}&\begin{array}{ccccccccc}
  0&  0&  1& 0& 0& 0& 0& 0& 0  \\
  1&  -1& -1& 0& 2& 1& 1& 2& 1  \\
 -1&  1& 0& 1& 1& 2& 2& 2& 4 
\end{array}&\begin{array}{cccc}
 0& -1&  0&  0 \\
 -3& -1& -2&  0 \\
 -2& -4& -3& -3 \\
\end{array}\\
\hline
\end{array}\end{aligned}\label{bundle2a.4m}\end{equation}

\item Repeated entry $\cO(3,1)$

\begin{equation}\begin{aligned}&\begin{array}{|c|c| |c|c|}
\hline
x_i & \Gamma^j & \Lambda^a & p_l \\ \noalign{\hrule height 1pt}\begin{array}{cccccccc}
0& 0& 0& 0& 0& 0& 1& 1  \\
1& 1& 0& 0& 0& 0& 0& 0 \\
0& 0& 1& 1& 1& 1& 0& 0  
\end{array}&\begin{array}{cc}
 -1& -1  \\
 -2&  0  \\
 -3& -1 
\end{array}&\begin{array}{ccccccccc}
  1&  0& 0& 0& 0& 0& 0& 0& 0  \\
  1& -1& 0& 0& 2& 1& 1& 2& 3  \\
 -2&  1& 1& 1& 1& 2& 2& 2& 2  
\end{array}&\begin{array}{cccc}
 -1&  0&  0&  0 \\
 -3& -1& -2& -3 \\
 -2& -4& -3& -1 
\end{array}\\
\hline
\end{array}\end{aligned}\label{bundle2a.5m}\end{equation}

\item Repeated entry $\cO(1,3)$

\begin{equation}\begin{aligned}&\begin{array}{|c|c| |c|c|}
\hline
x_i & \Gamma^j & \Lambda^a & p_l \\ \noalign{\hrule height 1pt}\begin{array}{cccccccc}
0& 0& 0& 0& 0& 0& 1& 1  \\
1& 1& 0& 0& 0& 0& 0& 0  \\
0& 0& 1& 1& 1& 1& 0& 0 
\end{array}&\begin{array}{cc}
 -1& -1  \\
 -2&  0  \\
 -3& -1 
\end{array}&\begin{array}{ccccccccc}
  0&  1& 0& 0& 0& 0& 0& 0& 0 \\
  1& -1& 0& 0& 2& 1& 1& 2& 1  \\
 -1&  0&  1& 1& 1& 2& 2& 2& 4 
\end{array}&\begin{array}{cccc}
  0& -1&  0&  0 \\
 -3& -1& -2& -1 \\
 -2& -4& -3& -3 
\end{array}\\
\hline
\end{array}\end{aligned}\label{bundle2a.6m}\end{equation}

\item Repeated entry $\cO(2,0)\oplus \cO(2,0)$

{\footnotesize \begin{equation}\begin{aligned}&\begin{array}{|c|c| |c|c|}
\hline
x_i & \Gamma^j & \Lambda^a & p_l \\ \noalign{\hrule height 1pt}\begin{array}{cccccccc}
0& 0& 0& 0& 0& 0& 1& 1 \\
1& 1& 0& 0& 0& 0& 0& 0  \\
0& 0& 1& 1& 1& 1& 0& 0 
\end{array}&\begin{array}{cc}
 -1& -1 \\
 -1& -1 \\
 -2& -2 
\end{array}&\begin{array}{cccccccccc}
  0&  0& 0& 0& 0& 0& 0& 0&  1& 0 \\
  1& -1& 0& 0& 2& 1& 1& 2&  1& 3  \\
 -1&  1&  1& 1& 1& 2& 2& 2& -2& 2  
\end{array}&\begin{array}{ccccc}
 -1&  0&  0&  0&  0 \\
 -3& -1& -2& -2& -2 \\
 -2& -4& -3&  0&  0 
\end{array}\\
\hline
\end{array}\end{aligned}\label{bundle2a.7m}\end{equation}}

\item Repeated entry $\cO(1,1)\oplus \cO(1,1)$

{\footnotesize
\begin{equation}\begin{aligned}&\begin{array}{|c|c| |c|c|}
\hline
x_i & \Gamma^j & \Lambda^a & p_l \\ \noalign{\hrule height 1pt}\begin{array}{cccccccc}
0& 0& 0& 0& 0& 0& 1& 1 \\
1& 1& 0& 0& 0& 0& 0& 0 \\
0& 0& 1& 1& 1& 1& 0& 0 
\end{array}&\begin{array}{cc}
 -1& -1 \\
 -1& -1  \\
 -2& -2 
\end{array}&\begin{array}{cccccccccc}
  0&  0& 0& 0& 0& 0& 0& 0&  1& 0 \\
  1& -1& 0& 0& 2& 1& 1& 2&  0& 2  \\
 -1&  1&  1& 1& 1& 2& 2& 2& -1& 3 
\end{array}&\begin{array}{ccccc}
  0&  0& -1&  0&  0 \\
 -3& -1& -2& -1& -1 \\
 -2& -4& -3& -1& -1 
\end{array}\\
\hline
\end{array}\end{aligned}\label{bundle2a.8m}\end{equation}}

\item Repeated entry $\cO(0,2)\oplus \cO(0,2)$

{\footnotesize
\begin{equation}\begin{aligned}&\begin{array}{|c|c| |c|c|}
\hline
x_i & \Gamma^j & \Lambda^a & p_l \\ \noalign{\hrule height 1pt}\begin{array}{cccccccc}
0& 0& 0& 0& 0& 0& 1& 1 \\
1& 1& 0& 0& 0& 0& 0& 0  \\
0& 0& 1& 1& 1& 1& 0& 0 
\end{array}&\begin{array}{cc}
 -1& -1 \\
 -1& -1 \\
 -2& -2 
\end{array}&\begin{array}{cccccccccc}
  0&  0& 0& 0& 0& 0& 0& 0&  1& 0 \\
  1& -1& 0& 0& 2& 1& 1& 2&  -1& 1 \\
 -1&  1&  1& 1& 1& 2& 2& 2&  0& 4 
\end{array}&\begin{array}{ccccc}
  0& -1&  0&  0&  0 \\
 -3& -1& -2&  0&  0 \\
 -2& -4& -3& -2& -2 
\end{array}\\
\hline
\end{array}\end{aligned}\label{bundle2a.9m}\end{equation}}

\item Repeated entry $\cO(1,3)$. 
\begin{equation}\begin{aligned}&\begin{array}{|c|c| |c|c|}
\hline
x_i & \Gamma^j & \Lambda^a & p_l \\ \noalign{\hrule height 1pt}\begin{array}{cccccccc}
0& 0& 0& 0& 0& 0& 1& 1  \\
1& 1& 0& 0& 0& 0& 0& 0  \\
0& 0& 1& 1& 1& 1& 0& 0  
\end{array}&\begin{array}{cc}
 -1& -1  \\
 -1& \ -1 \\
 -2& -2 
\end{array}&\begin{array}{ccccccccc}
  0&  1&  0& 0  & 0& 0& 0& 0& 0  \\
  1&  -1& -1& 1& 2& 1& 1& 2& 1  \\
 -1& -1& 1& 3  & 1& 2& 2& 2& 3 
\end{array}&\begin{array}{cccc}
  0&  0& 0&  -1 \\
 -3& -1& -2& -1 \\
 -2& -4& -3& -3 
\end{array}\\
\hline
\end{array}\end{aligned}\label{bundle2a.10m}\end{equation}

\item Repeated entry $\cO(1,3)$. Note that this CY 3-fold is known to be equivalent to the initial $\{2,4\}$ manifold in \eref{bundle2a_again} by CICY equivalences \cite{Candelas:1987kf}. See Section \ref{isomorphic_CYs} for a discussion.

\begin{equation}\begin{aligned}&\begin{array}{|c|c| |c|c|}
\hline
x_i & \Gamma^j & \Lambda^a & p_l \\ \noalign{\hrule height 1pt}\begin{array}{cccccccc}
0& 0& 0& 0& 0& 0& 1& 1  \\
1& 1& 0& 0& 0& 0& 0& 0 \\
0& 0& 1& 1& 1& 1& 0& 0 
\end{array}&\begin{array}{cc}
 -1& -1  \\
 -1& -1  \\
 -4& \ 0 
\end{array}&\begin{array}{ccccccccc}
  1&  0& 0& 0& 0& 0& 0& 0& 0  \\
  0& -1& 0& 0& 2& 1& 1& 2& 2  \\
 -1&  1& 1& 1& 1& 2& 2& 2& 3 
\end{array}&\begin{array}{cccc}
  0&  0& -1&  0 \\
 -3& -1& -2& -1 \\
 -2& -4& -3& -3 
\end{array}\\
\hline
\end{array}\end{aligned}\label{bundle2a.11m}\end{equation}

\item Repeated entry $\cO(0,2)\oplus \cO(2,0)$
{\footnotesize
\begin{equation}\begin{aligned}&\begin{array}{|c|c| |c|c|}
\hline
x_i & \Gamma^j & \Lambda^a & p_l \\ \noalign{\hrule height 1pt}\begin{array}{cccccccc}
0& 0& 0& 0& 0& 0& 1& 1 \\
1& 1& 0& 0& 0& 0& 0& 0  \\
0& 0& 1& 1& 1& 1& 0& 0 
\end{array}&\begin{array}{cc}
 -1& -1 \\
  0& -2 \\
 -3& -1 
\end{array}&\begin{array}{cccccccccc}
  0&  0& 0& 0& 0& 0& 0& 0&  1& 0 \\
  1& -1& 0& 0& 2& 1& 1& 2&  0& 2  \\
 -1&  1& 1& 1& 1& 2& 2& 2& -1& 3 
\end{array}&\begin{array}{ccccc}
  0&  0& -1&  0&  0 \\
 -3& -1& -2& -2& 0 \\
 -2& -4& -3&  0& -2 
\end{array}\\
\hline
\end{array}\end{aligned}\label{bundle2a.12m}\end{equation}}

\item Repeated entry $\cO(1,2)$
\begin{equation}\begin{aligned}&\begin{array}{|c|c| |c|c|}
\hline
x_i & \Gamma^j & \Lambda^a & p_l \\ \noalign{\hrule height 1pt} \begin{array}{cccccccc}
0& 0& 0& 0& 0& 0& 1& 1 \\
1& 1& 0& 0& 0& 0& 0& 0 \\
0& 0& 1& 1& 1& 1& 0& 0 
\end{array}&\begin{array}{cc}
 -1& -1  \\
 0&  -2 \\
 -3& -1 
\end{array}&\begin{array}{ccccccccc}
  1&  0&  0& 0& 0& 0& 0& 0 &0 \\
 -1& 1&  0& 0& 2& 1& 1& 2 &1 \\
 -2&  2& 1& 1& 1& 2& 2& 2 &2
\end{array}&\begin{array}{cccc}
  0&  0& 0 & -1 \\
 -3& -1& -2 & -1\\
 -2& -4& -3 & -2
\end{array}\\
\hline
\end{array}\end{aligned}\label{bundle2a.13m}\end{equation}

\item Repeated entry $\cO(1,0)\oplus \cO(1,0)\oplus \cO(2,2)$
{\scriptsize
\begin{equation}\begin{aligned}&\begin{array}{|c|c| |c|c|}
\hline
x_i & \Gamma^j & \Lambda^a & p_l \\ \noalign{\hrule height 1pt} \begin{array}{cccccccc}
0& 0& 0& 0& 0& 0& 1& 1 \\
1& 1& 0& 0& 0& 0& 0& 0 \\
0& 0& 1& 1& 1& 1& 0& 0 
\end{array}&\begin{array}{cc}
 -1& -1  \\
 -1& -1 \\
 -2& -2 
\end{array}&\begin{array}{ccccccccccc}
  0&  0&  0& 0& 0& 0& 0& 0 & 0 &1 &0 \\
 1& -1&  0& 0& 2& 1& 1& 2 & 2 &0 &2\\
 -1&  1& 1& 1& 1& 2& 2& 2 & 2 & -2& 2
\end{array}&\begin{array}{cccccc}
  0&  0& 0 & -1 & 0 &0 \\
 -3& -1& -2 & -2 & -1 & -1\\
 -2& -4& -3 & -2 & 0& 0
\end{array}\\
\hline
\end{array}\end{aligned}\label{bundle2a.14m}\end{equation}}

\item Repeated entry $\cO(1,1)\oplus \cO(2,2)$
{\footnotesize
\begin{equation}\begin{aligned}&\begin{array}{|c|c| |c|c|}
\hline
x_i & \Gamma^j & \Lambda^a & p_l \\ \noalign{\hrule height 1pt} \begin{array}{cccccccc}
0& 0& 0& 0& 0& 0& 1& 1 \\
1& 1& 0& 0& 0& 0& 0& 0 \\
0& 0& 1& 1& 1& 1& 0& 0 
\end{array}&\begin{array}{cc}
 -1& -1  \\
 -1& -1 \\
 -3& -1 
\end{array}&\begin{array}{cccccccccc}
  1&  0&  0& 0& 0& 0& 0& 0 &0 &0  \\
 0& -1&  0& 0& 2& 1& 1& 2 &2 &2  \\
 -2&  1& 1& 1& 2& 2& 2& 2 &2 &2
\end{array}&\begin{array}{ccccc}
  0&  0& 0& -1 &0\\
 -3& -1& -2 &-2 & -1\\
 -2& -4& -3 &-2 & -1
\end{array}\\
\hline
\end{array}\end{aligned}\label{bundle2a.15m}\end{equation}}

\item Repeated entry $\cO(1,3)\oplus \cO(1,3)$
{\footnotesize
\begin{equation}\begin{aligned}&\begin{array}{|c|c| |c|c|}
\hline
x_i & \Gamma^j & \Lambda^a & p_l \\ \noalign{\hrule height 1pt} \begin{array}{cccccccc}
0& 0& 0& 0& 0& 0& 1& 1 \\
1& 1& 0& 0& 0& 0& 0& 0 \\
0& 0& 1& 1& 1& 1& 0& 0 
\end{array}&\begin{array}{cc}
 -1& -1  \\
 -2&  0 \\
 -2& -2 
\end{array}&\begin{array}{cccccccccc}
  0&  1&  0& 0& 0& 0& 0& 0 &0 &0  \\
 1& -1&  0& 0& 2& 1& 1& 2 &1 &1 \\
 -1&  -1& 1& 1& 1& 2& 2& 2 & 3 &3
\end{array}&\begin{array}{ccccc}
  0&  0& 0& -1&0 \\
 -3& -1& -2 & -1&-1\\
 -2& -4& -3 &-3&-1
\end{array}\\
\hline
\end{array}\end{aligned}\label{bundle2a.16m}\end{equation}}
\end{enumerate}

\subsection{Target space dual chain to a new description of the monad}
Here we provide the complete list of $10$ target space duals beginning from the bundle given in \eref{bundle2b_again}. In Section \ref{sequivalence_sec}, it was argued that this bundle shares a stability wall with the bundle given in \eref{bundle2a_again} in the previous subsection, and is in fact isomorphic to $V$ given in \eref{branch2a}, though described as a new monad. This alternative description of the same geometry leads to yet another chain of target space dual geometries. This list can contain different representations of the same geometries listed in Section \ref{branch_in_dual} as well as new pairs $(\widetilde{X},\widetilde{V})$.
To begin we consider:
\begin{equation}\begin{aligned}&\begin{array}{|c|c| |c|c|}
\hline
x_i & \Gamma^j & \Lambda^a & p_l \\ \noalign{\hrule height 1pt}\begin{array}{cccccc}
1& 1& 0& 0& 0& 0  \\
0& 0& 1& 1& 1& 1  \\
\end{array}&\begin{array}{c}
 -2  \\
 -4 
\end{array}&\begin{array}{ccccccc}
 0& 0& 0& 0&  1& 1& 1  \\
 1& 1& 1& 2& -1& 1& 2 
\end{array}&\begin{array}{cc}
 -1& -2 \\
  -4& -3 
\end{array}\\
\hline
\end{array}\end{aligned}\label{bundle2b_again2}\end{equation}
On the stability wall given by $\mu(\cO(1,-1))=0$, this bundle is equivalent to the reducible bundle given in \eref{decomp_on_wall}.

Once again, it is possible to generate a chain of inequivalent target space duals by adding repeated line bundle entries to the monad. The relevant set of line bundles is
\beq
\{\cO(1,0), \cO(2,0), \cO(1,1), \cO(2,1), \cO(0,2), \cO(1,2), \cO(2,2), \cO(0,3), \cO(1,3), \cO(0,4)\}
\eeq

The list of $10$ target space duals for which the 3-fold is a CICY in products of project of projective spaces is given below.

\begin{enumerate}
\item Repeated entry $\cO(1,0)$

\begin{equation}\begin{aligned}&\begin{array}{|c|c| |c|c|}
\hline
x_i & \Gamma^j & \Lambda^a & p_l \\ \noalign{\hrule height 1pt}\begin{array}{cccccccc}
0& 0& 0& 0& 0& 0& 1& 1 \\
1& 1& 0& 0& 0& 0& 0& 0  \\
0& 0& 1& 1& 1& 1& 0& 0 
\end{array}&\begin{array}{cc}
 -1& -1 \\
 -1& -1  \\
 -1& -3 
\end{array}&\begin{array}{cccccccc}
  0& 0& 0& 0&  0& 0&  1& 0  \\
 0& 0& 0& 0&   1& 1&  0&  2  \\
 1& 1& 1&  2&  -1& 1& -1& 3 
\end{array}&\begin{array}{ccc}
  0& -1&  0 \\
 -1& -2& -1 \\
 -4& -3&  0 
\end{array}\\
\hline
\end{array}\end{aligned}\label{bundle2b.1m}\end{equation}

\item Repeated entry $\cO(0,3)$

\begin{equation}\begin{aligned}&\begin{array}{|c|c| |c|c|}
\hline
x_i & \Gamma^j & \Lambda^a & p_l \\ \noalign{\hrule height 1pt}\begin{array}{cccccccc}
0& 0& 0& 0& 0& 0& 1& 1  \\
1& 1& 0& 0& 0& 0& 0& 0  \\
0& 0& 1& 1& 1& 1& 0& 0 
\end{array}&\begin{array}{cc}
 -1& -1  \\
 -1& -1  \\
 -3& -1 
\end{array}&\begin{array}{cccccccc}
 0& 0&  1& 0&  0& 0& 0&  0  \\
 0& 0&  -1& 0&  1& 1& 1&  1  \\
 1& 1&  0& 2&  -1& 1& 2&  4 
\end{array}&\begin{array}{ccc}
  -1&  0&  0 \\
 -1& -2&  0 \\
 -4& -3& -3 
\end{array}\\
\hline
\end{array}\end{aligned}\label{bundle2b.2m}\end{equation}

\item Repeated entry $\cO(2,0)$

\begin{equation}\begin{aligned}&\begin{array}{|c|c| |c|c|}
\hline
x_i & \Gamma^j & \Lambda^a & p_l \\ \noalign{\hrule height 1pt}\begin{array}{cccccccc}
0& 0& 0& 0& 0& 0& 1& 1  \\
1& 1& 0& 0& 0& 0& 0& 0  \\
0& 0& 1& 1& 1& 1& 0& 0 
\end{array}&\begin{array}{cc}
 -1& -1 \\
 -2&  0 \\
 -1& -3 
\end{array}&\begin{array}{cccccccc}
 0& 0& 0&  1&  0& 0& 0&  0  \\
 0& 0& 0&  0&  1& 1& 1&  2  \\
 1& 1& 1& -1& -1& 1& 2&  3 
\end{array}&\begin{array}{ccc}
    0& -1&  0 \\
  -1& -2& -2 \\
  -4& -3&  0 
\end{array}\\
\hline
\end{array}\end{aligned}\label{bundle2b.3m}\end{equation}

\item Repeated entry $\cO(1,1)$

\begin{equation}\begin{aligned}&\begin{array}{|c|c| |c|c|}
\hline
x_i & \Gamma^j & \Lambda^a & p_l \\ \noalign{\hrule height 1pt}\begin{array}{cccccccc}
0& 0& 0& 0& 0& 0& 1& 1  \\
1& 1& 0& 0& 0& 0& 0& 0  \\
0& 0& 1& 1& 1& 1& 0& 0  
\end{array}&\begin{array}{cc}
 -1& -1  \\
 -1& -1  \\
 -2& -2 
\end{array}&\begin{array}{cccccccc}
 0& 0& 0& 0&  0&  1& 0&  0  \\
 0& 0& 0& 0&  1&  0& 1&  2  \\
 1& 1& 1& 2& -1& -1& 2&  3 
\end{array}&\begin{array}{cccc}
   0& -1&  0 \\
 -1& -2& -1 \\
  -4& -3& -1 
\end{array}\\
\hline
\end{array}\end{aligned}\label{bundle2b.4m}\end{equation}

\item Repeated entry $\cO(0,2)$

\begin{equation}\begin{aligned}&\begin{array}{|c|c| |c|c|}
\hline
x_i & \Gamma^j & \Lambda^a & p_l \\ \noalign{\hrule height 1pt}\begin{array}{cccccccc}
0& 0& 0& 0& 0& 0& 1& 1  \\
1& 1& 0& 0& 0& 0& 0& 0 \\
0& 0& 1& 1& 1& 1& 0& 0 
\end{array}&\begin{array}{cc}
 -1& -1 \\
 -1& -1  \\
 -2& -2 
\end{array}&\begin{array}{cccccccc}
 0& 0& 0&  1&  0& 0& 0&  0  \\
 0& 0& 0&  -1& 1& 1& 1&  1  \\
 1& 1& 1&  0&  -1& 1& 2&  4 
\end{array}&\begin{array}{ccc}
  -1&  0&  0 \\
  -1& -2&  0 \\
  -4& -3& -2 
\end{array}\\
\hline
\end{array}\end{aligned}\label{bundle2b.5m}\end{equation}

\item Repeated entry $\cO(1,3)$

\begin{equation}\begin{aligned}&\begin{array}{|c|c| |c|c|}
\hline
x_i & \Gamma^j & \Lambda^a & p_l \\ \noalign{\hrule height 1pt}\begin{array}{cccccccc}
0& 0& 0& 0& 0& 0& 1& 1  \\
1& 1& 0& 0& 0& 0& 0& 0  \\
0& 0& 1& 1& 1& 1& 0& 0 
\end{array}&\begin{array}{cc}
 -1& -1  \\
 -1& -1  \\
 -2& -2 
\end{array}&\begin{array}{cccccccc}
  1& 0& 0& 0&  0& 0& 0&  0  \\
 -1& 1& 0& 0&  1& 1& 1&  1  \\
 -1& 3& 1& 2&  -1& 1& 2&  3 
\end{array}&\begin{array}{cccc}
    0&  0& -1 \\
  -1& -2& -1 \\
  -4& -3& -3 
\end{array}\\
\hline
\end{array}\end{aligned}\label{bundle2b.6m}\end{equation}

\item Repeated entry $\cO(2,1)$

\begin{equation}\begin{aligned}&\begin{array}{|c|c| |c|c|}
\hline
x_i & \Gamma^j & \Lambda^a & p_l \\ \noalign{\hrule height 1pt}\begin{array}{cccccccc}
0& 0& 0& 0& 0& 0& 1& 1  \\
1& 1& 0& 0& 0& 0& 0& 0  \\
0& 0& 1& 1& 1& 1& 0& 0 
\end{array}&\begin{array}{cc}
 -1& -1  \\
 -2&  0  \\
 -2& -2 
\end{array}&\begin{array}{cccccccc}
 0& 0&  1& 0&  0& 0& 0&  0  \\
 0& 0&  0& 0&  1& 1& 1&  2  \\
 1& 1& -1& 2& -1& 1& 2&  3 
\end{array}&\begin{array}{ccc}
    0& -1&  0 \\
  -1& -2& -2 \\
  -4& -3& -1 
\end{array}\\
\hline
\end{array}\end{aligned}\label{bundle2b.8m}\end{equation}

\item Repeated entry $\cO(1,3)$

\begin{equation}\begin{aligned}&\begin{array}{|c|c| |c|c|}
\hline
x_i & \Gamma^j & \Lambda^a & p_l \\ \noalign{\hrule height 1pt}\begin{array}{cccccccc}
0& 0& 0& 0& 0& 0& 1& 1  \\
1& 1& 0& 0& 0& 0& 0& 0  \\
0& 0& 1& 1& 1& 1& 0& 0 
\end{array}&\begin{array}{cc}
 -1& -1  \\
 -1& -1  \\
 -4&  0 
\end{array}&\begin{array}{cccccccc}
  0& 0& 0& 0&  1& 0& 0&  0 \\
 0& 0& 0& 0&   0& 1& 1&  2  \\
 1& 1& 1&  2&  -1& 1& 2&  3 
\end{array}&\begin{array}{ccc}
   0& -1&  0 \\
  -1& -2& -1 \\
  -4& -3& -3 
\end{array}\\
\hline
\end{array}\end{aligned}\label{bundle2b.11m}\end{equation}

\item Repeated entry $\cO(2,2)$

\begin{equation}\begin{aligned}&\begin{array}{|c|c| |c|c|}
\hline
x_i & \Gamma^j & \Lambda^a & p_l \\ \noalign{\hrule height 1pt} \begin{array}{cccccccc}
0& 0& 0& 0& 0& 0& 1& 1 \\
1& 1& 0& 0& 0& 0& 0& 0 \\
0& 0& 1& 1& 1& 1& 0& 0 
\end{array}&\begin{array}{cc}
 -1& -1  \\
 -1& -1 \\
 -3& -1 
\end{array}&\begin{array}{cccccccc}
  0&  0&  0& 0& 1& 0& 0& 0 \\
  0&  0&  0& 0& 0& 2& 1& 2 \\
  1&  1&  1& 2&-2& 2& 2& 2
\end{array}&\begin{array}{ccc}
    0& 0 & -1 \\
  -1& -2 & -2\\
  -4& -3 & -2
\end{array}\\
\hline
\end{array}\end{aligned}\label{bundle2b.12}\end{equation}

\item Repeated entry $\cO(1,0) \oplus \cO(1,0)\oplus \cO(2,2)$

{\scriptsize \begin{equation}\begin{aligned}&\begin{array}{|c|c| |c|c|}
\hline
x_i & \Gamma^j & \Lambda^a & p_l \\ \noalign{\hrule height 1pt} \begin{array}{cccccccc}
0& 0& 0& 0& 0& 0& 1& 1 \\
1& 1& 0& 0& 0& 0& 0& 0 \\
0& 0& 1& 1& 1& 1& 0& 0 
\end{array}&\begin{array}{cc}
 -1& -1  \\
 -1& -1 \\
 -2& -2 
\end{array}&\begin{array}{cccccccccc}
  0&  0&  0& 0& 0& 0& 0& 1& 0& 0 \\
  0&  0&  0& 0& 1& 1& 1 &0& 2& 2 \\
  1&  1&  1& 2&-1& 1& 2 &-2& 2& 2
\end{array}&\begin{array}{ccccc}
   0& 0 & 0 & 0 & -1 \\
 -1& -2 & -1& -1& -2\\
 -4& -3 & 0& 0& -2
\end{array}\\
\hline
\end{array}\end{aligned}\label{bundle2b.13}\end{equation}}

\end{enumerate}



\end{document}